\documentclass[a4paper,fleqn,usenatbib]{mnras}

\usepackage{etoolbox}
\makeatletter
\newcount\c@additionalboxlevel
\newcount\c@maxboxlevel
\patchcmd\@combinedblfloats{\box\@outputbox}{%
  \stepcounter{additionalboxlevel}%
  \box\@outputbox
}{}{\errmessage{\noexpand\@combinedblfloats could not be patched}}

\AtBeginShipout{%
  \ifnum\value{additionalboxlevel}>\value{maxboxlevel}%
    \typeout{Warning: maxboxlevel might be too small, increase to %
      \the\value{additionalboxlevel}%
    }%
  \fi 
  \@whilenum\value{additionalboxlevel}<\value{maxboxlevel}\do{%
    \typeout{* Additional boxing of page `\thepage'}%
    \setbox\AtBeginShipoutBox=\hbox{\copy\AtBeginShipoutBox}%
    \stepcounter{additionalboxlevel}%
  }%
  \setcounter{additionalboxlevel}{0}%
}
\makeatother
\usepackage[T1]{fontenc}
\usepackage{ae,aecompl}
\usepackage{ulem}

\usepackage{graphicx,rotating,pdflscape,geometry}	% Including figure files
\usepackage{amsmath}	% Advanced maths commands
\usepackage{amssymb}	% Extra maths symbols

\title[The Cas A dust mass from a Herschel analysis]{The dust mass in Cassiopeia A from a spatially resolved \textit{Herschel} analysis}

\author[Ilse De Looze et al.]{I. De Looze,$^{1}$\thanks{E-mail: idelooze@star.ucl.ac.uk}
M.~J. Barlow,$^{1}$
B.~M. Swinyard,$^{1,2}$\thanks{Deceased May 22nd 2015}
J. Rho,$^{3}$
H.~L. Gomez,$^{4}$
 \newauthor M. Matsuura$^{4}$ \& R. Wesson$^{1}$ 
\\
$^{1}$Dept. of Physics \& Astronomy, University College London, Gower Street, London WC1E 6BT, UK\\
$^{2}$RAL Space, Rutherford Appleton Laboratory, Chilton, Didcot, Oxfordshire, OX11 0QX, UK\\
$^{3}$SETI Institute, 189 Bernardo Ave, Mountain View, CA 94043\\
$^{4}$School of Physics \& Astronomy, Cardiff University, The Parade, Cardiff, CF24 3AA, UK\\
}
\date{Accepted 2016 November 1. Received 2016 October 21; in original form 2016 September 13.}

\pubyear{2016}

\begin{document}
\label{firstpage}
\pagerange{\pageref{firstpage}--\pageref{lastpage}}
\maketitle

\begin{abstract}
Theoretical models predict that core-collapse supernovae (CCSNe) can be efficient dust producers (0.1-1.0 M$_{\odot}$), potentially accounting for most of the dust production in the early Universe. Observational evidence for this dust production efficiency is however currently limited to only a few CCSN remnants (e.g., SN~1987A, Crab Nebula). In this paper, we revisit the dust mass produced in Cassiopeia\,A (Cas\,A), a $\sim$330-year old O-rich Galactic supernova remnant (SNR) embedded in a dense interstellar foreground and background. We present the first spatially resolved analysis of Cas\,A based on \textit{Spitzer} and \textit{Herschel} infrared and submillimetre data at a common resolution of $\sim$0.6\,$\arcmin$ for this 5\,$\arcmin$ diameter remnant following a careful removal of contaminating line emission and synchrotron radiation. We fit the dust continuum from 17 to 500\,$\mu$m with a four-component interstellar medium (ISM) and supernova (SN) dust model. We find a concentration of cold dust in the unshocked ejecta of Cas\,A and derive a mass of 0.3-0.5\,M$_{\odot}$ of silicate grains freshly produced in the SNR, with a lower limit of $\geq$0.1-0.2\,M$_{\odot}$. For a mixture of 50$\%$ of silicate-type grains and 50$\%$ of carbonaceous grains, we derive a total SN dust mass between 0.4\,M$_{\odot}$ and 0.6\,M$_{\odot}$. These dust mass estimates are higher than from most previous studies of Cas\,A and support the scenario of supernova dominated dust production at high redshifts. We furthermore derive an interstellar extinction map for the field around Cas\,A which towards Cas\,A gives average values of A$_{\text{V}}$\,=\,6-8\,mag, up to a maximum of $A_{\text{V}}$ = 15\,mag.
 \end{abstract}


%%%%%%%%%%%%%%%%%%%%%%%%%%%%%%%%%%%%%%%%%%%%%%%%%%

%%%%%%%%%%%%%%%%% BODY OF PAPER %%%%%%%%%%%%%%%%%%

\section{Introduction}
The large reservoirs of dust observed in some high redshift galaxies (e.g., \citealt{2003A&A...406L..55B,2003MNRAS.344L..74P,2014MNRAS.441.1040R,2015Natur.519..327W}) have been hypothesised to originate from dust produced by supernovae from massive stars. Some theoretical studies (e.g., \citealt{1991A&A...249..474K,2001MNRAS.325..726T}) have supported a high efficiency of dust production (0.1-1.0\,M$_{\odot}$) in core-collapse supernovae (CCSNe) which would suffice to account for the dust mass budget observed in dusty high-redshift sources \citep{2003MNRAS.343..427M,2007ApJ...662..927D}. However, the dust reservoirs ($\leq$ 10$^{-2}$\,M$_{\odot}$) that were detected at mid-IR wavelengths during the first 1000 days in a number of CCSNe remained several orders of magnitude below these theoretical predictions \citep{2006Sci...313..196S,2007ApJ...665..608M,2009ApJ...704..306K,2011MNRAS.418.1285F}. With the recent advent of far-infrared (FIR) and submillimetre (submm) observing facilities (e.g., \textit{Herschel}, ALMA), the ability to also detect the emission from colder dust in CCSN remnants opened up and resulted in the detection of dust masses on the order of 0.1-1.0\,M$_{\odot}$ \citep{2010A&A...518L.138B,2011Sci...333.1258M,2012ApJ...760...96G,2014ApJ...782L...2I,2015ApJ...800...50M} in some nearby supernova remnants (SN 1987A, Crab Nebula, Cassiopeia A). Some supernova remnants show evidence for dust formation in the supernova ejecta once the ejecta material has sufficiently cooled after expansion to allow grain growth to take place (e.g., \citealt{2016MNRAS.457.3241A}). Recent work by \citet{2014Natur.511..326G}, \citet{2015MNRAS.446.2089W} and \citet{2016MNRAS.456.1269B} suggest that the dust mass in CCSN ejecta grows in time possibly due to accretion of material onto and coagulation of grain species. Of particular interest for studies of the mechanisms responsible for dust formation is the Galactic supernova remnant Cassiopeia A (hereafter, Cas\,A), which shows evidence for dust in the shocked outer supernova ejecta as well as in the inner, un-shocked regions of the remnant \citep{2008ApJ...673..271R,2010A&A...518L.138B,2014ApJ...786...55A}. 

\begin{figure}
	\includegraphics[width=8.5cm]{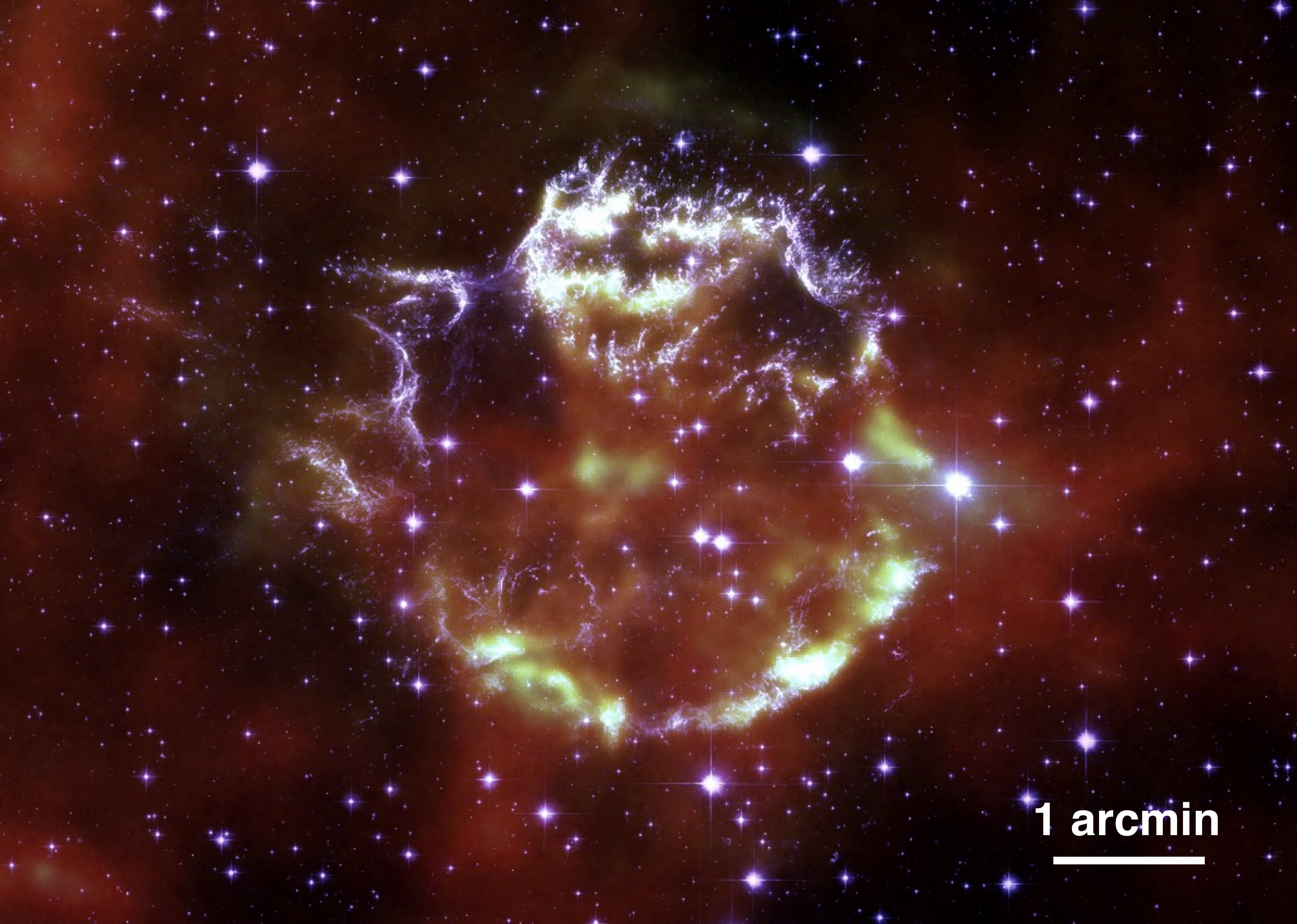}
    \caption{Cassiopeia\,A as viewed by the \textit{Herschel} PACS instrument at 70\,$\mu$m (red to green colours) and the \textit{Hubble Space Telescope} (HST) in the F625W, F775W and F850LP bands (white to purple colours).  }
    \label{CasA_composite}
\end{figure}

In this paper, we study \textit{Spitzer} and \textit{Herschel} infrared and submm maps of the supernova remnant Cas\,A on spatially resolved scales of 0.6\,pc to constrain the mass and position of formed dust species. Cas\,A (Fig.\,\ref{CasA_composite}) is the remnant of a supernova explosion of a massive progenitor about 330 years ago \citep{2006ApJ...645..283F}. Based on spectra of optical light echoes, Cas\,A was, more specifically, characterised as a hydrogen-poor Type IIb CCSN \citep{2008Sci...320.1195K}. Due to the relatively young age of the remnant, the mass of swept-up material is small compared to the mass in the supernova ejecta, which makes it still possible to separate supernova dust from any swept-up circumstellar material. Early \textit{IRAS}/\textit{ISO} studies detected 10$^{-4}$-10$^{-2}$\,M$_{\odot}$ of warm ($T_{\text{d}}$ $\sim$ 50-100 K) dust (e.g., \citealt{1987A&A...171..233B,1987ApJ...315..571D,1989ApJS...70..181A,1999ApJ...521..234A,2001A&A...369..589D}). Based on SCUBA observations at submm wavelengths, the presence of a cold ($T_{\text{d}}$ $\sim$ 15-20\,K) dust reservoir of 2-4 M$_{\odot}$ was inferred from the level of excess emission after subtraction of the non-thermal synchrotron emission component \citep{2003Natur.424..285D}. This large dust mass was, however, questioned and much of the excess submm emission was attributed to foreground interstellar dust \citep{2004Natur.432..596K}. \citet{2009MNRAS.394.1307D} interpreted the high level of polarisation at 850\,$\mu$m as due to the alignment of 1\,M$_{\odot}$ of dust with the magnetic field in the SNR. Several analyses of 3.6-160\,$\mu$m \textit{Spitzer} data of Cas\,A found warm dust masses (3$\times$10$^{-3}$\,M$_{\odot}$, \citealt{2004ApJS..154..290H}; 0.02-0.054\,M$_{\odot}$, \citealt{2008ApJ...673..271R}; $\sim$ 0.04\,M$_{\odot}$, \citealt{2014ApJ...786...55A}) significantly lower compared to the submm-derived cold dust masses. By including the \textit{Spitzer} IRS spectra in the dust spectral energy distribution (SED) modelling, \citet{2008ApJ...673..271R} and \citet{2014ApJ...786...55A} showed that the composition of warm dust grains in Cas\,A could be studied in more detail. While the spectral characteristics of most of the dust in the bright ejecta knots and X-ray emitting shocked ejecta (associated with bright [Ar~{\sc{ii}}] and [Ar~{\sc{iii}}] line features) were found to be consistent with a magnesium silicate composition (with varying relative abundance ratios of Mg and Si), a smooth spectral component associated with [Ne~{\sc{ii}}] emitting regions does not show any silicate features and was best reproduced by a Al$_{2}$O$_{3}$ (or carbonaceous) dust composition. The largest dust mass component was associated with an inner cold dust reservoir $\lesssim$0.1\,M$_{\odot}$ with unidentified dust composition \citep{2014ApJ...786...55A}. \citet{2010ApJ...719.1553S} inferred a dust mass of 0.06\,M$_{\odot}$, with an average $T_{\text{d}}$ $\sim$ 33\,K, from \textit{AKARI} and \textit{BLAST} observations covering the 50 to 500\,$\mu$m wavelength range, but their large beam size (1.3$\arcmin$, 1.6$\arcmin$, 1.9$\arcmin$ at 250, 350 and 500\,$\mu$m, respectively) hampered a clear separation of the interstellar and supernova dust material. The higher angular resolution of \textit{Herschel} (18.2$\arcsec$, 24.9$\arcsec$, 36.3$\arcsec$ at 250, 350 and 500\,$\mu$m, respectively) enabled \citet{2010A&A...518L.138B} to carry out a global fit to the insterstellar and supernova far-infrared dust emission. They derived an SN dust mass of 0.075~M$_\odot$, emitting at T$\sim$35~K, but were unable to determine whether any cooler dust was present in Cas~A due to the difficulty to distinguish between ISM and SN dust emission at longer wavelengths (160-350\,$\mu$m). The above results based on \textit{Spitzer}, \textit{BLAST} and \textit{Herschel} data were consistent with the dust evolution models of \citet{2010ApJ...713..356N} who predicted 0.08\,M$_{\odot}$ of new grain material in Cas\,A at a temperature of $\sim$40 K, of which 0.072\,M$_{\odot}$ was predicted to reside in the inner remnant regions unaffected by the reverse shock.  

In this paper, we combine \textit{Spitzer}, \textit{Herschel}, \textit{WISE} and \textit{Planck} photometric data from mid-infrared (MIR) to millimetre (mm) wavelengths and \textit{Spitzer} and \textit{Herschel} spectroscopic observations. We make a detailed study of the supernova dust emission on spatially resolved scales, which allows us to more accurately separate the intrinsic supernova dust emission from the non-thermal emission component and from the continuum emission by cold interstellar dust material. In Section \ref{Sect_data}, we present an overview of the observational datasets used for this analysis. Section \ref{ModelComp} outlines the modelling technique for the various emission components (synchrotron radiation, ISM and SN dust emission). In Section \ref{DustSED}, the multi-wavelength SED is modelled on resolved scales in order to derive the distribution of temperatures and masses of newly formed dust grains in Cas\,A. Section \ref{Discussion} presents the SN dust masses and their uncertainties resulting from the SED modelling and discusses them in light of previous results. Our main conclusions are presented in Section \ref{Conclusions}. 

In the Appendix, we provide an overview of various methods used for the analysis presented in this paper. Appendix \ref{Planck_flux} compares the different \textit{Planck} measurements for Cas\,A. Appendix \ref{Sect_line} explains how we modelled the contribution of line emission to the different broadband images. Appendix \ref{Sect_RF} discusses the photo-dissociation region (PDR) modelling technique that was used to constrain the interstellar radiation field (ISRF) along the sightline of Cas\,A. In Appendix \ref{Sect_globalfit}, we present the results of a global SED fitting analysis. Appendix \ref{Sect_modelverification} verifies the applicability of our models by directly comparing the models to observations (\ref{Discuss_comparison}) and discusses the effect of small variations in the ISRF on the SN dust masses (\ref{Discuss_IS}). In Appendix \ref{Sec_modelprediction}, we apply the model results to predict the relative contribution of line emission, synchrotron radiation, ISM and SN dust emission at IR-submm wavelengths (\ref{Sect_CasA_aftercorr}), to present a model image at 850\,$\mu$m (\ref{Discuss_850mu}), to estimate the interstellar and SN visual extinction (\ref{AVmodel}).

\section{Observational data}
\label{Sect_data}

\subsection{Herschel}
\textit{Herschel} \citep{2010A&A...518L...1P} observations of Cas\,A were obtained using Guaranteed Time (GT) contributed by the SPIRE Consortium Specialist Astronomy Group 6 (SAG-6) and the German PACS consortium, as part of the MESS GT programme (PI: M. Groenewegen, \citealt{2011A&A...526A.162G}). Table \ref{CasA_Herscheldata} gives an overview of the different \textit{Herschel} datasets, observation identification numbers (ObsIDs), observing dates and integration times for the various \textit{Herschel} instruments. 

\begin{table*}
\centering
\caption{Overview of the observation identification numbers (ObsIDs), observing dates, central coordinate positions and total observing times for the \textit{Herschel} PACS and SPIRE photometric, and PACS integral field unit (IFU) and SPIRE FTS spectroscopic observations of Cas\,A.}
\label{CasA_Herscheldata}
\begin{tabular}{|l|ccccc|} % four columns, alignment for each
\hline
Object & ObsID & Date & RA (J2000) & DEC (J2000) & Obs Time \\
  & & [y-m-d] & [$^{h}$$^{m}$$^{s}$] & [$^{\circ}$ $\arcmin$ $\arcsec$] & [$s$] \\
\hline
\multicolumn{6}{|c|}{PACS photometry} \\ 
\hline
Cas\,A & 1342188204 & 2009-12-17 & 23:23:22.72 & 58:48:53.38 & 1889 \\
Cas\,A & 1342188205 & 2009-12-17 & 23:23:23.11 & 58:48:53.01 & 1889 \\
Cas\,A & 1342188206 & 2009-12-17 & 23:23:19.01 & 58:48:51.25 & 1889 \\
Cas\,A & 1342188207 & 2009-12-17 & 23:23:20.23 & 58:48:56.86 & 1889 \\
\hline
\multicolumn{6}{|c|}{SPIRE photometry} \\ 
\hline
Cas\,A & 1342183681 & 2009-09-12 & 23:23:19.48 & 58:49:23.66 & 5005 \\
Cas\,A & 1342188182 & 2009-12-17 & 23:23:21.94 & 58:49:59.49 & 5010 \\
\hline
\multicolumn{6}{|c|}{PACS IFU spectroscopy} \\ 
\hline
Cas\,A--SP1 & 1342212249 & 2011-01-01 & 23:23:28.61 & 58:48:59.17 & 2267 \\
Cas\,A--SP1 & 1342212250 & 2011-01-01 & 23:23:28.20 & 58:49:05.10 & 1139 \\
Cas\,A--SP2 & 1342212253 & 2011-01-01 & 23:23:24.94 & 58:51:26.98 & 2267 \\
Cas\,A--SP2 & 1342212254 & 2011-01-01 & 23:23:24.50 & 58:51:33.31 & 1139 \\
Cas\,A--SP3 & 1342212257 & 2011-01-01 & 23:23:12.76 & 58:49:12.26 & 2267 \\
Cas\,A--SP3 & 1342212258 & 2011-01-01 & 23:23:13.19 & 58:49:18.37 & 1139 \\
Cas\,A--SP4 & 1342212245 & 2011-01-01 & 23:23:32.82 & 58:47:48.39 & 2267 \\
Cas\,A--SP4 & 1342212246 & 2011-01-01 & 23:23:32.41 & 58:47:54.44 & 1139 \\
Cas\,A--SP5 & 1342212251 & 2011-01-01 & 23:23:27.40 & 58:47:23.04 & 2267 \\
Cas\,A--SP5 & 1342212252 & 2011-01-01 & 23:23:26.99 & 58:47:28.89 & 1139 \\
Cas\,A--SP6 & 1342212243 & 2011-01-01 & 23:23:40.49 & 58:48:52.93 & 2267 \\
Cas\,A--SP6 & 1342212244 & 2011-01-01 & 23:23:39.61 & 58:49:05.82 & 1139 \\
Cas\,A--SP7 & 1342212247 & 2011-01-01 & 23:23:30.45 & 58:50:10.21 & 2267 \\
Cas\,A--SP7 & 1342212248 & 2011-01-01 & 23:23:30.03 & 58:50:16.42 & 1139 \\
Cas\,A--SP8 & 1342212255 & 2011-01-01 & 23:23:16.84 & 58:47:41.01 & 2267 \\
Cas\,A--SP8 & 1342212256 & 2011-01-01 & 23:23:16.43 & 58:47:46.89 & 1139 \\
Cas\,A--SP9 & 1342212259 & 2011-01-01 & 23:23:12.87 & 58:48:15.45 & 2267 \\
Cas\,A--SP9 & 1342212260 & 2011-01-01 & 23:23:12.46 & 58:48:21.36 & 1139 \\
\hline
\multicolumn{6}{|c|}{SPIRE FTS spectroscopy} \\ 
\hline
Cas\,A-centre & 1342202265 & 2010-08-08 & 23:23:29 & 58:48:54 & 3476 \\ 
Cas\,A-north & 1342204034 & 2010-08-23 & 23:23:25 & 58:50:55 & 3476 \\ 
Cas\,A-north-west & 1342204033 & 2010-08-23 & 23:23:14 & 58:49:08 & 3476 \\ 
\hline
\end{tabular}
\end{table*}

\subsubsection{PACS photometry}
The Photodetector Array Camera and Spectrometer (PACS, \citealt{2010A&A...518L...2P}) observed Cas\,A on December 17, 2009 and January 1, 2011, respectively. The PACS photometry data were obtained in parallel scan-map mode with two orthogonal scans of length 22$\arcmin$ observed at the nominal scan speed of 20 $\arcsec$ s$^{-1}$ in the blue+red and green+red filters. The total on-source integration in the blue (70\,$\mu$m) and green (100\,$\mu$m) filters was 2376s, while the integration time in the red filter (160\,$\mu$m) was 4752s. The FWHM of the PACS beam corresponded to 5.6$\arcsec$, 6.8$\arcsec$ and 11.4$\arcsec$ at 70, 100 and 160\,$\mu$m, respectively (see PACS Observers' Manual).

The PACS photometry data have been reduced with the latest \texttt{HIPE} v14.0.0 \citep{2010ASPC..434..139O} using the standard script which allows us to combine the scan and cross scans for a single field into one map. The script takes the Level 1 data from the \textit{Herschel} Science Archive (HSA), masks glitches, subtracts the baselines for separate scan legs, applies a drift correction and finally merges all scan and cross scan pairs to a final output map with default pixel sizes of 1.6$\arcsec$ for the blue and green filters, and 3.2$\arcsec$ for the red filter.

To correct for the shape of the spectrum, we apply colour corrections to the PACS maps (see the PACS calibration document PICC-ME-TN-038). With a dominant contribution of warm ($T_{\text{d}}$ $\sim$ 80\,K) supernova dust emission at PACS 70\,$\mu$m (see Table \ref{Table_Fluxfraction}), we apply a colour correction factor of 0.989. The PACS\,160\,$\mu$m emission is shown later to be dominated by emission from ISM dust irradiated by a radiation field  $G\sim$0.6$G_{\text{0}}$\footnote{$G$ corresponds to the average FUV interstellar radiation normalised to the units of the \citet{1968BAN....19..421H} field, i.e. $G_{\text{0}}$=1.6$\times$10$^{-3}$ erg s$^{-1}$ cm$^{-2}$ \citep{2005pcim.book.....T}. A normalisation to the \citet{1978ApJS...36..595D} field (indicated as $\chi_{0}$) is frequently used and is related to the \citet{1968BAN....19..421H} field as $G_{\text{0}}$ = 1.7\,$\chi_{\text{0}}$.}. The colour correction for a blackbody with temperature $T$ = 17.6\,K is 0.967 at 160\,$\mu$m. For the PACS 100\,$\mu$m band, we find an equal contribution by dust emission from ISM and SN dust. The colour correction factor for the PACS 100\,$\mu$m image is, therefore, calculated as the average (1.038) of the colour correction factors for blackbodies with temperatures of $T$ = 17.6\,K (1.069) and $T$ = 100\,K (1.007). The latter correction factors, and any factors mentioned in the remainder of this paper are multiplicative factors. The PACS maps are assumed to have a calibration uncertainty of 5\,$\%$ \citep{2014ExA....37..129B}. 

\subsubsection{PACS IFU spectroscopy}
\begin{figure}
	\includegraphics[width=8.5cm]{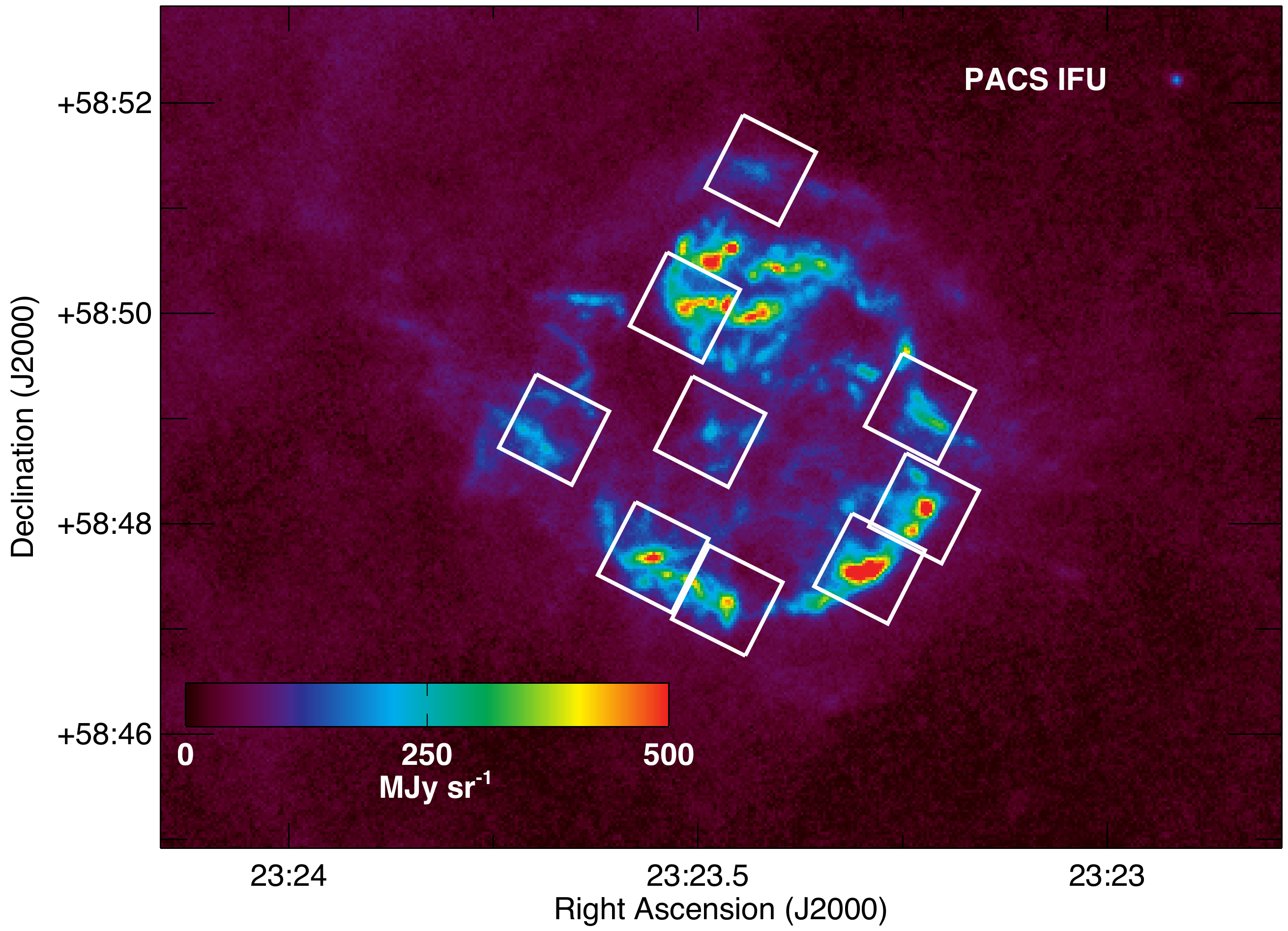}
    \caption{PACS\,70\,$\mu$m image of Cas A, overlaid with the on-source regions targeted with the PACS IFU spectrometer in the chopping mode. The off-source chopped positions were offset by 6$\arcmin$ at a PA of $\sim$240 degrees.}
    \label{CasA_AOR_PACS_IFU}
\end{figure}

PACS spectroscopy data in PACS-IFU mode (FoV $\sim$ 47$\arcsec$) were obtained at nine different positions in Cas\,A, mainly targeting the shocked and the central regions of the remnant (see Figure \ref{CasA_AOR_PACS_IFU}). Each position was observed in the wavelength ranges 51-72\,$\mu$m and 102-146\,$\mu$m (Range Mode SED B2A + Short R1) and in the ranges 70-105\,$\mu$m and 140-220\,$\mu$m (Range Mode SED B2B + Long R1). The PACS spectra were reduced to level 2 using the standard PACS chopped large range scan and SED pipeline in \texttt{HIPE} v14.0.0 using the PACS$\_$CAL$\_$32$\_$0 calibration file. The PACS IFU line measurements are assumed to have a calibration uncertainty of 13\,$\%$ and 16\,$\%$ short- and longwards of 150\,$\mu$m, respectively, based on a combination of the absolute calibration uncertainty of 12\,$\%$ and the relative uncertainty due to spaxel variations of 5\,$\%$ ($\leq$150\,$\mu$m) and 10\,$\%$ ($>$150\,$\mu$m).

\subsubsection{SPIRE photometry}

The Spectral and Photometric Imaging Receiver (SPIRE, \citealt{2010A&A...518L...3G}) observed Cas\,A on August 8th and 23rd, 2010, and on September 12th and December 17th, 2009, respectively. The SPIRE observations consisted of two orthogonal scans observed at the nominal scan speed of 20$\arcsec$ s$^{-1}$ simultaneously at each wavelength (250, 350 and 500\,$\mu$m) covering a 32$\arcmin$ $\times$ 32$\arcmin$ region centred on Cas A with an integration time of 2876s for each field. The FWHM of the SPIRE beam corresponded to 18.2$\arcsec$, 24.9$\arcsec$ and 36.3$\arcsec$ at 250, 350 and 500\,$\mu$m, respectively (see SPIRE Observers' Manual). The data were processed using HIPE version v14.0.0 using the standard pipeline for the SPIRE Large Map Mode with extended source calibration. From level 0.5 to level 1, an electrical crosstalk, temperature drift and bolometer time response correction is applied and a wavelet deglitching algorithm is run for all building blocks. To process the level 1 building blocks, we use a script to combine data from different scans. On the combined data set, we ran the destriper (instead of the baseline subtraction) to obtain an optimum fit between all timelines. The \textit{Planck} HFI maps at 857 and 545\,GHz (350 and 550\,$\mu$m) were, furthermore, used to determine the absolute scaling of the SPIRE maps with extended emission\footnote{Global fluxes for Cas\,A are 7.6$\%$ and 0.9$\%$ lower at 250\,$\mu$m and 350\,$\mu$m and 0.5$\%$ higher at 500\,$\mu$m compared to the original SPIRE flux calibration (without including the \textit{Planck} maps to determine the absolute calibration of SPIRE images).}. We applied colour correction factors to the SPIRE 250\,$\mu$m (0.9875), 350\,$\mu$m (0.9873), and 500\,$\mu$m (0.9675) maps, appropriate for a spectrum with $F_{\nu}$ $\propto$ $\nu^{-2}$. The calibration uncertainties for the SPIRE images are assumed to be 4\,$\%$, resulting from the quadratic sum of the 4\,$\%$ absolute calibration error from the assumed models used for Neptune (SPIRE Observers' manual\footnote{http://herschel.esac.esa.int/Docs/SPIRE/html/spire$\_$om.html}) and the random uncertainty of 1.5\,$\%$ on the repetitive measurements of Neptune \citep{2013MNRAS.433.3062B}.  

\subsubsection{SPIRE FTS spectroscopy}

\begin{figure}
	\includegraphics[width=8.5cm]{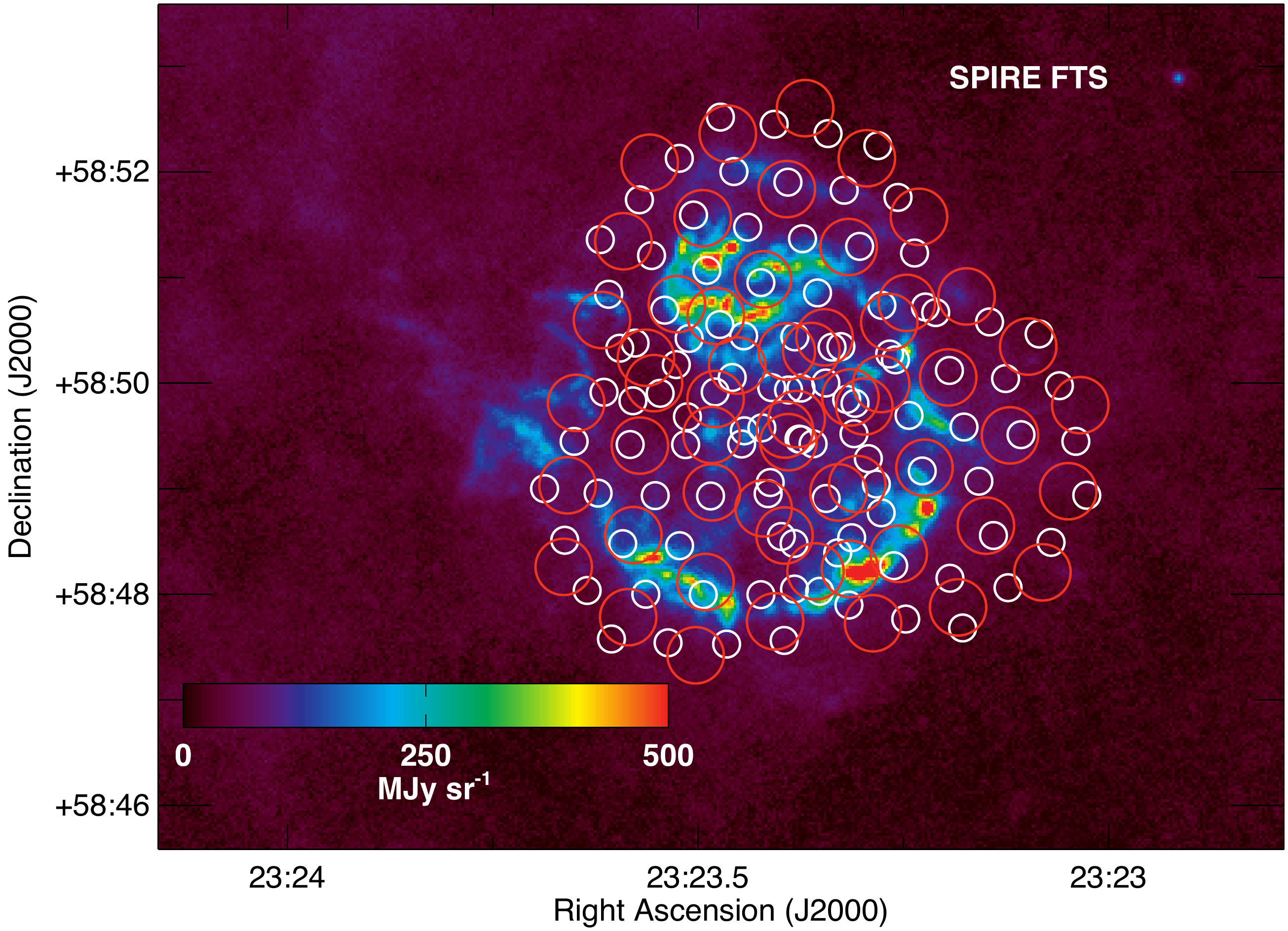}
    \caption{PACS\,70\,$\mu$m image of Cas A, overlaid with the three SPIRE FTS pointings with individual positions for the SSW and SLW detectors indicated as white and orange circles, respectively. }
    \label{CasA_AOR_SPIRE_FTS}
\end{figure}

The SPIRE Fourier Transform Spectrometer (FTS) spectra were obtained in sparse spatial sampling and high-spectral resolution mode, covering the 194-671\,$\mu$m wavelength range. Three different regions (centre, north, north-west) were targeted (see Figure \ref{CasA_AOR_SPIRE_FTS}) with the two arrays of the bolometer detectors, each with 24 repetitions. The 35 detectors of the SSW (SPIRE Short Wavelength) array covered the 194-313\,$\mu$m range, while the SLW (SPIRE Long Wavelength) array of 19 detectors covered the 303-671\,$\mu$m wavelength range. The SSW and SLW detectors have an average FWHM of 19$\arcsec$ and 34$\arcsec$, respectively \citep{2013ApOpt..52.3864M}. 

The SPIRE FTS data were reduced in \texttt{HIPE} v14.0.0, with version SPIRE$\_$CAL$\_$14$\_$3 of the calibration files including the latest corrections that match the FTS extended calibration with the SPIRE photometer. We used the standard pipeline in \texttt{HIPE} for the reduction of single pointing SPIRE spectrometer observations, with extended source calibration and without apodisation. The standard pipeline included a first and second order deglitching procedure, non-linearity and phase corrections, baseline subtraction, and corrections for the telescope and instrument emission. The spectral lines in the SPIRE FTS data were fitted with the SPIRE Spectrometer Line Fitting algorithm in \texttt{HIPE} using a sinc function to model the instrumental line shape \citep{2014SPIE.9143E..2DN}. In addition to the formal uncertainties from line fitting, we add a 10$\%$ calibration uncertainty \citep{2014MNRAS.440.3658S} to the line flux uncertainties. 

\subsection{Ancillary data}

\subsubsection{Spitzer}

The Infrared Array Camera (IRAC, \citealt{2004ApJS..154...10F}), Multi-band Imaging Photometer (MIPS, \citealt{2004ApJS..154...25R}) and Infrared Spectrograph (IRS, \citealt{2004SPIE.5487...62H}) on board the \textit{Spitzer} Space Telescope \citep{2004ApJS..154....1W} have all targeted Cas\,A. The MIPS data at 24, 70 and 160\,$\mu$m were observed as part of the \textit{Spitzer} Early Release Observation program (ID 718, PI: G. Rieke) on November 30, 2003. Cas\,A was mapped with MIPS at medium scan speed over a total area of 12.7$\arcmin$ $\times$ 30$\arcmin$. More details about the observing strategy and a detailed analysis of the MIPS data for Cas\,A are presented in \citet{2004ApJS..154..290H}. We used the IRAC data for Cas\,A observed as part of the \textit{Spitzer} program \textit{The Evolution of Dust in Cassiopeia A} (ID 3310, PI: L. Rudnick) on January 18, 2005 (see \citealt{2006ApJ...652..376E} for a description of the observations). We retrieved the IRAC 3.6 and 8\,$\mu$m, and MIPS 24\,$\mu$m final data products (with 10.4s, 10.4s and 3.67s integration time per pixel, respectively) from the \textit{Spitzer} Heritage archive\footnote{http://sha.ipac.caltech.edu/applications/Spitzer/SHA/}. Extended source correction factors of 0.91 and 0.74 have been applied to the IRAC\,3.6\,$\mu$m and IRAC\,8\,$\mu$m images, respectively, following the recommendations of the IRAC Instrument Handbook\footnote{http://irsa.ipac.caltech.edu/data/SPITZER/docs/irac/\-iracinstrumenthandbook/}. 

The IRAC\,3.6\,$\mu$m map (dominated by synchrotron emission from Cas\,A) did not require a correction for the shape of the spectrum (assumed to be a power law spectrum $F_{\nu}$ $\propto$ $\nu_{\alpha}$ with $\alpha$ anywhere between $-1$ and $0$, see Section \ref{Sect_synchr}). We did correct the IRAC 8\,$\mu$m emission (arising primarily from hot SN dust in Cas\,A) with a colour correction factor of 0.9818 assuming a blackbody spectrum with a temperature of 400\,K. The IRAC\,3.6\,$\mu$m is contaminated by many field stars, which prevents the determination of the fainter synchrotron emission at those wavelengths. We therefore manually selected sixty-four bright targets in the field of Cas\,A and replaced the emission within an aperture with radius $R$ (selected to encompass the star's emission) with random background noise. The mean value of the background and background variation has been measured within an annulus with inner radius $R$ and outer radius $2*R$\footnote{This radius was chosen to encompass all of the star's emission and ranges from values of 2 to 7$\arcsec$ across all stars.}. The flux calibration uncertainties in the IRAC\,3.6\,$\mu$m and 8\,$\mu$m, and MIPS\,24\,$\mu$m maps are assumed to be 10$\%$ (recommended by the IRAC Instrument Handbook) and 4\,$\%$ \citep{2007PASP..119..994E}, respectively. 

We also used the \textit{Spitzer} IRS spectra processed by \citet{2008ApJ...673..271R}. More details about the observations and analysis can be retrieved from their paper. The \textit{Spitzer} IRS observations were taken as part of the same \textit{Spitzer} program on January 13, 2005 during a total observing time of 11.5 hours. Nearly the entire remnant was mapped with the IRS Short-Low (SL: 5-15\,$\mu$m) and Long-Low (LL: 15-40\,$\mu$m) filters using 16$\times$360 and 4$\times$91 pointings with spectra obtained every 5$\arcsec$ and 10$\arcsec$, respectively. The spectra were reduced using the CUBISM package \citep{2003PASP..115..928K,2007PASP..119.1133S} with the standard IRS pipeline (version S12), including background subtraction and corrections for extended source emission.

We derived a global IRS spectrum for Cas\,A by extracting the emission within an aperture with radius R~=~165$\arcsec$. The SL and LL spectra were scaled to have similar emission in overlapping wavebands. The global spectrum was multiplied by a scaling factor of 1.12 to match the global photometry of the continuum emission in the IRAC\,8\,$\mu$m, WISE\,12\,$\mu$m and 22\,$\mu$m and MIPS\,24\,$\mu$m wavebands. Based on this global spectrum, we derived the continuum emission for Cas\,A at 17 and 32\,$\mu$m to constrain the SED fitting procedure in Section \ref{DustSED}. The total flux uncertainties (about 12$\%$) for the IRS spectra and the continuum fluxes were derived by combining the uncertainties in the scaling factors to match the emission in the broadband filters (11\,$\%$) and the absolute calibration uncertainty (5\,$\%$).

\subsubsection{WISE}

The Wide-field Infrared Survey Explorer (\textit{WISE}, \citealt{2010AJ....140.1868W}) observed Cas\,A in four photometric bands at 3.4, 4.6, 11.6 and 22.0\,$\mu$m with a resolution of FWHM = 6.1$\arcsec$, 6.4$\arcsec$, 6.5$\arcsec$, and 12$\arcsec$, respectively. We retrieved the \textit{WISE} 11.6 and 22.0\,$\mu$m maps for Cas\,A from the NASA/IPAC Infrared Science Archive\footnote{http://irsa.ipac.caltech.edu/frontpage/}. The \textit{WISE} images (in units of DN) were converted to Vega magnitudes using the photometric zero point magnitudes indicated in the image headers (with the MAGZP keyword). To convert the \textit{WISE} images to flux densities, we applied the zero magnitude flux densities reported in the Explanatory Supplement to the NEOWISE Data Release Products\footnote{http://wise2.ipac.caltech.edu/docs/release/neowise/}. Three different types of correction factors needed to be applied to extended sources (Jarrett et al. 2013). The first aperture correction factor corrected for the PSF profile fitting that was used for the \textit{WISE} absolute photometric calibration, which required corrections of -0.030 mag and 0.029 mag for the WISE 11.6 and 22.0\,$\mu$m bands. A second correction factor is the colour correction, which accounts for the shape of the spectrum of Cas\,A. Since the emission at mid-infrared wavelengths is mostly dominated by warm SN dust, we applied the correction factors applicable for blackbodies with temperatures of 200\,K and 100\,K to the \textit{WISE} 11.6\,$\mu$m (1.0006) and 22\,$\mu$m (1.0032) bands. The dust temperatures that dominate the emission at those wavelengths have been inferred from mid-infrared studies of Cas\,A \citep{2004ApJS..154..290H,2008ApJ...673..271R,2014ApJ...786...55A}. The last correction factor (0.92), which accounts for the calibration discrepancy between \textit{WISE} photometric standard blue stars and red galaxies, only needed to be applied to the \textit{WISE} 22\,$\mu$m band. We assume calibration uncertainties of 4.5\,$\%$ and 5.7\,$\%$ \citep{2013AJ....145....6J} for the \textit{WISE} 12 and 22\,$\mu$m filters.

\subsubsection{Planck}
\label{Planck_sec}

\textit{Planck} \citep{2011A&A...536A...1P} observed the entire sky in nine submm and mm wavebands during the mission lifetime (2009-2013). The \textit{Planck} satellite had two instruments, with the High Frequency Instrument (HFI, \citealt{2010A&A...520A...9L}) operating at 857, 545, 353, 217, 143 and 100 GHz (or wavelengths of 350, 550, 850 $\mu$m and 1.38, 2.1 and 3 mm, respectively) and the Low Frequency Instrument (LFI, \citealt{2010A&A...520A...4B}) covering the 70, 40 and 33 GHz frequencies (or wavelengths of 4.3, 6.8 and 10 mm). For this analysis, we have used the customised \textit{Planck} flux measurements from \citet{2016A&A...586A.134P} derived from aperture photometry within an aperture optimised for the angular size of Cas\,A and the beam size at every \textit{Planck} frequency (see Table \ref{Table_Planckflux}, last row). In Appendix \ref{Planck_flux}, we give an overview of the various \textit{Planck} measurements available for Cas\,A and discuss our choice for these flux measurements.

\subsection{Image preparation}
\label{ImaPrep_sec}                         
                                                                       
The background in each of the maps has been determined by measuring the backgrounds in a sufficiently large number of apertures with radius = 4$\times$FWHM at each waveband. The number of background apertures depended on the size of the background region available, and was typically about 20. The IRAC and MIPS images have a small field-of-view, while the \textit{Herschel} images are dominated by emission from interstellar material in the Perseus arm which makes it hard to find emission-free regions. We selected regions off the remnant with the final background level chosen to be the mean value from all background apertures. We have chosen to subtract these backgrounds from the images before convolution of all images to the 500\,$\mu$m resolution. 

To compare the emission of Cas\,A across all wavebands in an unbiased way, we convolved all IR/submm images to the same resolution, which was chosen to be the resolution of the SPIRE 500\,$\mu$m image (FWHM = 36.3$\arcsec$). We used the convolution kernels from \citet{2011PASP..123.1218A} for the convolution of these images to the SPIRE\,500\,$\mu$m resolution. All convolved images were rebinned to the pixel grid of the SPIRE 500\,$\mu$m map with pixel size of 14$\arcsec$ (or 0.23 pc at the adopted distance of 3.4\,kpc for Cas A, \citealt{1995ApJ...440..706R}). We have furthermore verified that a convolution with kernels produced based on the more recent \textit{Herschel} PACS and SPIRE PSF models derived from dedicated Vesta and Mars observations and published by \citet{2016A&A...591A.117B} would not affect the results published in this work. We produced such convolution kernels for the specific \textit{Herschel} observations of Cas\,A that account for the position angle of the Z-axis of the telescope during the observations of the source and target image, respectively, using the \texttt{Pypher} (Python-based PSF Homogenization kERnels production\footnote{https://pypher.readthedocs.io/en/latest/}) software package. With variations in the global fluxes of Cas\,A of less than 1$\%$ (except at 250\,$\mu$m with a 2$\%$ offset) between the two sets of convolved images, we conclude that the type of kernels used for the convolution of the \textit{Herschel} maps of Cas\,A will not affect the determination of SN dust masses in Cas\,A.       

For the SED modelling procedure described in Section \ref{DustSED}, we need to know the uncertainty on the flux in every pixel. These uncertainties determine how well the model fits the observations, and play an important role in setting the uncertainties on the contributions from different emission components. The main uncertainties are due to errors on the determination of the background levels, calibration uncertainties and uncertainties in the synchrotron subtraction. The background uncertainties are driven by both large scale background variations and pixel-by-pixel noise. The two independent background errors are computed as the standard deviation of the mean background values derived for different background regions and the mean of the standard deviation of the pixel-by-pixel variation in different background regions, respectively. The uncertainty associated with the subtraction of the synchrotron emission is calculated based on the uncertainties associated with the spectral index and normalisation factor determined from the \textit{Planck} data (see Section \ref{Sect_synchr}). 

\subsection{Flux comparisons}
\label{FluxComp_sec}

Table \ref{Table_Fluxfraction} (first column) provides an overview of the global photometric measurements for Cas\,A. Several other works have studied the infrared emission from Cas\,A, which makes it possible to compare our global photometry to other published fluxes. \citet{2010A&A...518L.138B} used the same set of \textit{Herschel} observations, but their data were reduced with an earlier version of the \texttt{HIPE} data reduction pipeline. It is, therefore, of interest to compare the flux measurements reported by \citet{2010A&A...518L.138B} to this work for the same aperture. While the PACS\,70\,$\mu$m photometry is consistent within the error bars ($F_{70\,\mu m}$=179$\pm$11 Jy here and $F_{70\,\mu m}$=169$\pm$17 Jy from \citealt{2010A&A...518L.138B}), the other PACS photometric measurements published by \citet{2010A&A...518L.138B} ($F_{100\,\mu m}$=192$\pm$19 Jy, $F_{160\,\mu m}$=166$\pm$17 Jy) are 17.7$\%$ and 29.7$\%$ lower compared to the values derived here. In the 160\,$\mu$m channel we might expect an opposite trend due to the optical field distortion that was not applied in \texttt{HIPE} 12 and all earlier versions (which caused an overestimate of the flux by 6-7$\%$). We believe the flux difference can mainly be attributed to the different data reduction techniques that were used to process the data. While the first \texttt{HIPE} versions only allowed one to reduce data with the standard map making tool \texttt{PhotProject}, the latest version of the data reduction was performed with \texttt{Scanamorphos}. While \texttt{PhotProject} applied a high-pass filtering technique to remove the 1/f noise (mainly due to data with low spatial frequencies or large scale emission features in the maps), \texttt{Scanamorphos} does not make any specific assumptions to model the low frequency noise and takes into account the redundancy of observations. \texttt{Scanamorphos} has been shown to give a better estimate of the background level in PACS maps compared to the \texttt{PhotProject} map-making algorithm which can remove or underestimate the emission of extended structures. Given that the field around Cas\,A is populated with extended emission originating from ISM dust, we believe that the discrepancy between fluxes is largely due to different estimates of the background levels.

In the SPIRE wavebands, our integrated fluxes are also higher by 8.3\,$\%$, 17.8\,$\%$ and 12.2\,$\%$ compared to the values published by \citet{2010A&A...518L.138B} ($F_{250}$=168$\pm$17 Jy, $F_{350}$=92$\pm$10 Jy, $F_{500}$=52$\pm$7 Jy), but close to being consistent within the error bars. Based on the SPIRE beam areas that were assumed during the early days of \textit{Herschel} observations (501, 944 and 1924 arcsec$^{2}$ at 250, 350 and 500\,$\mu$m, respectively, \citealt{2010A&A...518L...4S}) compared to the more recent estimates (464, 822 and 1768 arcsec$^{2}$), we would expect the latest flux densities to be lower. The opposite trend suggests that variations in the background level determination play a prominent role in explaining the different photometric measurements. 

Other than by \textit{Herschel}, Cas\,A has been observed by several other space and ground-based facilities at similar wavelengths. Based on \textit{Spitzer} MIPS data, \citet{2004ApJS..154..290H} reported a flux density $F_{\text{70}}$=107$\pm$22 Jy. Observations by \textit{AKARI} yielded flux densities of $F_{\nu}$=71$\pm$20\,Jy, 105$\pm$21\,Jy and 92$\pm$18\,Jy at 65, 90 and 140\,$\mu$m \citep{2010ApJ...719.1553S}, respectively. \citet{2010ApJ...719.1553S} also remeasured the \textit{ISO}\,170\,$\mu$m flux ($F_{\text{170}}$=101$\pm$20\,Jy), and reported \textit{BLAST} 250, 350 and 500\,$\mu$m photometry ($F_{\text{250}}$=76$\pm$16\,Jy, $F_{\text{350}}$=49$\pm$10\,Jy, $F_{\text{500}}$=42$\pm$8\,Jy). With SCUBA, \citet{2003Natur.424..285D} measured flux densities of $F_{\text{450}}$=69.8$\pm$16.1\,Jy and $F_{\text{850}}$=50.8$\pm$5.6\,Jy. While the SCUBA\,450\,$\mu$m flux is consistent with the SPIRE\,500\,$\mu$m photometry (see Table \ref{Table_Fluxfraction}), the MIPS, \textit{AKARI}, \textit{ISO} and \textit{BLAST} measurements are all significantly lower (up to a factor of 2.5) compared to the \textit{Herschel} PACS and SPIRE fluxes derived in this paper. \citet{2010A&A...518L.138B} came to a similar conclusion and attributed the flux discrepancies to the better resolution in the \textit{Herschel} maps which allows a more accurate background subtraction. Due to the highly structured emission of the ISM material in the surroundings of Cas\,A, it can become difficult to determine the background level in an image with poor spatial resolution, due to ISM dust confusion. While the high resolution of \textit{Herschel} images allows us to better resolve the ISM emission from the true background in the IR/submm images, resulting in the recovery of higher flux densities for Cas\,A, the main difference of our work compared to previous studies, analysing the same set of \textit{Herschel} observations, is the detailed modelling on spatially resolved scales of the different emission components contributing along the sight line towards Cas\,A. In support of this claim, we repeated the modelling based on the \textit{Herschel} fluxes published by \citet{2010A&A...518L.138B} and retrieved a total SN dust mass higher by a factor of three compared to the global analysis of the \textit{Herschel} data of Cas\,A by \citet{2010A&A...518L.138B} (see Section \ref{CompareBefore.sec}).

\begin{table*}
	\centering
			\caption{Overview of the aperture photometry results for Cas\,A in the IRAC\,8\,$\mu$m, \textit{WISE} 12 and 22\,$\mu$m, IRS\,17 and 32\,$\mu$m continuum, MIPS\,24\,$\mu$m, PACS\,70, 100 and 160\,$\mu$m, SPIRE\,250, 350 and 500\,$\mu$m, and SCUBA\,850\,$\mu$m wavebands. Flux densities, $F_{\nu}$, in units of Jy have been measured within an aperture with radius R=165$\arcsec$ centred on the position (RA, DEC)~=~(350.86311$^{\circ}$,58.813292$^{\circ}$). The second column lists the total flux measured within this aperture, while the third and fourth column list the flux attributed to line and synchrotron emission, respectively. Columns five to ten list the ISM and SN dust emission in every waveband for an ISM dust model with ISRF scaling factors $G$\,=\,0.3\,$G_{\text{0}}$, 0.6\,$G_{\text{0}}$ and 1.0\,$G_{\text{0}}$, respectively, based on the best fitting four-component SED model (one interstellar and three SN dust components). The values in parentheses represent the contributions of the different emission components relative to the total flux. }
	\begin{tabular}{|l|c|c|c|cc|cc|cc|} % four columns, alignment for each
		\hline
                Waveband & Total & Line  & Synchrotron  & ISM dust & SN dust & ISM dust & SN dust  & ISM dust & SN dust \\
                 & $F_{\nu}$ [Jy] & $F_{\nu}$ [Jy] &$F_{\nu}$ [Jy] & (G=0.3$G_{\text{0}}$) & (G=0.3$G_{\text{0}}$) & (G=0.6$G_{\text{0}}$) & (G=0.6$G_{\text{0}}$) &  (G=1.0$G_{\text{0}}$) & (G=1.0$G_{\text{0}}$) \\
		\hline 
               IRAC\,8\,$\mu$m               & 11.3$\pm$1.1     & 4.7$\pm$0.5 & 1.6$\pm$0.6 & 3.8$\pm$0.4 & 0.2$\pm$0.0 & 6.5$\pm$0.4 & \textbf{0.2$\pm$0.0} & 9.1$\pm$0.4 & 0.1$\pm$0.0 \\               
                & & [41.6$\%$] & [14.2$\%$] & [33.6$\%$] & [1.8$\%$] & [57.5$\%$] & [\textbf{1.8$\%$}] & [80.5$\%$] & [0.9$\%$] \\
               \textit{WISE}\,12\,$\mu$m  & 20.0$\pm$0.8     & 3.4$\pm$0.4  & 2.1$\pm$0.7 & 2.4$\pm$0.2 & 3.5$\pm$0.3 & 4.0$\pm$0.3 & \textbf{3.4$\pm$0.3} & 5.7$\pm$0.3 & 2.9$\pm$0.3 \\  
                & & [17.0$\%$] & [10.5$\%$] & [12.0$\%$] & [17.5$\%$] & [20.0$\%$] & [\textbf{17.0$\%$}] & [28.5$\%$] & [14.5$\%$] \\
	       IRS\,17                               & 68.6$\pm$8.0    &   -                   & 2.4$\pm$0.9 & 1.5$\pm$0.1 & 63.5$\pm$5.4 & 2.6$\pm$0.2 & \textbf{63.3$\pm$6.0} & 3.6$\pm$0.2 & 60.1$\pm$6.0 \\  
                & & & [3.5$\%$] &  [2.2$\%$] & [92.6$\%$] & [3.8$\%$] & [\textbf{92.3$\%$}] & [5.2$\%$] & [87.6$\%$] \\
               \textit{WISE}\,22                 & 208.1$\pm$11.6 & 2.8$\pm$0.3 & 3.0$\pm$1.1 & 2.0$\pm$0.2 & 204.1$\pm$17.6 & 3.4$\pm$0.2 & \textbf{202.0$\pm$19.3} & 4.8$\pm$0.2 & 202.5$\pm$20.8 \\  
                & & [1.3$\%$] & [1.4$\%$] &  [1.0$\%$] & [98.1$\%$] & [1.6$\%$] & [\textbf{97.1$\%$}] & [2.3$\%$] & [97.3$\%$] \\
               MIPS\,24                            & 205.6$\pm$6.3   & 45.5$\pm$5.5 & 3.1$\pm$1.1 & 2.2$\pm$0.2 & 155.2$\pm$13.6 & 3.8$\pm$0.2 & \textbf{153.4$\pm$15.0} & 5.3$\pm$0.2 & 151.6$\pm$16.7 \\  
                & & [22.1$\%$] & [1.5$\%$]  &  [1.1$\%$] & [75.5$\%$] & [1.8$\%$] & [\textbf{74.6$\%$}] & [2.6$\%$] & [73.7$\%$] \\
	       IRS\,32                              & 179.2$\pm$21.1 & -                     & 3.6$\pm$1.2 & 2.9$\pm$0.3 & 174.2$\pm$15.7 & 5.0$\pm$0.3 & \textbf{168.5$\pm$17.3} & 7.0$\pm$0.3 & 166.2$\pm$21.3 \\  
                & & & [2.0$\%$] & [1.6$\%$] & [97.2$\%$] & [2.8$\%$] & [\textbf{94.0$\%$}] & [3.9$\%$] & [92.7$\%$] \\
               PACS\,70                           & 178.8$\pm$10.7 & -                     & 6.5$\pm$2.0 & 8.9$\pm$0.9 & 160.9$\pm$16.6 & 21.8$\pm$1.4 & \textbf{149.5$\pm$20.1} & 40.5$\pm$1.9 & 130.4$\pm$26.3 \\  
                & & & [3.6$\%$] &  [5.0$\%$] & [90.0$\%$] & [12.2$\%$] & [\textbf{83.6$\%$}] & [22.7$\%$] & [72.9$\%$] \\
               PACS\,100                         & 233.4$\pm$15.3 & 16.0               & 8.1$\pm$2.5 & 36.0$\pm$3.6 & 173.4$\pm$19.9 & 80.2$\pm$5.3 & \textbf{125.8$\pm$19.9} & 130.7$\pm$6.2 & 73.5$\pm$17.6 \\  
                & & [6.9$\%$] & [3.5$\%$] & [15.4$\%$] & [74.3$\%$] & [34.4$\%$] & [\textbf{53.9$\%$}] & [56.0$\%$] & [31.5$\%$] \\
               PACS\,160                         & 236.2$\pm$19.2  & -                     & 11.0$\pm$3.3 & 99.9$\pm$9.9 & 132.4$\pm$15.9 & 165.9$\pm$10.9 & \textbf{69.9$\pm$12.0} & 219.6$\pm$10.4 & 25.2$\pm$6.9 \\  
                & & & [4.7$\%$] & [42.3$\%$] & [56.1$\%$] & [70.2$\%$] & [\textbf{29.6$\%$}] & [93.0$\%$] & [10.7$\%$] \\    
               SPIRE\,250                       & 183.2$\pm$10.0   & -                    & 14.5$\pm$4.2 & 102.2$\pm$10.1 & 61.5$\pm$7.5 & 139.9$\pm$9.2 & \textbf{27.3$\pm$4.8} & 162.8$\pm$7.7 & 7.5$\pm$2.1 \\  
                & & & [7.9$\%$] &  [55.8$\%$] & [33.6$\%$] & [76.4$\%$] & [\textbf{14.9$\%$}] & [89.0$\%$] & [4.1$\%$] \\
               SPIRE\,350                       & 111.9$\pm$5.4      &  -                   & 18.2$\pm$5.0 & 69.6$\pm$6.9 & 26.6$\pm$3.3 & 86.8$\pm$5.7 & \textbf{10.9$\pm$1.9} & 94.9$\pm$4.5 & 2.6$\pm$0.8 \\  
                & & & [16.3$\%$] & [62.2$\%$] & [23.8$\%$] & [77.6$\%$] & [\textbf{9.7$\%$}] & [84.8$\%$] & [2.3$\%$] \\
               SPIRE\,500                       & 59.2$\pm$2.2        & -                    & 23.5$\pm$2.3 & 31.1$\pm$3.1 & 6.9$\pm$0.8 & 35.1$\pm$2.3 & \textbf{2.6$\pm$0.5} & 36.0$\pm$1.7 & 0.5$\pm$0.2 \\  
                & & & [39.7$\%$]  &  [52.5$\%$] & [11.7$\%$] & [59.3$\%$] & [\textbf{4.4$\%$}] & [60.8$\%$] & [0.8$\%$] \\
               SCUBA\,850                      & 50.8$\pm$5.6          & -                   & 32.3$\pm$8.1 & 8.1$\pm$0.8 & 1.1$\pm$0.1 & 8.6$\pm$5.7 & \textbf{0.4$\pm$0.1} & 8.5$\pm$0.4 & 0.1$\pm$0.0 \\  
                & & &  [63.6$\%$]  & [15.9$\%$] & [2.2$\%$] & [16.9$\%$] & [\textbf{0.8$\%$}] & [16.7$\%$] & [0.2$\%$] \\
		\hline
	\end{tabular}
		\label{Table_Fluxfraction}
\end{table*}

\section{Modelling the various emission components}
\label{ModelComp}

\subsection{Modelling the synchrotron component}
\label{Sect_synchr}
%Figure 3
\begin{figure}
	\includegraphics[width=8.5cm]{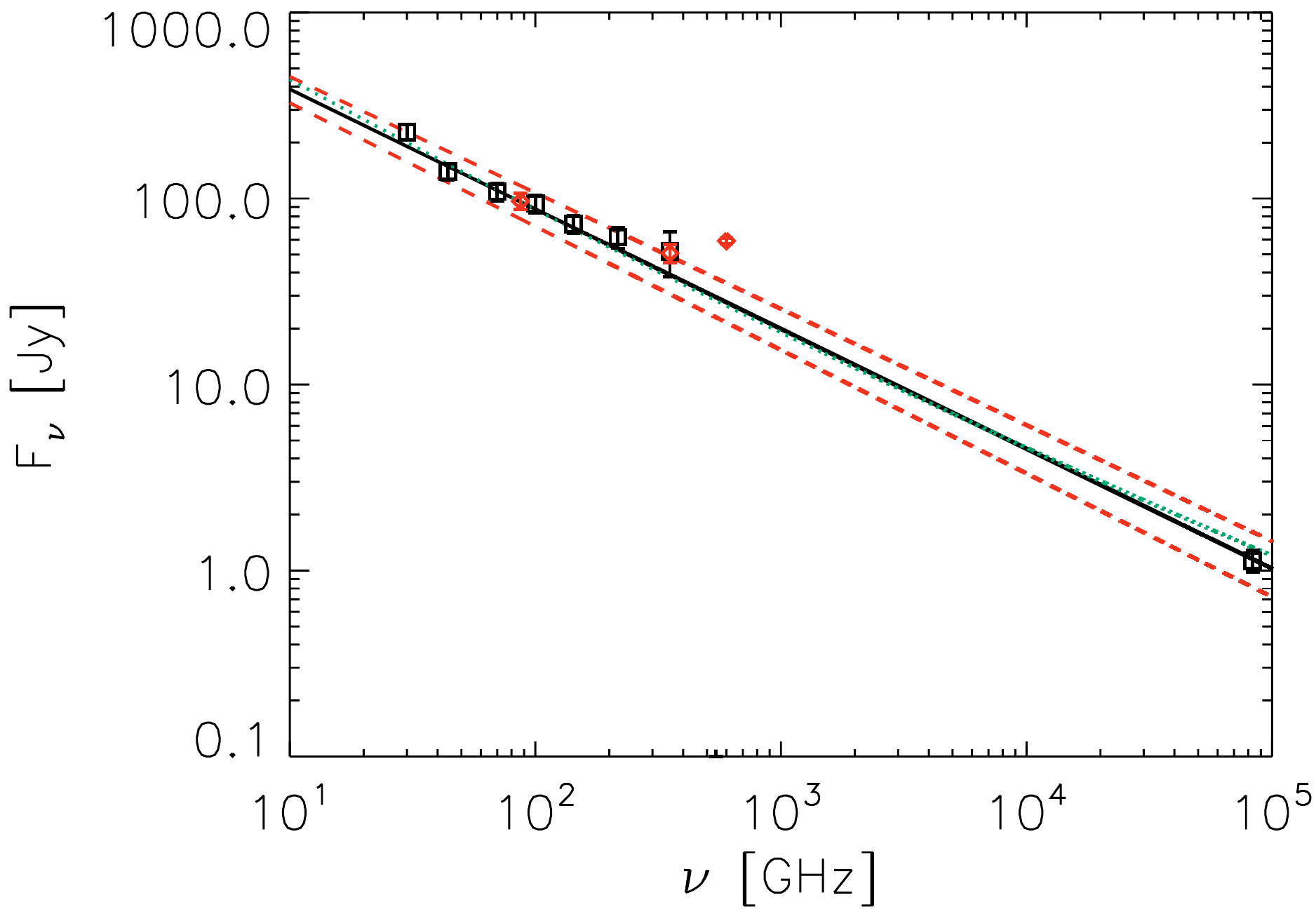}
    \caption{Global flux measurements for Cas A based on \textit{Planck} data \citep{2016A&A...586A.134P} and \textit{Spitzer} IRAC\,3.6\,$\mu$m observations (black squares). \textit{Planck} fluxes at frequencies 44\,GHz$\leq$$\nu$$\leq$143\,GHz and the IRAC\,3.6\,$\mu$m flux are used to constrain the spectral index and normalisation of the synchrotron spectrum, while the \textit{Planck} measurements at $\nu$$\gtrsim$217\,GHz already have a non-negligible contribution of interstellar and/or supernova dust emission. Previous observations of Cas\,A have been indicated as red diamonds for comparison (see Appendix \ref{Planck_flux}). The power law that best describes the synchrotron power spectrum at lower frequencies is indicated as a solid black line. The red dashed lines indicate the upper and lower limits to the synchrotron spectrum, determined from the uncertainties on the spectral index fitting. The green dashed curve shows an extrapolation of the synchrotron spectrum with spectral curvature from \citet{2015ApJ...805..119O} which greatly overestimates the IRAC\,3.6\,$\mu$m flux. }
    \label{Fig_Planck_synchr}
\end{figure}

Based on the latest \textit{Planck} measurements (see Section \ref{Planck_sec}), we determined the best fitting spectral index and normalisation factor that reproduces the synchrotron emission from Cas\,A detected by \textit{Planck} from 143 to 44 GHz (or 2 to 7 mm). We refrained from using bands at higher frequencies ($\nu$ $\geq$ 217 GHz) due to a possible contribution of the emission and polarisation from supernova dust \citep{2003Natur.424..285D,2009MNRAS.394.1307D} or emission from ISM dust (Krause et al. 2004). We also neglect the 30\,GHz flux measurement since the 30\,GHz flux for Cas\,A and for other SNRs reported by \citet{2016A&A...586A.134P} shows an offset from the other \textit{Planck} measurements. In addition to the four \textit{Planck} measurements, we include the IRAC\,3.6\,$\mu$m flux measured within an aperture of radius R=165$\arcsec$ (after masking the stars in the IRAC\,3.6\,$\mu$m image) to constrain the synchrotron spectrum at shorter wavelengths.

Using the \citet{2016A&A...586A.134P} aperture flux measurements at 44, 70, 100 and 143\,GHz, a spectral index $\alpha$ and normalisation factor C were derived by fitting a function F(C,$\alpha$) = C $\times$ $\nu^{\alpha}$ ($\nu$ in GHz) with spectral index $\alpha$ and scaling factor $C$ to these \textit{Planck} and IRAC\,3.6\,$\mu$m fluxes. Relying on the Levenberg-Marquardt least-square fitting minimisation routine MPFIT in IDL, we derived a best fitting spectral index $\alpha$ = -0.644$\pm$0.020 and normalisation factor C=1706.7$\pm$199.2\,Jy. Figure \ref{Fig_Planck_synchr} shows the \textit{Planck} fluxes reported by \citet{2016A&A...586A.134P} indicated as black squares, with the red diamonds corresponding to flux measurements for Cas\,A from other facilities. The best fitting power law is indicated as a solid black line. The data point at the highest \textit{Planck} frequency is significantly higher compared to the extrapolation of this power law, due to an increasing contribution of thermal dust emission towards higher frequencies. The dashed red lines correspond to the lower and upper limits of this synchrotron spectrum calculated based on the uncertainties of the spectral index fitting results. These uncertainties on the synchrotron spectrum are considered in the synchrotron models at every photometric wavelength.

Although a spectral index of $\alpha$=-0.644 provides a good fit to the global synchrotron power spectrum in Cas A, local variations in spectral index between different knots have been observed \citep{1996ApJ...456..234A,1999ApJ...518..284W,2014ApJ...785....7D}. The shocked ejecta with the brightest synchrotron emission are mostly consistent with this spectral index derived on global scales, while $\alpha$ variations are observed in some knots outside of the reverse shock \citep{2014ApJ...785....7D}. A spectral index of $\alpha$=-0.644 is also somewhat shallower compared to the values used in several other studies of Cas\,A: $\alpha$ = -0.69 \citep{2004ApJS..154..290H}, $\alpha$ = -0.70 \citep{2010A&A...518L.138B}, $\alpha$ = -0.71 \citep{2003ApJ...592..299R,2008ApJ...673..271R,1999ApJ...521..234A}, which were based on spectral index measurements at lower frequencies. A flattening of the synchrotron spectrum was suggested by \citet{2015ApJ...805..119O} and attributed to non-linear particle acceleration. The best fitting synchrotron spectrum including spectral curvature ($S_{\nu}$=$S_{\text{1GHz}}$ $\nu^{-\alpha+a\log \nu}$ $e^{-\tau_{0}\nu^{-2.1}}$ with $\alpha$=0.760, $a$=0.020, $\tau_{0}$=8.559$\times$10$^{-5}$, see green dotted curve in Figure \ref{Fig_Planck_synchr}) reported by \citet{2015ApJ...805..119O} (where the scaling relations of \citealt{2014ARep...58..626V} were used to account for secular fading) is however consistent with the best-fitting synchrotron spectrum derived for Cas\,A based on \textit{Planck} data in this work.

\begin{figure}
	\includegraphics[width=8.5cm]{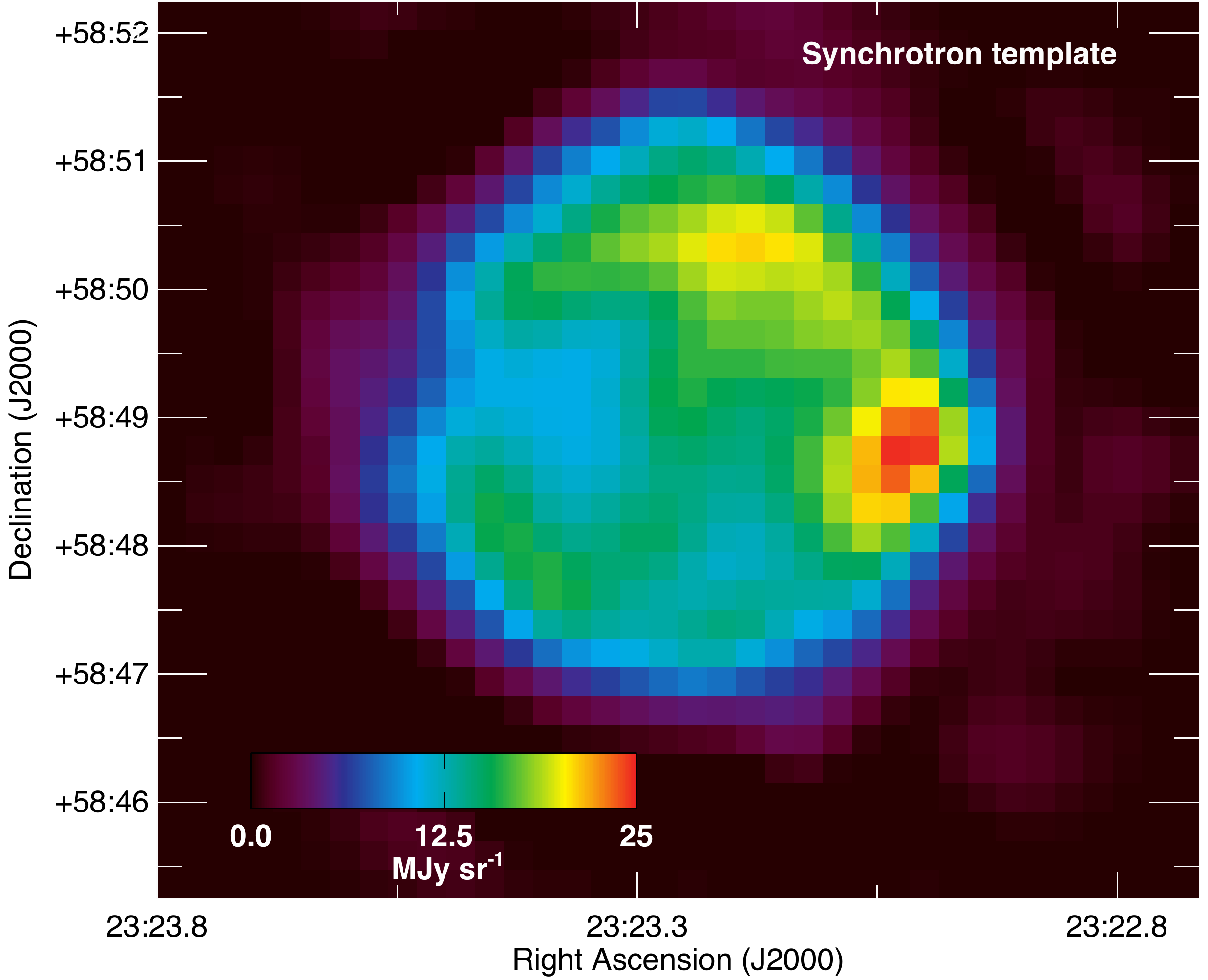}
    \caption{The synchrotron image for the SPIRE\,500\,$\mu$m waveband based on an extrapolation of the 3.7\,mm map of \citet{1999ApJ...518..284W} using the best fitting spectral index and normalisation to the \textit{Planck} fluxes (see Section \ref{Sect_synchr}).}
    \label{Ima_CasA_synchrotron}
\end{figure}

The BIMA array map of Cas\,A at 3.7 mm from \citet{1999ApJ...518..284W} was used to determine the spatial distribution of synchrotron emission on subarcsec scales at IR/submm wavelengths by extrapolation (see Figure \ref{Ima_CasA_synchrotron}). The 3.7 mm BIMA map is a multifrequency synthesis (MFS) mosaic obtained by combining 16 frequency bands between 77 and 85 GHz data to obtain a single map with mean frequency of 83.1 GHz and final resolution of 6.5$\arcsec$ $\times$ 6.2$\arcsec$. The original flux in the BIMA 3.7 mm map was determined by summing the emission within a 165$\arcsec$ radius aperture and a background annulus between 165$\arcsec$ and 320$\arcsec$. The flux calibration of the original 3.7 mm data was updated to $F_{\text{83.1 GHz}}$ =  100.5$^{+21.9}_{-19.1}$ Jy\footnote{The flux uncertainty accounts for the errors on the best fitting parameters, i.e., the spectral index $\alpha$ and normalisation factor $C$.} (or F$_{\nu}$ = C $\times$ 83.1$^{\alpha}$) using the scaling factor C (1706.68 $\pm$ 199.20 Jy) and spectral index $\alpha$ (-0.644 $\pm$ 0.020) that best reproduces the spectrum of the synchrotron emission in Cas\,A as determined based on \textit{Planck} and IRAC\,3.6\,$\mu$m data. 

While a power law slope of -0.644 is assumed for the current synchrotron model based on fits to the 44 to 143\,GHz \textit{Planck} and IRAC\,3.6\,$\mu$m data (see Section \ref{Sect_synchr}), the synchrotron spectrum slope may not remain constant across the millimeter to near-infrared frequency range. A power law fit restricted to the \textit{Planck} data would result in a power law index of -0.54. If we extrapolate this synchrotron spectrum to submm wavelengths, we find synchrotron contribution which are higher by 3$\%$, 5$\%$ and 9$\%$ of the total flux at SPIRE\,250, 350 and 500\,$\mu$m wavelengths compared to the synchrotron model presented here. Rerunning the same sets of SED models as presented in Section \ref{DustSED} using the updated synchrotron model resulted in minor changes to the SN dust masses and remained within the limits of uncertainty. The effect of variations in the spectral index on the synchrotron contribution at longer wavelengths is thus negligible compared to the model and observational uncertainties, and small variations in the spectral index will not affect the results presented in this paper.

\subsection{The interstellar dust model}
\label{Sect_ISdust}

\begin{figure}
	\includegraphics[width=8.5cm]{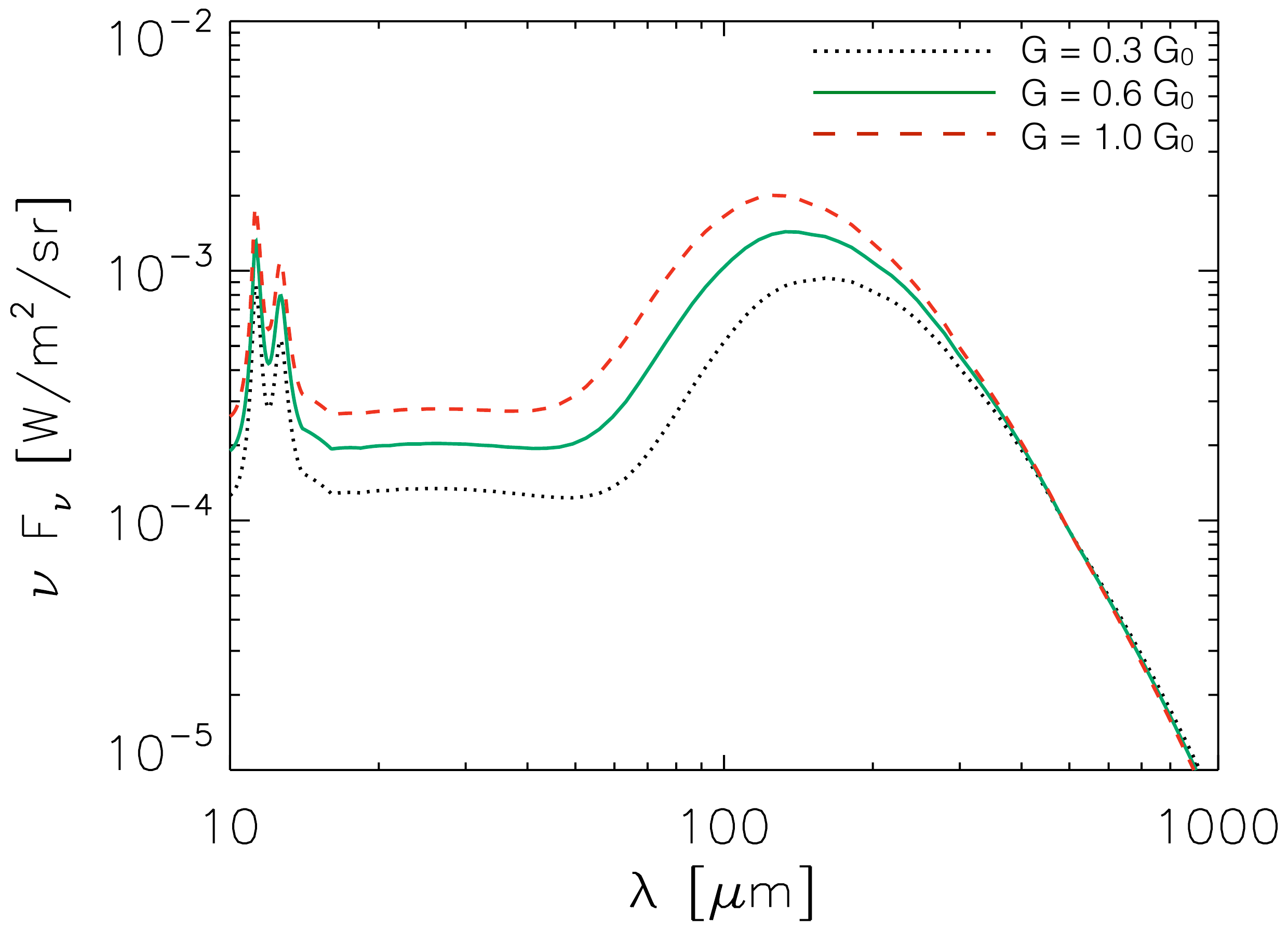}
    \caption{The SED models from the \citet{2013A&A...558A..62J} dust model with ISRF scaling factors G\,=\,0.3\,$G_{\text{0}}$ (dotted black curve), 0.6\,$G_{\text{0}}$ (solid green curve) and 1.0\,$G_{\text{0}}$ (dashed red curve), all normalised to the same SPIRE\,500\,$\mu$m flux. }
    \label{Ima_SED_Jones}
\end{figure}

\begin{figure}
	\includegraphics[width=8.5cm]{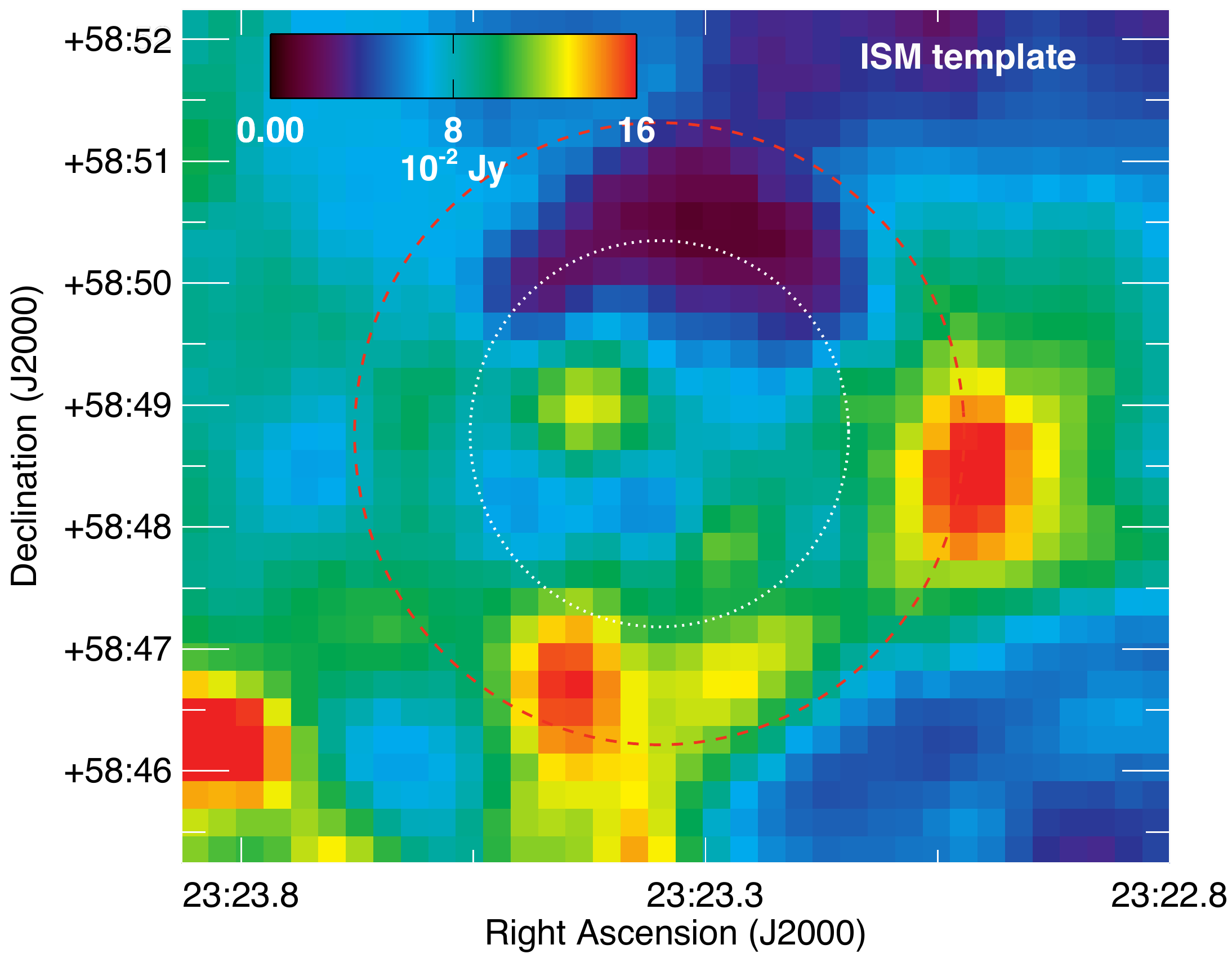}
    \caption{The SPIRE\,500\,$\mu$m map, after subtraction of the synchrotron radiation component, which was used to model the spatial distribution of ISM dust along the line of sight of Cas\,A. For reference, the position of the forward and reverse shock are indicated as dashed red and dotted white circles, respectively.}
    \label{Ima_CasA_SPIRE500temp}
\end{figure}

To model the emission of interstellar dust towards Cas\,A, we apply the THEMIS (The Heterogeneous dust Evolution Model for Interstellar Solids) dust model \citep{2013A&A...558A..62J,2014A&A...565L...9K} which includes a set of amorphous hydrocarbon grains (a-C(:H)) and silicates with iron nano-particle inclusions (a-Sil$_{\text{Fe}}$), for which the optical properties were derived from laboratory studies, and the size distribution and abundances of grain species were calibrated to reproduce the extinction and emission observed in the diffuse interstellar regions in the Milky Way. The shape of the ISM dust emission spectrum depends on the temperature of the emitting dust grains (and thus the strength of the ISRF that is heating these grains) and will determine the peak wavelength of the ISM dust emission (see Figure \ref{Ima_SED_Jones}). To normalise the spectrum, we use the SPIRE\,500\,$\mu$m map (after subtraction of the synchrotron emission, see Fig. \ref{Ima_CasA_SPIRE500temp}) as a tracer of the ISM dust mass. 

To determine the average temperature of ISM dust along the line of sight to Cas\,A, we need to derive the scaling factor of the ISRF that is heating the dust. To derive this ISRF scaling factor, we fit the PACS\,100 and 160\,$\mu$m, and SPIRE\,250, 350 and 500\,$\mu$m emission in the ISM dust-dominated regions surrounding Cas\,A with a physically motivated SED model. We exclude a circular patch with radius of 165$\arcsec$ centred on the SNR and retain a sample of 5912 $``$interstellar dust" pixels (of size 14$\times$14\,arcsec$^2$) with $\geq$5$\sigma$ detections in the PACS 160\,$\mu$m and SPIRE\,250, 350 and 500\,$\mu$m wavebands. We use the SED fitting tool \texttt{DustEm} \citep{2011A&A...525A.103C} to model the dust emission for a prefixed composition of dust grains with a given size distribution, optical properties and dust emissivity. By deriving the dust emissivity for every single grain species of a given size and composition (based on a dust temperature distribution), the non-local thermal equilibrium emission for grain species of all sizes has been taken into account. We create a library of SED models with $G$ $\in$ [0.1, 2.5] $G_{\text{0}}$ (with steps of 0.1) and $N_{\text{H}}$ $\in$ [2.5, 76] $\times$ 10$^{21}$ cm$^{-2}$ (with stepwise increase by a factor of 1.05) to find the SED model that best fits the observations. The shape of the ISRF is chosen to be similar to the radiation field in the solar neighbourhood \citep{1983A&A...128..212M}. The best fitting SED model is determined from a least-square fitting procedure to the PACS and SPIRE data points. We performed the fitting on the PACS and SPIRE images without applying colour corrections since the model SED was convolved with the filter response curves.

\begin{figure*}
	\includegraphics[width=18cm]{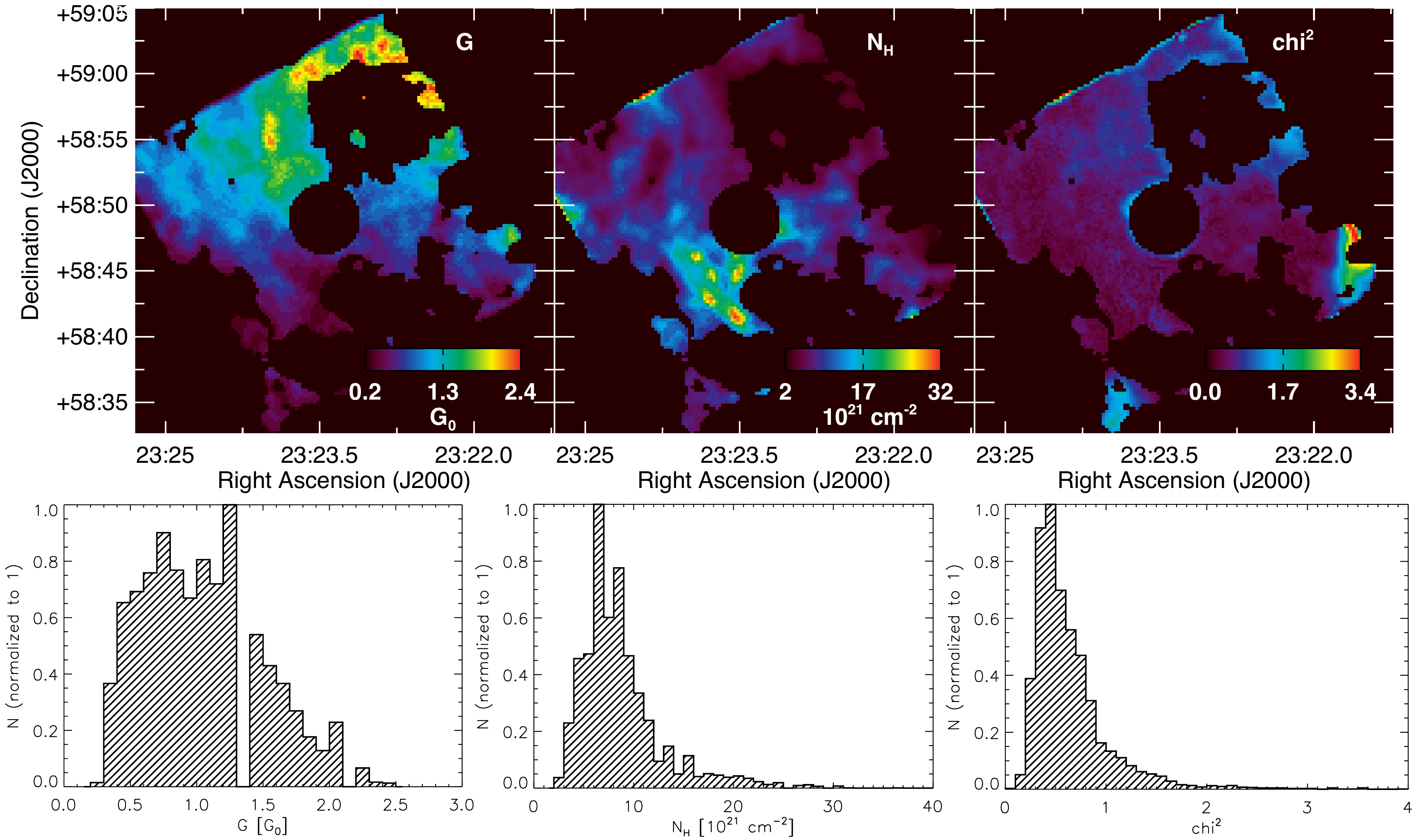}
    \caption{Top row: Maps of the best fitting $G$ and $N_{\text{H}}$ model parameters, and reduced $\chi^{2}$ values. Bottom row: normalised histograms displaying the distribution of $G$, $N_{\text{H}}$ and $\chi^{2}$ values for all 5912 ISM dust pixels.}
    \label{Ima_CasA_interstellar}
\end{figure*}

Figure \ref{Ima_CasA_interstellar} (top panel) shows the resulting maps of the best fitting $G$ and $N_{\text{H}}$ parameters and corresponding $\chi^{2}$ values for the best fitting model in every pixel, while the bottom panels shows the distribution of $G$ and $N_{\text{H}}$ parameter values, and the reduced $\chi^{2}$ values for all ISM dust pixels. With a reduced $\chi^{2}$ peaking below 1 for most pixels, we are confident that the \citet{2013A&A...558A..62J} dust model is adequate to fit the ISM dust component surrounding Cas\,A. The scaling factor ranges between values of G\,=\,0.2\,$G_{\text{0}}$ and G\,=\,2.4\,$G_{\text{0}}$ corresponding to dust temperatures between $T_{\text{d}}$=13.7 K and 20.6 K \footnote{These dust temperatures were calculated by taking the average of the mean equilibrium temperatures for amorphous silicate grains with a forsterite-type and enstatite-type chemical composition based on the temperatures derived by \texttt{DustEm} for each grain species of a given size. To determine the mean equilibrium temperature for forsterite and enstatite grains, we have weighted the individual grain temperatures with the density for grains in a given size bin.}. In the immediate surroundings of Cas\,A (see Fig. \ref{Ima_CasA_interstellar}), the $G$ values range between 0.3\,$G_{\text{0}}$ up to 1.5\,$G_{\text{0}}$. 

To constrain the variations in the radiation field $G$, we require a method to derive the ISRF scaling factor along the sightline of Cas\,A. Due to the contribution of supernova dust emission at infrared wavelengths, we cannot derive the ISRF scaling factor $G$ based on SED modelling\footnote{We have experimented with using the SPIRE\,250\,$\mu$m-to-SPIRE\,350\,$\mu$m colour to constrain the scaling factor of the interstellar radiation field, but this technique proved unsuccessful due to the contribution from supernova dust to the SPIRE wavebands and the dispersion in the correlation between the colour ratio and $G$.}. In Appendix \ref{Sect_RF}, we present a PDR modelling technique to derive the ISRF scaling factor based on PDR model parameters derived from [C{\sc{i}}] 1-0, 2-1 and CO(4-3) line intensities originating from ISM material along the line of sight to Cas\,A. For most of the sight lines towards Cas\,A, we retrieve an ISRF scaling factor of 0.6\,$G_{\text{0}}$ with some exceptions where ISRF scaling factors of 0.3\,$G_{\text{0}}$ and 1.0\,$G_{\text{0}}$ seem to fit better. We have, therefore, performed the SED fitting procedure described in Section \ref{DustSED} assuming an ISM dust model with scaling factor of $G$=0.6\,$G_{\text{0}}$. The difference between assuming ISRF scaling factors of $G$\,=\,0.3\,$G_{\text{0}}$ and $G$\,=\,1.0\,$G_{\text{0}}$ can result in a factor of 2 variation in the ISM dust emission at PACS\,100 and 160\,$\mu$m wavelengths (see Figure \ref{Ima_SED_Jones}), and will thus be an important constraint in accurately recovering the dust mass within Cas\,A. In Appendix \ref{Discuss_IS}, we compare the SN dust masses and temperatures retrieved from SED modelling using three different ISM dust models with scaling factors $G$=0.3\,$G_{\text{0}}$, 0.6\,$G_{\text{0}}$ and 1.0\,$G_{\text{0}}$, or ISM dust temperatures of $T_{\text{d}}$~=~14.6\,K, 16.4\,K and 17.9\,K, respectively. 

\begin{figure*}
	\includegraphics[width=5.9cm]{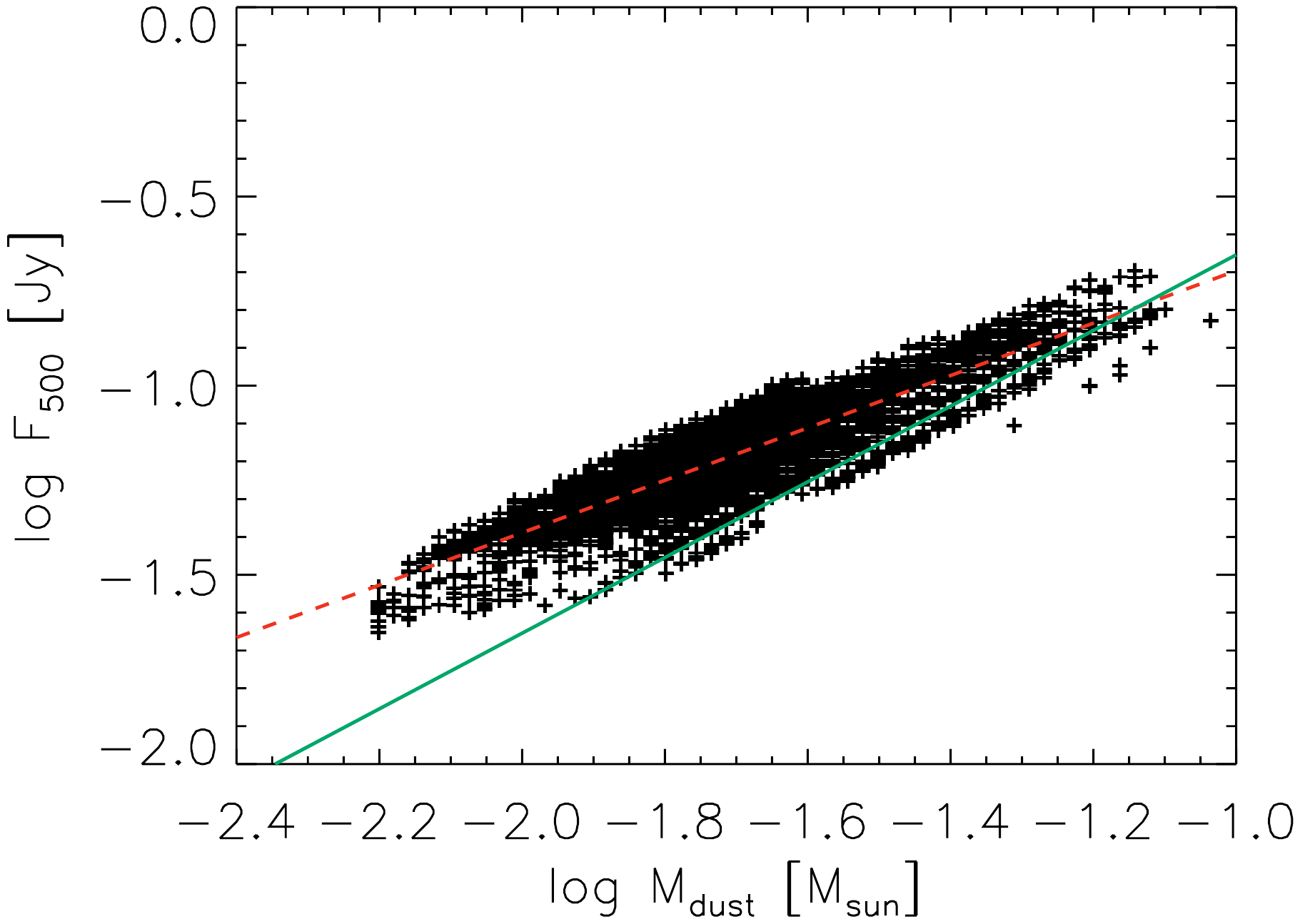} 
	\includegraphics[width=5.7cm]{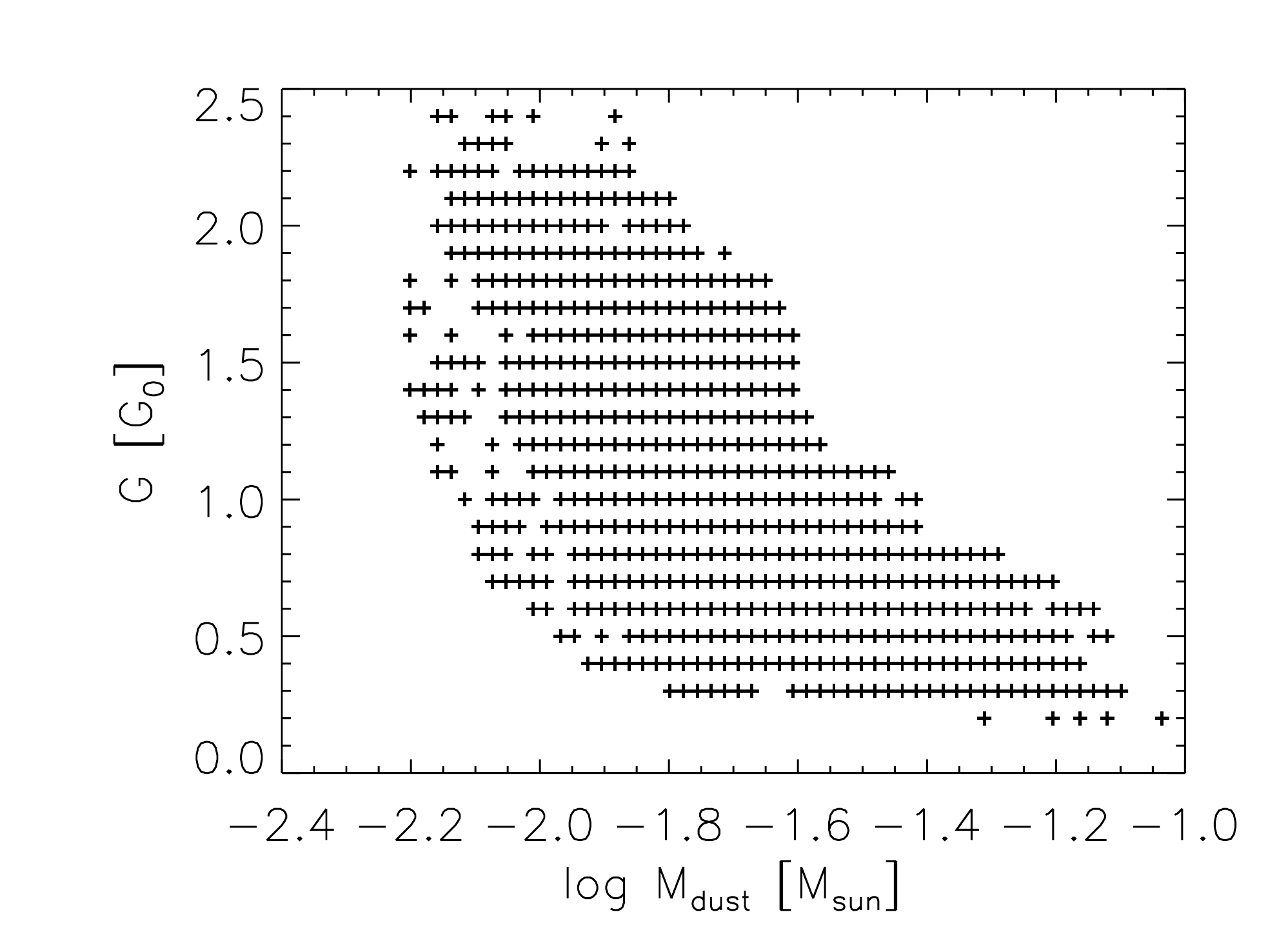} 	
	\includegraphics[width=5.9cm]{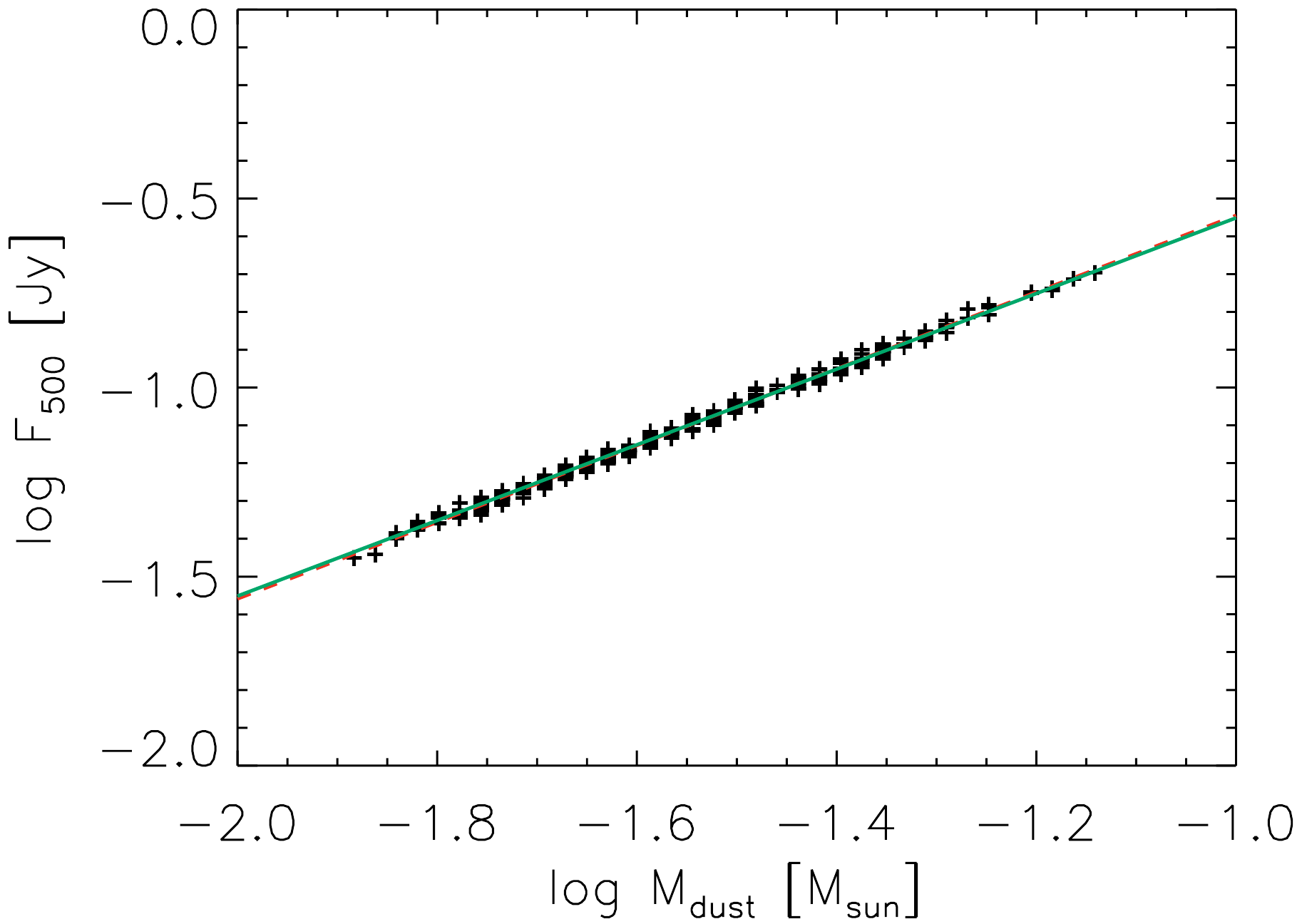} \\
    \caption{Left: Correlation between the SPIRE\,500\,$\mu$m flux density and the ISM dust mass derived by fitting the dust SED with \texttt{DustEm}. The dashed red line corresponds to the best fitting relation y = $a$x + $b$ with slope $a$ = 0.692, while a linear correlation is indicated as a solid green curve for comparison. The dashed red line is the best fitting trend while the solid green curve represents a linear correlation. Middle: Correlation between the ISM dust mass and scaling factor of the radiation field. Right: Correlation between the SPIRE\,500\,$\mu$m flux density and the ISM dust mass for pixels with G\,=\,0.6\,$G_{\text{0}}$.  }
    \label{Ima_CasA_F500_Mdust}
\end{figure*}

Other than the strength of the radiation field illuminating the ISM dust, we need to estimate the column density of ISM material along the sightline of Cas\,A. We assume that the SPIRE\,500\,$\mu$m band is mostly dominated by ISM dust emission (after subtraction of the synchrotron component) and thus is a good tracer of the ISM material. Figure \ref{Ima_CasA_F500_Mdust} (left panel) shows the relation between the SPIRE\,500\,$\mu$m and the ISM dust mass derived from SED fits for the 5912 ISM pixels in the regions surrounding Cas\,A. The dashed red line shows the best fitting trend, with a slope of 0.692, while the solid green line shows a linear correlation with slope of 1 for comparison. The relatively small dispersion (0.074 dex or an uncertainty of 19$\%$) around the best fitting trend makes us confident that the SPIRE\,500\,$\mu$m flux density is a good tracer of the ISM dust mass. The main uncertainty on the ISM dust model, thus, arises from the ISRF scaling factor ($G$\,=\,0.3, 0.6 or 1.0\,$G_{\text{0}}$) that illuminates the dust. 

% Figure 6

% Figure 7

While we might expect a relation between the SPIRE\,500\,$\mu$m flux density and ISM dust mass that is closer to linear, the trend in Figure \ref{Ima_CasA_F500_Mdust} shows a clear deviation from linearity. Moving to larger dust masses, the SPIRE\,500\,$\mu$m flux shows a smaller increase and does not increase proportionally to the dust mass. We believe this trend is driven by the anti-correlation between the dust mass and the scaling factor of the radiation field, i.e., regions with higher dust mass are exposed to softer radiation fields (see Fig. \ref{Ima_CasA_F500_Mdust}, middle panel). The latter trend is not surprising given that less UV photons will be able to penetrate regions of higher density, which results in lower $G$ values in regions of high density. Since the same dust mass illuminated by a radiation field with $G$\,=\,2.4\,$G_{\text{0}}$ (which is the maximum $G$ found among the 5912 interstellar dust regions) will be 2.5 times more luminous at 500\,$\mu$m compared to the same dust content irradiated by a $G_{\text{0}}$=0.2 radiation field (which corresponds to the minimum $G$), the deviation from non-linearity (which corresponds to more or less a factor of $\sim$2.5 across the covered range in $F_{\text{500}}$) seems driven by the variation in radiation field that is heating the dust. Although the dust emission might become optically thick at higher dust column densities, this does not seem to play a major role in the flattening of the trend here.

To take into account the dependence on $G$ in the relation between the SPIRE\,500\,$\mu$m flux and dust mass, we fit individual relations of the form $\log$ $F_{\text{500}}$ = $a\times \log M_{\text{d}}$ + $b$ for pixels with best fitting scaling factors $a$ and $b$ for $G$\,=\,0.3\,$G_{\text{0}}$, 0.6\,$G_{\text{0}}$, and 1.0\,$G_{\text{0}}$ (i.e., the radiation fields that are explored in our SED fitting analysis). The best fitting relations (depending on the assumed value for $G$) were found to be:
\begin{equation}
\begin{split}
\label{Eq1_label}
F_{\text{500}}~[Jy]~ &=~2.093 \times M_{\text{d}}^{1.008} [\text{M}_{\odot}]~\text{for}~G~=~0.3\,G_{\text{0}} \\
			       &=~2.943 \times M_{\text{d}}^{1.014} [\text{M}_{\odot}]~\text{for}~G~=~0.6\,G_{\text{0}} \\
                                &=~3.204 \times M_{\text{d}}^{0.979} [\text{M}_{\odot}]~\text{for}~G~=~1.0\,G_{\text{0}} \\
\end{split}
\end{equation}
These relations for a given value of $G$ show a trend that is very close to linear, and have significantly smaller dispersions (0.002 dex, 0.003 dex, and 0.002 dex, respectively) corresponding to uncertainties smaller than 1\,$\%$ (see Fig. \ref{Ima_CasA_F500_Mdust}, right panel for the trend for $G$\,=\,0.6\,$G_{\text{0}}$). To model the interstellar dust emission, the ISM dust models with $G$\,=\,0.3\,$G_{\text{0}}$, 0.6\,$G_{\text{0}}$, and 1.0\,$G_{\text{0}}$ are scaled to the dust mass derived based on the SPIRE\,500\,$\mu$m flux density and the above relations. More specifically, the ISM scaling factor is allowed to vary between [0.75,1.0] $\times$ $F_{\text{500}}$(thermal) $\pm$ $\sigma_{\text{500}}$ where $F_{\text{500}}$(thermal) corresponds to the SPIRE\,500\,$\mu$m flux density after subtraction of the synchrotron radiation. The maximum contribution of 25$\%$ from SN dust emission at 500\,$\mu$m is an arbitrary upper limit, and is never reached during the SED fitting procedures with typical values of a few $\%$ up to a maximum of 19$\%$. 

\subsection{The SN dust model}
\label{Sect_SN}
We use a three component SN dust model for Cas\,A with hot, warm and cold SN dust components with dust temperatures $T_{\text{d}}$ in the ranges [100\,K, 200\,K], [40\,K, 100\,K] and [10\,K, 40\,K], respectively\footnote{While the temperature distribution of the dust in Cas\,A might be better approximated by a continuous distribution, the three different dust components give an idea of the average temperatures to which the different dust components in Cas\,A are heated.}. The dust composition for the hot SN dust component (Mg$_{0.7}$SiO$_{2.7}$) was derived based on studies of the \textit{Spitzer} IRS spectra by \citet{2008ApJ...673..271R} and \citet{2014ApJ...786...55A}. The unusual peaks in Cas A's dust emission spectrum at 9\,$\mu$m and 21\,$\mu$m suggest that most of the hotter dust in the ejecta is not composed of the silicates present in the general ISM, but instead corresponds to magnesium silicate grains with low Mg-to-Si ratios (Mg$_{0.7}$SiO$_{2.7}$, \citealt{1999ApJ...521..234A,2008ApJ...673..271R,2014ApJ...786...55A}).   

The composition of the warm and cold dust in Cas\,A is less well constrained due to the absence of obvious dust features. We therefore explore the effect on the dust SED modelling and mass determinations of using different dust species to model the warm and cold dust component in Cas\,A. In order to identify the particular type of silicates and/or other grain compositions that produce each of the characteristic SEDs, we assembled a set of grain absorption efficiencies from published optical constants (see Table \ref{Table_DustSpecies}). The selected dust composition is consistent with grain species that result from the spectral fitting to the IRS spectra and \textit{Herschel} PACS data by \citet{2014ApJ...786...55A}. In addition to the amorphous carbon grains from \citet{1991ApJ...377..526R}, we explore the results of SED fitting with H-poor carbonaceous solids with a band gap of 0.1\,eV \citep{2012A&A...540A...1J,2012A&A...540A...2J,2012A&A...542A..98J}\footnote{The latter aromatic carbonaceous grains are also used in the THEMIS dust model to reproduce the ISM dust emission in our Galaxy.}, which allows us to show the effect of variations in optical grain properties on the carbonaceous dust mass. The optical properties of these a-C grains with a narrow band gap have been calibrated on laboratory data. Dust mass absorption coefficients have been calculated from the complex refractive index for each dust species using Mie theory \citep{1908AnP...330..377M} assuming spherical grains of size $a$=1\,$\mu$m\footnote{Because the dust mass absorption coefficients are very similar for the porous and compact Al$_{2}$O$_{3}$ dust species and amorphous carbon ``AC1" and ``BE1" type grains, we have only done the SED modelling for porous Al$_{2}$O$_{3}$ and amorphous carbon ``AC1" grains.}. For dust species which lacked dust optical properties up to 1000\,$\mu$m, we extrapolated the available dust mass absorption coefficients to mm wavelengths. For silicate-type grains, we assumed a $\lambda^{-2}$ power law behaviour, while for Al$_{2}$O$_{3}$ the variation of $\kappa_{\text{abs}}$ with wavelength was fit with a power law between 50 and 200\,$\mu$m in order to extrapolate the dust mass absorption coefficients to longer wavelengths. We adopted mass densities of 1.6 g cm$^{-3}$ for amorphous carbon grains and 2.5 g cm$^{-3}$ for all the silicate and oxygen-rich grains \citep{2013A&A...558A..62J}. The last two columns in Table \ref{Table_DustSpecies} present the maximum dust masses derived based on nucleosynthesis models for type II and type IIb events for progenitors with masses similar to Cas\,A, assuming that all produced metals are locked into dust grains. 

The progenitor of Cas\,A has been suggested to have been a Wolf-Rayet star with a high nitrogen abundance \citep{2001AJ....122.2644F} and an initial mass between 15 and 30 M$_{\odot}$ \citep{2001NuPhA.688..168K,2006ApJ...640..891Y}. A higher progenitor mass ($\sim$30 M$_{\odot}$) is favoured based on chemical abundance studies \citep{2002RMxAC..12...94P,2009A&A...506.1249P}, while other analyses have suggested a lower initial progenitor mass of 23\,M$_{\odot}$ \citep{2009A&A...506.1249P}. For a solar metallicity star with an initial mass of 30 M$_{\odot}$\footnote{We take an average of the 30A and 30B models of \citet{1995ApJS..101..181W} which have initial expansion velocities of $v$ = 12,700 km s$^{-1}$ and $v$ = 18,000 km s$^{-1}$, and $^{56}$Ni masses of 0 M$_{\odot}$ and 0.44 M$_{\odot}$. The observed $v_{max}$ (14,000 km s$^{-1}$, \citealt{2006ApJ...645..283F}) and $^{56}$Ni mass (0.058-0.16 M$_{\odot}$, \citealt{2009ApJ...697...29E}) for Cas\,A seem to lie in between these two models.}, the core-collapse supernova models of \citet{1995ApJS..101..181W} predict elemental yields for hydrogen (10.5 M$_{\odot}$), carbon (0.29 M$_{\odot}$), nitrogen (0.10 M$_{\odot}$), oxygen (3.65--4.88 M$_{\odot}$), neon (0.44--0.49 M$_{\odot}$), magnesium (0.27--0.35 M$_{\odot}$), aluminium (0.04--0.05 M$_{\odot}$) and silicon (0.14--0.38 M$_{\odot}$). For each of the grain species listed in Table \ref{Table_DustSpecies} we have calculated the maximum dust mass that is possible to condense in the ejecta of Cas A based on the elemental abundances predicted by the above supernova models for a 30 M$_{\odot}$ progenitor star. Given that Cas\,A resulted from a supernova Type IIb explosion (rather than Type II), we have also listed the maximum dust masses derived from the nucleosynthesis models of \citet{2010ApJ...713..356N} for a type IIb supernova with a progenitor mass of 18\,M$_{\odot}$, based on their total elemental abundances predicted for carbon (0.114\,M$_{\odot}$), oxygen (0.686\,M$_{\odot}$), magnesium (0.107\,M$_{\odot}$), aluminium (9.31$\times$10$^{-3}$\,M$_{\odot}$), silicon (0.107\,M$_{\odot}$), sulphur (3.33$\times$10$^{-2}$\,M$_{\odot}$) and other heavy elements (7.92$\times$10$^{-2}$\,M$_{\odot}$). The maximum dust masses derived from the supernova Type IIb model are factors of 3 to 5 lower compared to the \citet{1995ApJS..101..181W} model predictions. Since we only have the elemental yields predicted for a Type IIb event for 18\,M$_{\odot}$ progenitor, the latter maximum dust masses might be uncertain by factors of a few due to the uncertainties on the initial mass of Cas\,A's progenitor.

\begin{figure*}
	\includegraphics[width=17.5cm]{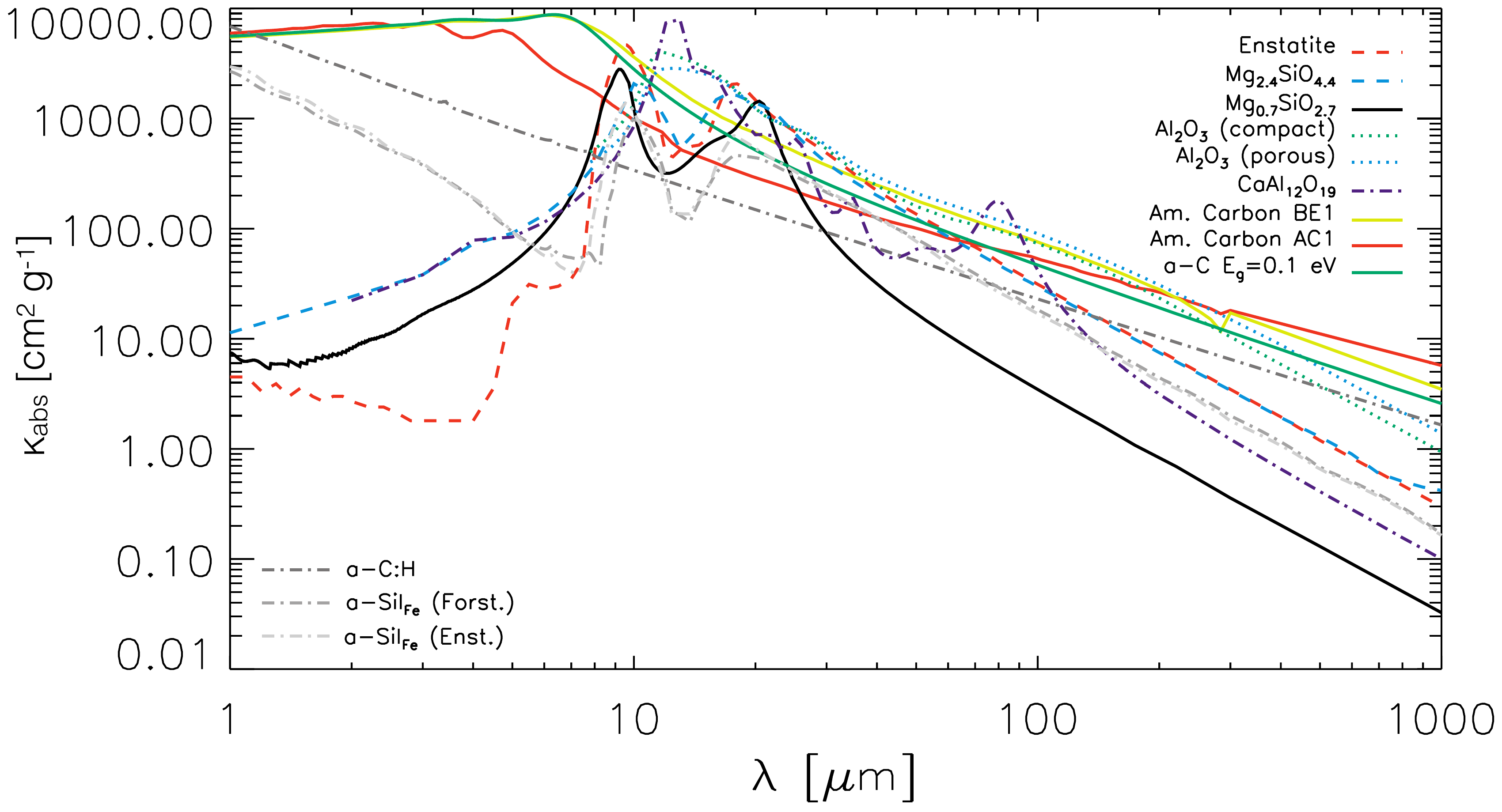}
    \caption{The variation as a function of wavelength in dust mass absorption coefficients, $\kappa_{\text{abs}}$, calculated based on Mie theory for spherical grains of size a = 1\,$\mu$m, for different dust species. The legend in the top right corner clarifies the composition of the different dust species, whose sources are listed in Table \ref{Table_DustSpecies}. For reference, the dust mass absorption coefficients for the grain species (large a-C(:H) grains and large carbon-coated, amorphous silicate grains with metallic iron nano-particle inclusions (a-Sil$_{\text{Fe}}$) and with a forsterite-type or enstatite-type chemical composition) constituting the THEMIS dust model typical of ISM dust in our Galaxy are indicated as dark, intermediate and light grey dashed-dotted lines (cf. legend on the bottom left).}
    \label{Fig_Dust_kabs}
\end{figure*}

Figure \ref{Fig_Dust_kabs} presents the dust mass absorption coefficients for different grain species as a function of wavelength. We restrict our SED fitting analysis to large grains of radius a\,=\,1\,$\mu$m, which has been shown to be a representative size for grains in nearby supernova remnants (e.g., \citealt{2014Natur.511..326G,2015MNRAS.446.2089W,2015ApJ...801..141O,2016MNRAS.456.1269B}). The assumed size of the grains does not however strongly affect the dust emissivity at IR/submm wavelengths\footnote{Assuming a grain size of a\,=\,0.1\,$\mu$m would only change the SN dust masses within the model uncertainties.}. The large variations in dust mass absorption coefficients at longer wavelengths imply that the dust mass derived from SED fitting will depend strongly on the assumed dust composition. The dust mass absorption coefficients for the different grain species explored in this work are also compared to the absorption efficiencies of typical ISM dust grains. For reference, we overlay the $\kappa_{\lambda}$ values for the large a-C(:H) grains and large carbon-coated, amorphous silicate grains with metallic iron nano-particle inclusions (a-Sil$_{\text{Fe}}$) and with a forsterite-type or enstatite-type chemical composition composing the THEMIS dust model that is representative for Galactic ISM dust at high Galactic latitude.

The SN dust emission is modelled by multiplying the dust mass absorption coefficient, $\kappa_{\lambda}$, with the emission spectrum of a modified blackbody of a given temperature. We assume optically thin dust emission. The dust mass is then derived from:
\begin{equation}
F_{\nu}~=~\frac{M_{\text{d}}}{D^{2}}~\kappa_{\nu}~B_{\nu}(T)
\end{equation}
for a single temperature SN dust model with $F_{\nu}$, the observed flux density at frequency $\nu$ in W m$^{-2}$ Hz$^{-1}$; $M_{\text{d}}$, the dust mass in g; $\kappa_{\nu}$, the dust mass absorption coefficient at frequency $\nu$ in cm$^{2}$ g$^{-1}$; D, the distance to Cas\,A in cm and $B_{\nu}$($T$), the Planck function describing the emission of a blackbody with temperature $T$. To model the IR-submm emission across the entire spectrum, we sum the three SN dust model components with different dust temperatures.

% Table 7
\begin{table*}
	\centering
	\caption{Overview of the different dust grain species that are explored in this paper to model the composition of the warm and cold SN dust components in Cas\,A. For each dust composition, the wavelength range and reference for the optical properties are given. To compare the dust emissivity of different grain species, we also include the dust mass absorption coefficients at 160\,$\mu$m, $\kappa_{\text{abs,160}}$, which were calculated for spherical grains of radius a = 1\,$\mu$m using Mie theory. The last two columns give the maximum dust masses for the different dust species that can be formed in Cas\,A, based on predicted elemental yields from nucleosynthesis models for a supernova Type II \citep{1995ApJS..101..181W} and Type IIb event \citep{2010ApJ...713..356N}. We rely on the elemental abundances predicted by \citet{1995ApJS..101..181W} for a 30 M$_{\odot}$ progenitor star, while \citet{2010ApJ...713..356N} calculates the elemental abundances for a progenitor with initial mass of 18\,M$_{\odot}$.}
	\label{Table_DustSpecies}
	\begin{tabular}{lccccc} % four columns, alignment for each
		\hline
                Dust species & $\lambda$ & Ref & $\kappa_{\text{abs,160}}$ & Type II $M_{\text{d,max}}$ & Type IIb $M_{\text{d,max}}$  \\
                   & [$\mu$m] &  & [cm$^{2}$ g$^{-1}$] & [M$_{\odot}$] & [M$_{\odot}$] \\
		\hline
		MgSiO$_{3}$ & 0.2-500 & 1 & 12.7 & 1.37 & 0.38 \\
		Mg$_{0.7}$SiO$_{2.7}$ & 0.2-470 & 2 & 1.3 & 1.21 & 0.34 \\
		Mg$_{2.4}$SiO$_{4.4}$ & 0.2-8,200 & 2 & 16.4 & 0.93  & 0.29  \\
		Al$_{2}$O$_{3}$-porous & 7.8-500 & 3 & 45.3 & 0.10 & 0.02 \\
		Al$_{2}$O$_{3}$-compact & 7.8-200 & 3 & 36.1 & 0.10 & 0.02 \\
		CaAl$_{12}$O$_{19}$ & 2-10,000 & 4 & 5.7 & 0.11 & 0.02 \\	
		Am. carbon ``AC1" & 0.01-9,400 & 5 & 33.0 & 0.29 & 0.11 \\
		Am. carbon ``BE1" &0.01-9,400 & 5 & 39.7 & 0.29 & 0.11 \\
		a-C ($E_{\text{g}}$=0.1eV) & 0.022-1,000,000 & 6 & 25.4 & 0.29 & 0.11 \\
		\hline 
	\end{tabular}

	References: 1: \citet{1995A&A...300..503D}; 2: \citet{2003A&A...408..193J}; 3: \citet{1997ApJ...476..199B}; \\ 4: \citet{2002A&A...392.1047M}; 5: \citet{1991ApJ...377..526R}; 6: \citet{2012A&A...540A...1J,2012A&A...540A...2J,2012A&A...542A..98J}.
%	\tablefoot{\\	
%	References: 1: \citet{1995A&A...300..503D}; 2: \citet{2003A&A...408..193J}; 3: \citet{1997ApJ...476..199B}; \\ 4: \citet{2002A&A...392.1047M}; 5: \citet{1991ApJ...377..526R}; 6: \citet{2012A&A...540A...1J,2012A&A...540A...2J,2012A&A...542A..98J}.}
\end{table*}

%Figure 19. 

%%%%%%%%%%%%
%SED modelling%%%
%%%%%%%%%%%%

\section{Dust SED modelling and results}
\label{DustSED}
We aim to derive the contribution of dust emission intrinsic to Cassiopeia\,A, and its mass. To do this we fit a multi-component SED model to the \textit{Spitzer} IRS continuum at 17\,$\mu$m and 32\,$\mu$m\footnote{We included the continuum emission from the \textit{Spitzer} IRS spectra at 17 and 32\,$\mu$m to constrain the continuum spectrum at wavelengths shortwards and longwards of the 21\,$\mu$m peak. Without these constraints, it was impossible to constrain the dust temperature and the exact contribution from the hot and warm SN dust components, which would affect the fitting of the colder SN dust.}, and to the \textit{WISE} 22\,$\mu$m, MIPS 24\,$\mu$m, PACS 70, 100 and 160\,$\mu$m and SPIRE 250, 350 and 500\,$\mu$m data points, which were corrected for line contamination (see Appendix \ref{Sect_line}) and synchrotron emission (see Section \ref{Sect_synchr}). We omit the IRAC\,8\,$\mu$m and WISE\,12\,$\mu$m data points from the SED fitting procedure to avoid biases introduced by the fitting of the mid-infrared emission features, which have been attributed to aromatic-rich nano-particles. At the same time, the dust emission originating from the reverse shock regions that dominates in those mid-infrared wavebands comes from a hot ($\sim$500 K) SN dust component. To avoid the addition of another SN dust component in the SED model, we restrict the SED fitting procedure to the wavelength domain from 17 to 500\,$\mu$m. %polycyclic aromatic hydrocarbons (PAHs)

To reproduce the multi-wavelength spectrum, we construct a four-component SED model with an ISM dust component and hot, warm and cold SN dust components. For the ISM dust model, we adopt the ISM dust model from \citet{2013A&A...558A..62J} for a radiation field of $G$=0.6\,$G_{\text{0}}$ (see Section \ref{Sect_ISdust}). We have modelled the SN dust emission in Cas\,A with a fixed dust composition of silicates with a low Mg/Si ratio of 0.7 (i.e., Mg$_{0.7}$SiO$_{2.7}$) for the hot dust component, and different dust compositions for the warm and cold dust in Cas A (see Section \ref{Sect_SN}). For every SED fit, we assume a single dust composition for the warm and cold SN dust components since the relative abundances of different dust species could not be constrained in our SED modelling procedure due to model degeneracies. The SED fits for a single dust composition provide the SN dust masses assuming the cold+warm SN dust component is entirely made up of these dust species. It is however likely that the warm+cold SN dust is composed of a combination of these various dust species\footnote{Several studies have indicated that the supernova explosion that produced Cas\,A was highly asymmetric and turbulent \citep{2001ApJS..133..161F,2006ApJ...645..283F,2010ApJ...725.2038D,2015Sci...347..526M,2016ApJ...822...22O}. The stratification of ejecta and mixing of heavy elements \citep{2010ApJ...714.1371H} will determine the composition and amount of dust formed in different parts of the remnant.} with a SN dust mass in between the values retrieved for the various dust compositions. 

We require a total of 7 parameters to fit the IR/submm SN dust emission from Cas\,A and the surrounding ISM dust. The 7 free parameters include the dust mass and temperature for the hot, warm and cold SN dust components, and the scaling of the ISM dust model in the range [$F_{\text{500}}$-$\sigma_{\text{500}}$-(0.25$\times$$F_{\text{500}}$),$F_{\text{500}}$+$\sigma_{\text{500}}$]\footnote{We do not consider the scaling factor of the radiation field $G$, that illuminates the ISM dust, as a free parameter, since we assume a fixed value (0.3, 0.6 or 1.0\,$G_{\text{0}}$) for the different SED fitting procedures.}. With ten different data points to fit the shape and intensity level of the SED in every pixel, we are able to constrain each of these parameters. The best fitting parameters are determined from a Levenberg-Marquardt least-square fitting procedure in IDL using the function \texttt{MPFIT}. 

%Figure 17
\begin{figure}
	\includegraphics[width=9cm]{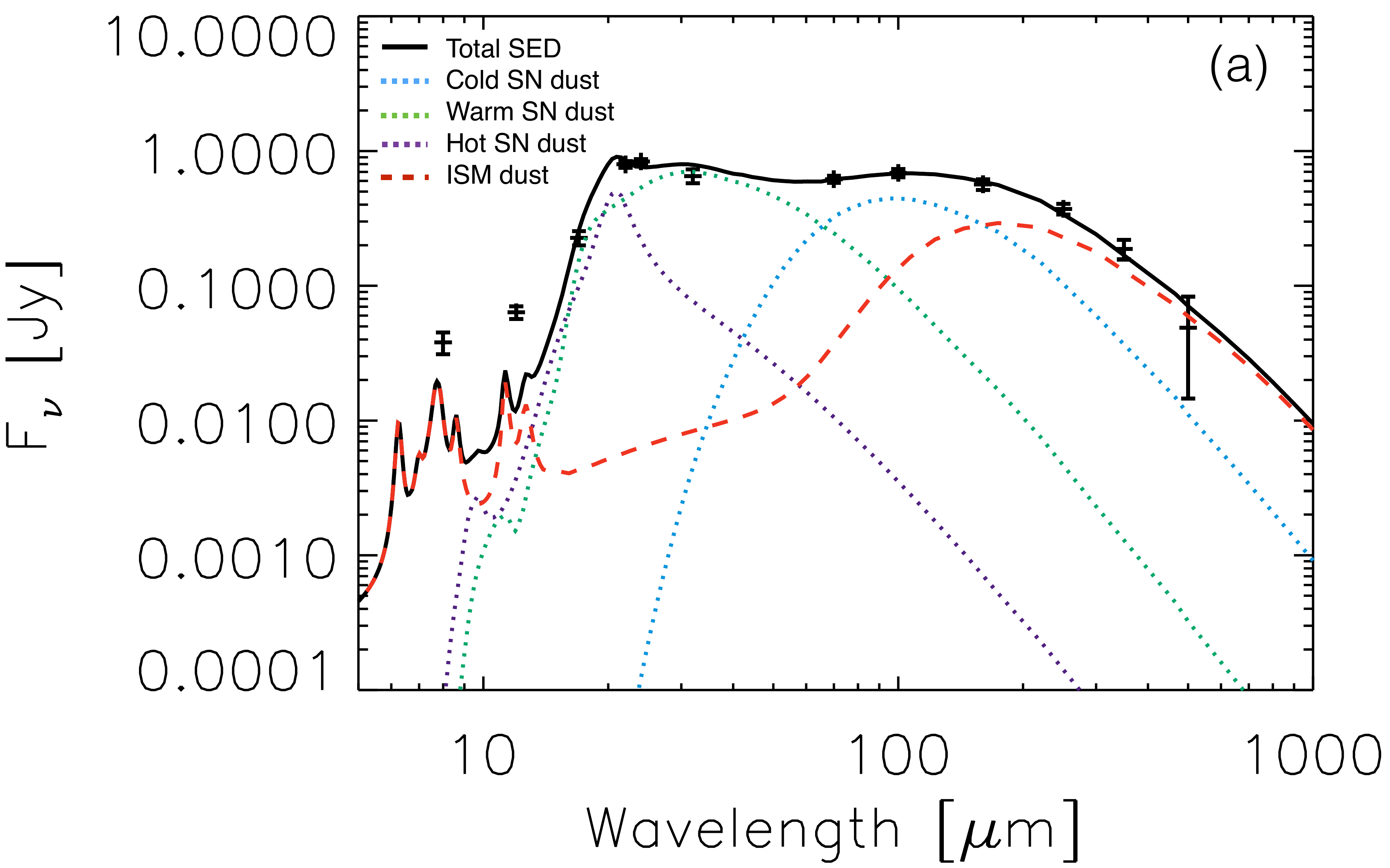} \\
		\includegraphics[width=9cm]{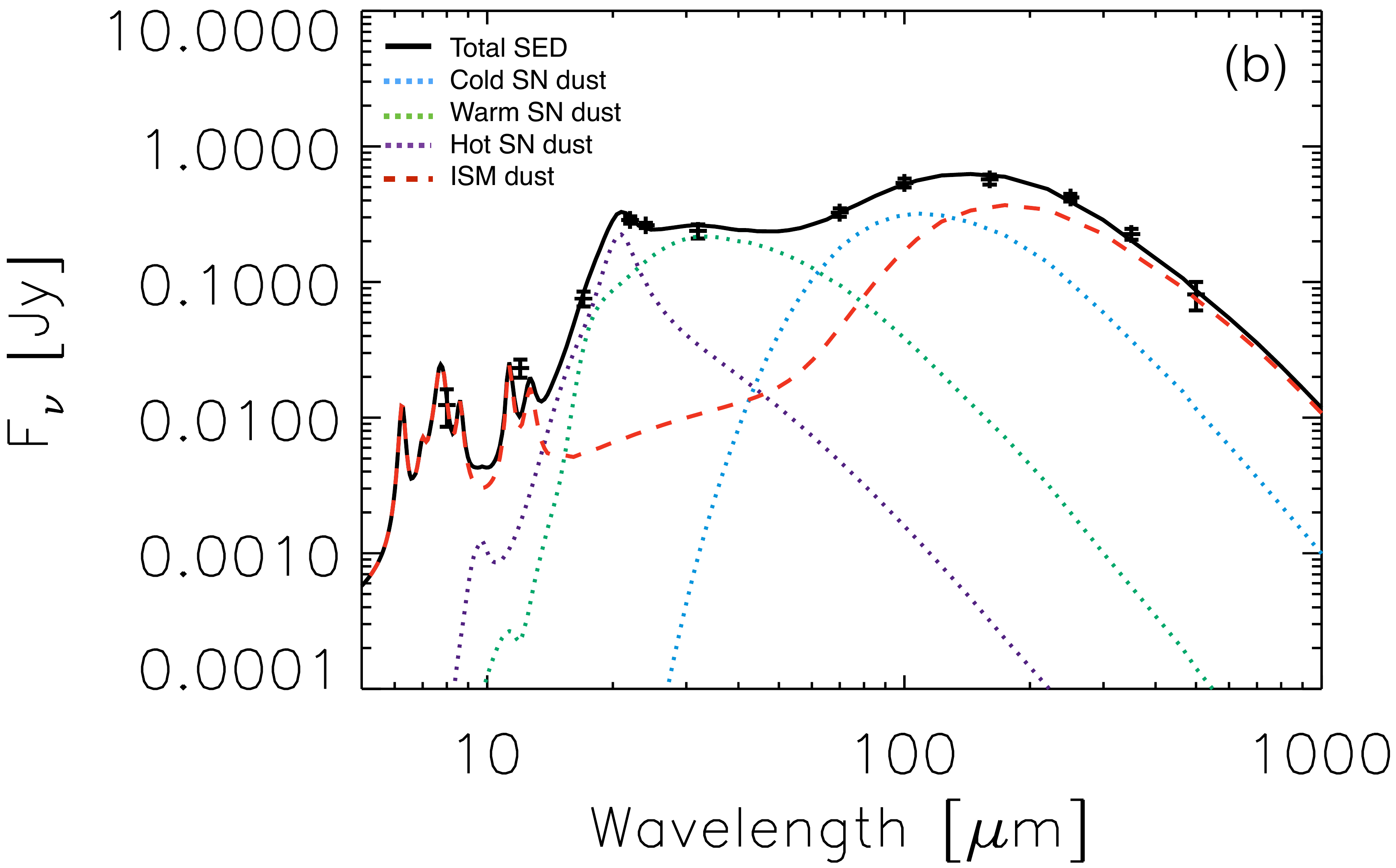} \\
			\includegraphics[width=9cm]{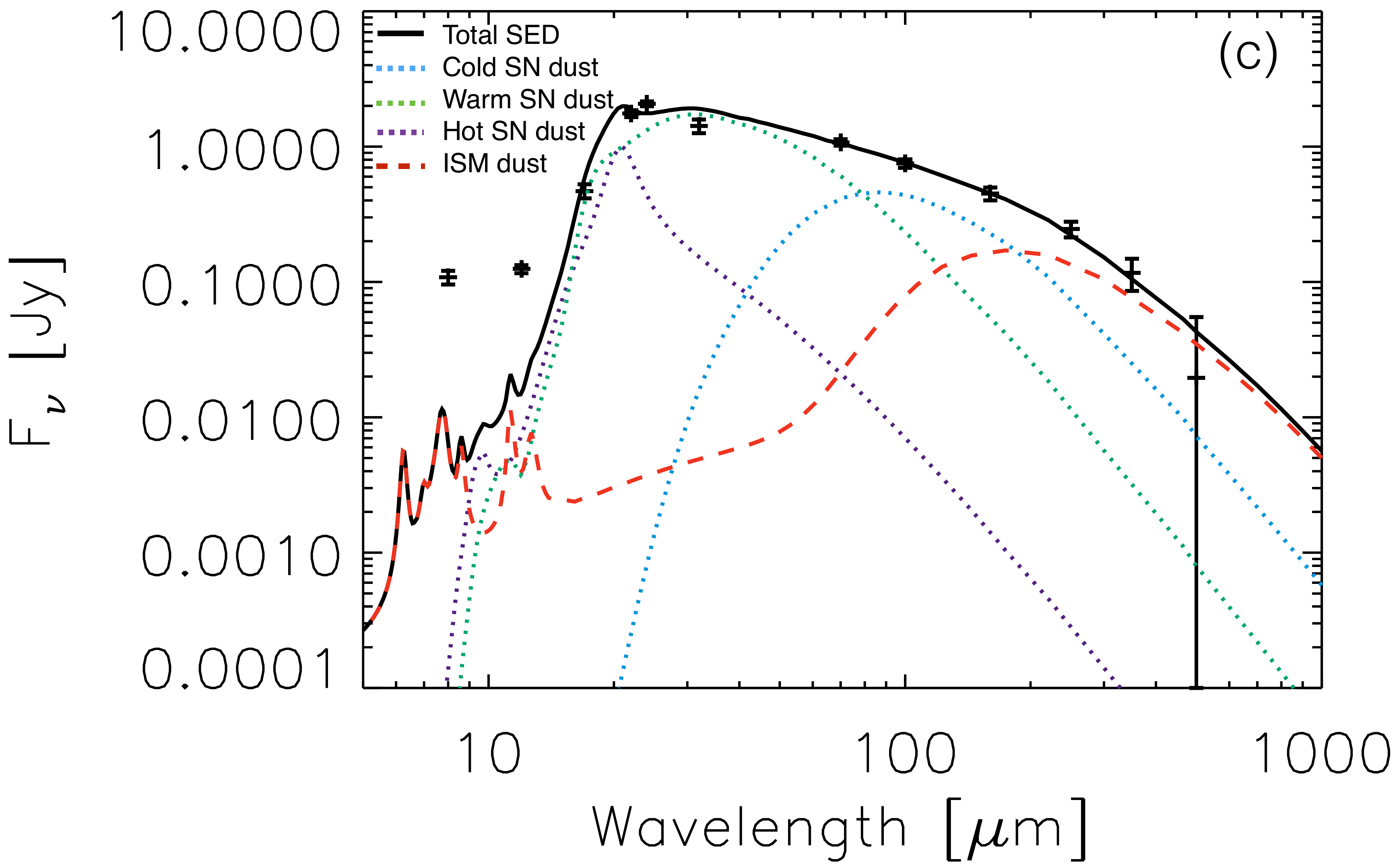}
    \caption{Spatially resolved SED fits constrained by the 17-500\,$\mu$m photometry for three pixels (whose locations are shown in Fig. \ref{Ima_CasA_resolvedSED_maps}) targeting different regions of the supernova remnant. The different SED components are outlined in the legend on the top left of every panel. }
    \label{Ima_CasA_resolvedSED_SED}
\end{figure}

%Figure 15
\begin{figure*}
	\includegraphics[width=11.75cm]{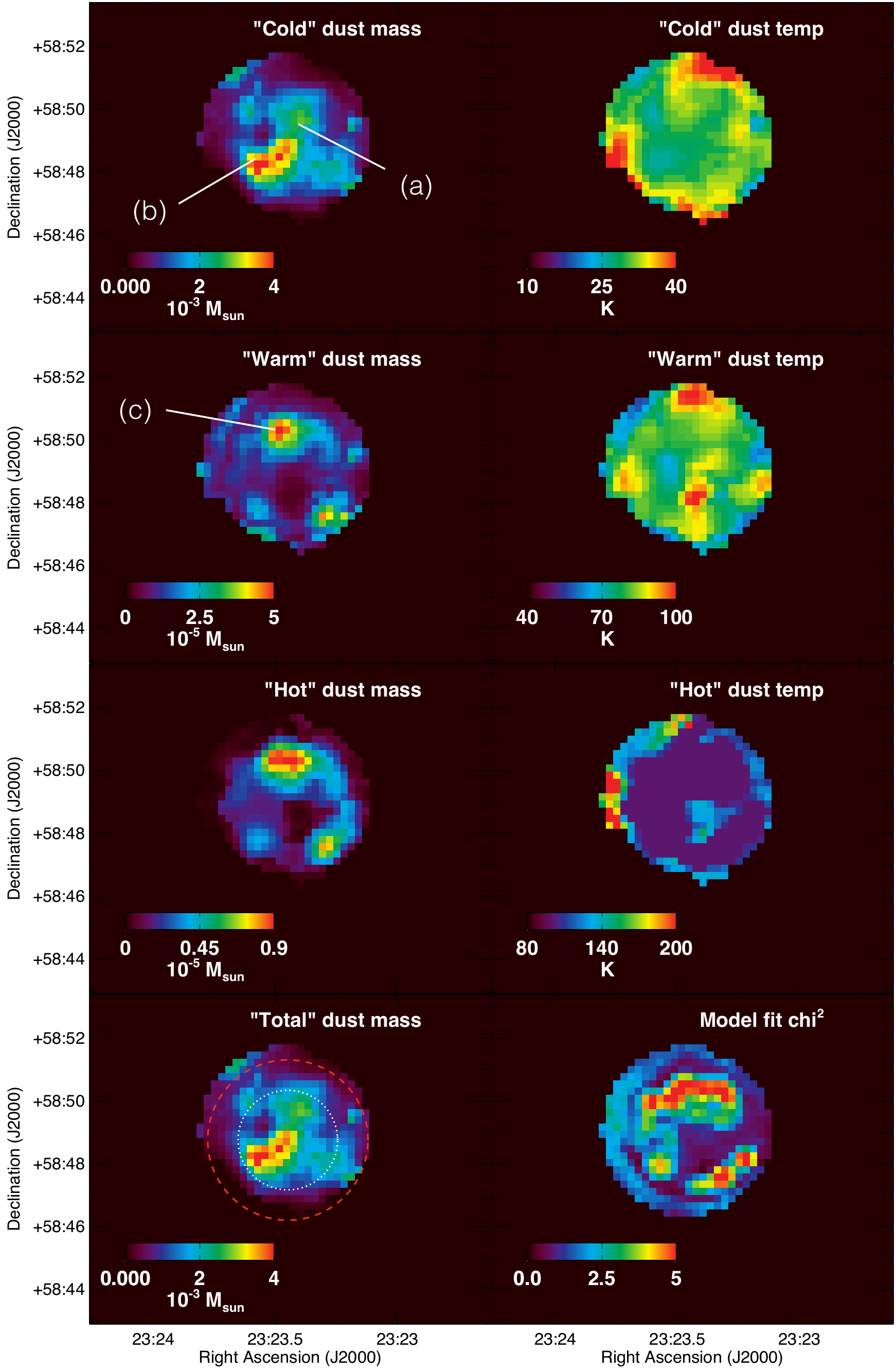}
     \caption{Dust mass maps (left) and temperature maps (right) for the hot, warm and cold SN dust components in Cas\,A. The bottom row shows the summed $"$Total$"$ dust mast map (left) and $\chi^{2}$ map (right). The individual SEDs of the pixels indicated as (a), (b) and (c) in the dust mass maps are presented in Figure \ref{Ima_CasA_resolvedSED_SED}. In the bottom left panel, the position of the forward and reverse shock are indicated as dashed red and dotted white circles, respectively.}
    \label{Ima_CasA_resolvedSED_maps}
\end{figure*}

To locate the position of dust grains that were formed in Cas\,A, we performed a spatially resolved fit to the IR/submm images. We did a similar global SED fitting analysis for which the results are presented in Appendix \ref{Sect_globalfit}\footnote{Due to strong variations in ISM contributions and SN dust temperatures across the field of Cas\,A, we believe that the average global spectrum does not capture the local variations within Cas\,A very well. We therefore prefer to rely on the SN dust masses and temperatures derived from the resolved SED fitting analysis.}. To test that pixels were independent, we rebinned the pixels to a size of 36$\arcsec$ which corresponds to the FWHM of the SPIRE 500\,$\mu$m beam. For aesthetic purposes we present the output maps from the SED fitting using the nominal pixel size of 14$\arcsec$ for the SPIRE 500\,$\mu$m waveband and verified that the results are similar to the lower resolution maps. We only fitted pixels within an aperture radius of 165$\arcsec$ centred on Cas\,A and with detections above 3$\sigma$ in both the PACS\,100\,$\mu$m and 160\,$\mu$m wavebands. For maps with pixel sizes of 14$\arcsec$ and 36$\arcsec$, this resulted in the determination of best fitting parameters for 438 and 79 pixels, respectively. We omitted some pixels with unrealistically low cold dust temperatures ($\leq$14\,K) at the edge of the map. 

Figure \ref{Ima_CasA_resolvedSED_SED} shows representative SEDs for three individual pixels in different regions of Cas\,A (Figure \ref{Ima_CasA_resolvedSED_maps} shows the location of these pixels). The first position coincides with a position inside the reverse shock region with a SN dust contribution representative for the unshocked ejecta. The second SED originates from a peak in the cold dust mass in the unshocked ejecta. Although the ISM contribution is dominant in the SPIRE wavebands, the SN dust contribution dominates at PACS\,70 and 100\,$\mu$m. The third SED shows a representative SED of a reverse shock region in the North. The hot and warm SN dust components clearly dominate in this region, with a smaller contribution from ISM dust emission (which is consistent with the lower SPIRE\,500\,$\mu$m flux in this region). While the IRAC\,8\,$\mu$m and WISE\,12\,$\mu$m fluxes were not used to constrain the SED model parameters, the ISM+SN dust model reproduces these two mid-infrared constraints of the SED in the middle panel. The SEDs in the top and bottom panels show an excess of 8 and 12\,$\mu$m emission relative to our best fitting model, which likely can be attributed to the presence of a hotter SN dust component in Cas\,A (aromatic-rich nano-particle variations in the ISM might play a role as well). \citet{2014ApJ...786...55A} show that part of the SN dust in Cas\,A reaches dust temperatures of $T_{\text{d}}$$\sim$400-500\,K which would emit at these near-infrared wavelengths.

% Table 6
\begin{table*}
	\centering
	\caption{The results of the \textbf{spatially resolved SED fitting} procedure for a variety of dust species. We present dust masses and temperatures for the hot (columns 2 and 3), warm (columns 4 and 5) and cold SN dust components (columns 6 and 7) for three different runs with ISRF scaling factors G = 0.3\,$G_{\odot}$, 0.6\,$G_{\odot}$ and 1.0\,$G_{\odot}$, respectively. Columns 8 and 9 list the total dust mass (i.e., the sum of the hot, warm and cold dust masses) and the reduced $\chi^{2}$ value that corresponds to the best fitting SED model. The uncertainties on the SN dust masses correspond to the 1$\sigma$ errors computed from the covariance matrix during the SED fitting based on the observational uncertainties of the flux in every waveband. Columns 10 and 11 present the lower limits on the SN dust temperature and mass. The lower limit on the total SN dust mass was calculated by scaling the ISM dust model to the extreme value of the SPIRE 500\,$\mu$m flux density (i.e., $F_{\text{500}}$ $+$ $\sigma_{500}$) to maximise the contribution from ISM dust emission to the global SED in every pixel (i.e., leaving no room for a SN dust contribution at 500\,$\mu$m). The dust temperature corresponds to the dust temperature of the coldest SN dust component obtained from this lower limit fit. The numbers in boldface correspond to the results from SED fitting    for an ISM model with G=0.6\,G$_{\text{0}}$ and a SN dust model with silicate-type grains (MgSiO$_{3}$, Mg$_{2.4}$SiO$_{4.4}$), and are believed to be representative of the SN dust conditions in Cas\,A.}
	\label{Table_SEDfit_resolved}
	\begin{tabular}{|lcccccccc|cc|} % four columns, alignment for each
	\hline 	
	Dust species & \multicolumn{2}{c}{Hot dust} & \multicolumn{2}{c}{Warm dust} & \multicolumn{2}{c}{Cold dust} & \multicolumn{2}{c}{Total dust} & \multicolumn{2}{|c|}{Total dust} \\
	                     & \multicolumn{2}{c}{Best fit} &\multicolumn{2}{c}{Best fit} &\multicolumn{2}{c}{Best fit} &\multicolumn{2}{c}{Best fit} & \multicolumn{2}{|c|}{Lower limit} \\
%	\hline 		
	 & $T_{\text{d}}$ & $M_{\text{d}}$ & $T_{\text{d}}$ & $M_{\text{d}}$ & $T_{\text{d}}$ & $M_{\text{d}}$ &  $M_{\text{d}}$ &  $\chi^{2}$ & $T_{\text{d}}$ & $M_{\text{d}}$(lower)  \\
	 & (K) & (10$^{-3}$ M$_{\odot}$) & (K) & (10$^{-2}$ M$_{\odot}$) & (K) & (M$_{\odot}$) & (M$_{\odot}$) & & (K) & (M$_{\odot}$)  \\
%	\hline 	 
	\hline 	
	\multicolumn{11}{c}{\textbf{$G$~=~0.3~$G_{\text{0}}$}} \\ 
%	\hline 
%	\hline 
	MgSiO$_{3}$ & 100$\pm$23 & 0.9$\pm$0.2 & 79$\pm$9 & 0.6$\pm$0.1 & 27$\pm$3 & 1.4$\pm$0.2 & 1.4$\pm$0.2 & 1.40 & 31$\pm$3 & 0.6$\pm$0.1    \\	
	Mg$_{0.7}$SiO$_{2.7}$ & 200$\pm$0 & 0.03$\pm$0.01 & 55$\pm$2 & 35.4$\pm$1.9 & 20$\pm$1 & 49.3$\pm$9.6 & 49.7$\pm$9.6 & 3.90 & 26$\pm$2 & 8.0$\pm$1.2 \\
	Mg$_{2.4}$SiO$_{4.4}$ & 130$\pm$41 & 0.4$\pm$0.2 & 78$\pm$10 & 0.8$\pm$0.1 & 28$\pm$4 & 0.9$\pm$0.1 & 0.9$\pm$0.1 & 1.71 & 33$\pm$4 & 0.5$\pm$0.1   \\	
	Al$_{2}$O$_{3}$ (porous) & 100$\pm$11 & 1.3$\pm$0.2 & 80$\pm$8 & 0.5$\pm$0.1 & 27$\pm$4 & 0.5$\pm$0.1 & 0.5$\pm$0.1 & 1.56 & 34$\pm$4 & 0.11$\pm$0.02  \\	
	CaAl$_{12}$O$_{19}$ & 100$\pm$12 & 1.1$\pm$0.1 & 78$\pm$6 & 1.0$\pm$0.1 & 16$\pm$1 & 41.7$\pm$5.0 & 41.7$\pm$5.0 & 4.08 & 19$\pm$2 & 5.6$\pm$0.9 \\	
	Am. carbon "AC1" & 100$\pm$19 & 1.1$\pm$0.2 & 94$\pm$7 & 0.6$\pm$0.1 & 28$\pm$3 & 0.6$\pm$0.1 & 0.6$\pm$0.1 & 1.98 & 39$\pm$2 & 0.08$\pm$0.02 \\	
	a-C ($E_{\text{g}}$=0.1eV) & 100$\pm$15 & 1.1$\pm$0.2 & 87$\pm$8 & 0.6$\pm$0.1 & 27$\pm$3 & 0.9$\pm$0.1 & 0.9$\pm$0.2 & 1.66 & 36$\pm$3 & 0.2$\pm$0.1 \\			
%	\hline 	 
%	\hline 	
	\multicolumn{11}{c}{\textbf{$G$~=~0.6~$G_{\text{0}}$}} \\ 
%	\hline 
%	\hline 
	MgSiO$_{3}$ & \textbf{100$\pm$17} & \textbf{0.9$\pm$0.2} & \textbf{79$\pm$10} & \textbf{0.6$\pm$0.1} & \textbf{30$\pm$4} & \textbf{0.5$\pm$0.1} & \textbf{0.5$\pm$0.1} & \textbf{1.76}  & \textbf{38$\pm$3} & \textbf{0.17$\pm$0.02}   \\	
	Mg$_{0.7}$SiO$_{2.7}$ & 200$\pm$0 & 0.03$\pm$0.01 & 56$\pm$3 & 34.4$\pm$1.9 & 21$\pm$2 & 21.1$\pm$9.2  & 21.4$\pm$9.2 & 3.82 & 30$\pm$7 & 1.9$\pm$0.5  \\
	Mg$_{2.4}$SiO$_{4.4}$ & \textbf{120$\pm$40} & \textbf{0.4$\pm$0.2} & \textbf{79$\pm$11} & \textbf{0.9$\pm$0.1} & \textbf{32$\pm$5} & \textbf{0.3$\pm$0.1} & \textbf{0.3$\pm$0.1} & \textbf{1.74} & \textbf{39$\pm$2} & \textbf{0.13$\pm$0.02}  \\			
	Al$_{2}$O$_{3}$ (porous) & 100$\pm$8 & 1.3$\pm$0.2 & 80$\pm$8 & 0.5$\pm$0.1 & 30$\pm$8 & 0.3$\pm$0.2 & 0.3$\pm$0.2 & 1.81 & 38$\pm$3 & 0.03$\pm$0.01  \\	
	CaAl$_{12}$O$_{19}$ & 100$\pm$7 & 1.1$\pm$0.2 & 78$\pm$6 & 0.9$\pm$0.1 & 16$\pm$2 & 30.6$\pm$6.5 & 30.6$\pm$6.5 & 4.02 & 22$\pm$5 & 1.6$\pm$0.5 \\	
	Am. carbon "AC1" & 100$\pm$17 & 1.1$\pm$0.2 & 93$\pm$8 & 0.6$\pm$0.1 & 28$\pm$6 & 0.5$\pm$0.2 & 0.5$\pm$0.2 & 1.82 & 38$\pm$4 & 0.03$\pm$0.01 \\		
	a-C ($E_{\text{g}}$=0.1eV) & 100$\pm$12 & 1.2$\pm$0.2 & 86$\pm$9 & 0.7$\pm$0.1 & 28$\pm$7 & 0.6$\pm$0.1 & 0.6$\pm$0.2 & 1.82 & 39$\pm$2 & 0.05$\pm$0.01 \\			
%	\hline 	 
%	\hline 	
	\multicolumn{11}{c}{\textbf{$G$~=~1.0~$G_{\text{0}}$}} \\ 
%	\hline 
%	\hline 
	MgSiO$_{3}$ & 100$\pm$16 & 1.0$\pm$0.2 & 79$\pm$11 & 0.7$\pm$0.1 & 32$\pm$13 & 0.20$\pm$0.03  & 0.20$\pm$0.03 & 2.05  & 39$\pm$2 & 0.06$\pm$0.01   \\	
	Mg$_{0.7}$SiO$_{2.7}$ & 200$\pm$0 & 0.03$\pm$0.01 & 56$\pm$2 & 32.0$\pm$1.7 & 13$\pm$8 & 5.6$\pm$9.2 & 5.8$\pm$6.5 & 3.78 & 35$\pm$7 & 0.4$\pm$0.1 \\
	Mg$_{2.4}$SiO$_{4.4}$ & 120$\pm$38 & 0.4$\pm$0.2 & 76$\pm$10 & 1.0$\pm$0.1 & 35$\pm$18 & 0.11$\pm$0.02 & 0.11$\pm$0.02 & 1.80 & 39$\pm$2 & 0.05$\pm$0.01  \\		
	Al$_{2}$O$_{3}$ (porous) & 100$\pm$10 & 1.3$\pm$0.2 & 78$\pm$7 & 0.5$\pm$0.1 & 20$\pm$12 & 0.4$\pm$0.3 & 0.4$\pm$0.3 & 1.92 & 37$\pm$6 & 0.010$\pm$0.002  \\		
	CaAl$_{12}$O$_{19}$ & 100$\pm$5 & 1.1$\pm$0.2 & 78$\pm$5 & 0.9$\pm$0.1 & 15$\pm$3 & 22.9$\pm$6.1 & 22.9$\pm$6.1 & 4.06 & 32$\pm$9 & 0.3$\pm$0.2 \\	
	Am. carbon "AC1" & 100$\pm$16 & 1.1$\pm$0.2 & 92$\pm$8 & 0.7$\pm$0.1 & 23$\pm$10 & 0.6$\pm$0.3 & 0.6$\pm$0.3 & 1.78 & 35$\pm$7 & 0.010$\pm$0.002 \\						
	a-C ($E_{\text{g}}$=0.1eV) & 100$\pm$13 & 1.2$\pm$0.2 & 84$\pm$9 & 0.7$\pm$0.1 & 23$\pm$11 & 0.7$\pm$0.2 & 0.7$\pm$0.2 & 1.90 & 37$\pm$6 & 0.02$\pm$0.01 \\			
	\hline
	\end{tabular}
\end{table*}

Table \ref{Table_SEDfit_resolved} presents the best fitting dust masses and temperatures derived from the spatially resolved SED fitting procedure. Dust masses are derived by summing the contributions from individual pixels. The dust temperatures and their uncertainties are determined as the median and standard deviation of the dust temperatures for individual pixels. Depending on the dust composition, the modelled mass of the cold SN dust component in Cas\,A can vary by more than two orders of magnitude. We derive dust masses of 0.3\,M$_{\odot}$ (Mg$_{2.4}$SiO$_{4.4}$) and 0.5\,M$_{\odot}$ (MgSiO$_{3}$) for silicate-type grains, while SN dust masses of 0.5-0.6\,M$_{\odot}$ and 0.3\,M$_{\odot}$ are derived for carbonaceous and Al$_{2}$O$_{3}$ grains, respectively. The highest dust masses (several tens of M$_{\odot}$) are retrieved for Mg$_{0.7}$SiO$_{2.7}$ and hibonite (CaAl$_{12}$O$_{19}$) dust compositions. However, based on the elemental abundances predicted from nucleosynthesis models (see Table \ref{Table_DustSpecies}), some of these high dust masses can already be ruled out due to the expected lack of material to form such grains. With unrealistically high dust masses needed to reproduce the IR-submm dust SED, we can exclude Mg$_{0.7}$SiO$_{2.7}$, CaAl$_{12}$O$_{19}$ and Al$_{2}$O$_{3}$ grains as the dominant dust species in Cas\,A. Also amorphous carbon grains might not be a plausible major dust component in Cas\,A due to the overwhelmingly oxygen-rich composition of the remnant \citep{1979ApJ...233..154C,2010A&A...509A..59D}. Due to the presence of a variety of metals in different parts of the SN ejecta (e.g., \citealt{2008ApJ...673..271R,2014ApJ...786...55A}), the condensation of SN dust with a range of different dust compositions throughout the remnant might be more realistic\footnote{The formation of carbonaceous grains is dependent on predictions of nucleosynthesis models. While \citet{2010ApJ...713..356N} predict that around 40$\%$ of the 0.2\,M$_{\odot}$ newly formed dust grains in their models have a carbonaceous composition, \citet{2016A&A...587A.157B} predict the formation of predominantly silicate-type grains with 0.1\,M$_{\odot}$ of carbon grains out of the 0.8\,M$_{\odot}$ modelled dust mass for Cas\,A.}. Assuming the condensation of SN dust composed for 50$\%$ of silicate-type grains and for 50$\%$ of carbonaceous grains (a similar relative dust fraction is favoured in \citealt{Bevan2016b} to reproduce the Cas\,A line profiles), we would require a total SN dust mass between 0.4\,M$_{\odot}$ and 0.6\,M$_{\odot}$ (which combine the lower and upper bounds of the results for carbonaceous and silicate-type grains) to reproduce the observations. 

To derive a lower limit to the dust mass present in Cas A, we restrict the scaling factor of the ISM dust model so that it reproduces all of the SPIRE 500\,$\mu$m flux (after subtraction of the synchrotron emission), equivalent to no supernova dust contribution at 500\,$\mu$m. While this scenario is unlikely, given that the SCUBA 850\,$\mu$m polarisation data imply a contribution from dust \citep{2009MNRAS.394.1307D}, we interpret the resulting SN dust masses as strict lower limits. For MgSiO$_3$ and Mg$_{2.4}$SiO$_{4.4}$ grain compositions, we derive lower limits of $>$0.2M$_\odot$ and $>$0.1M$_\odot$ based on spatially resolved SED fitting, respectively. A lower limit of $>$0.03M$_\odot$ of Al$_2$O$_{3}$ or $>$0.03-0.05M$_\odot$ of amorphous carbon grains would be sufficient to reproduce the IR/submm SEDs on spatially resolved scales, and not violate the predictions of nucleosynthesis models for metal production.

Figure \ref{Ima_CasA_resolvedSED_maps} presents the SN dust mass and temperature maps that were derived from SED modelling assuming a Mg$_{0.7}$SiO$_{2.7}$ composition for the hot component and a MgSiO$_{3}$ composition for the warm and cold SN dust components, respectively, for an ISM model with ISRF $G$ = 0.6\,$G_{\text{0}}$. The last row shows the total dust mass map and the reduced $\chi^{2}$ values representative for the goodness of the fit in every pixel. The positions of the forward and reverse shock as determined by \citet{2001ApJ...552L..39G} based on \textit{Chandra} X-ray data (at radii of 153$\arcsec$ and 95$\arcsec$, respectively) have been overlaid on the total dust mass map (bottom left) as dashed red and dotted white lines, respectively.  

The temperature maps in Figure \ref{Ima_CasA_resolvedSED_maps} make it clear that the dust mass components in Cas\,A are heated to different temperatures. The temperature of the $``$cold$"$ dust component in Cas\,A varies from $\sim$25\,K to $\sim$30\,K interior to the reverse shock, and reaches temperatures of $\sim$30\,K to $\sim$35\,K in the outer ejecta. The warm dust component has temperatures of around 80 K (similar to the temperature $T_{\text{d}}$=82 K derived by \citealt{2004ApJS..154..290H} to fit the \textit{Spitzer} IRAC and MIPS fluxes), and is distributed over the reverse shock regions. An additional hotter dust component (with $T_{\text{d}}$ around 100\,K) and a Mg$_{0.7}$SiO$_{2.7}$ dust component are required to fit the spectral peak at 21\,$\mu$m in the observed SED of Cas\,A. 

The sum of all these dust components yields a total dust mass map (bottom left panel in Figure \ref{Ima_CasA_resolvedSED_maps}). This total dust mass map shows a relatively smooth distribution in the unshocked inner regions, with a peak south-east of the centre of the remnant, which seems to suggest that dust formed more or less uniformly in the inner ejecta of Cas\,A. The average dust column density of the inner ejecta is about 0.0025 M$_{\odot}$ pixel$^{-1}$ with a peak of 0.004 M$_{\odot}$ pixel$^{-1}$ while the cold SN dust column density quickly drops to about 0.001 M$_{\odot}$ pixel$^{-1}$ (and lower) in the outer ejecta. Since the ISM dust contribution varies widely along the sight lines to Cas\,A (see Figure \ref{Ima_CasA_SPIRE500temp}), the confinement of the cold SN dust component to the unshocked ejecta suggests that the ISM dust emission has been modelled accurately and supports the inference that the residual emission can be attributed to cold dust formed in Cas\,A. The drop in SN dust mass outside the reverse shock regions may be consistent with the destruction of part of the freshly formed dust in Cas\,A by the reverse shock (e.g., \citealt{2010ApJ...713..356N}). Comparing the average dust column densities inside and outside the reverse shock, we estimate the dust destruction efficiency of the reverse shock in Cas\,A to be $\sim$70$\%$\footnote{To estimate the fraction of dust destroyed by the reverse shock, we assume that the ejecta are distributed homogeneously within a sphere confined by the forward shock. Based on this simple geometrical model, we can estimate the length of a column through the sphere's centre (306$\arcsec$=2*R) and the extent of a sightline at the position of the reverse shock (240$\arcsec$), which is taken into account when comparing the dust column density in- and outside the reverse shock.}. The latter value is lower compared to the 80\,$\%$, 99\,$\%$ and 88\,$\%$ dust destruction efficiencies estimated for small silicate-type grains by \citet{2010ApJ...715.1575S}, \citet{2016A&A...587A.157B} and \citet{2016arXiv160202754M}, respectively\footnote{The dust destruction efficiency based on the models presented in \citet{2016A&A...587A.157B} was estimated by comparing their best fitting dust mass (0.83\,M$_{\odot}$) for the current epoch to the final dust mass that will be ejected into the ISM (1.07$\times$10$^{-2}$) according to their models.}. Given the expected strong increase in heavy element abundances towards the inner regions of the SN ejecta and the consequent uncertainties on the geometrical distribution of SN dust and dust mass surface densities interior/exterior to the reverse shock, we do not rule out that dust destruction efficiencies maybe higher or lower than 70$\%$. Based on a dust destruction efficiency of 70$\%$, and comparing the volumes of ejecta that have been affected by the reverse shock (76$\%$) and the $\sim$0.4-0.5\,M$_{\odot}$ of dust interior to the reverse shock (24$\%$), we would estimate an initial dust mass up to 1.6-2.0\,M$_{\odot}$, under the debatable assumption that the dust condensation occurred homogeneously throughout the entire sphere of ejecta.
 
While most of the cold SN dust is confined to the unshocked ejecta, there are concentrations of cold SN dust that extend beyond the reverse shock on opposite sides of the remnant. Even though not perfectly aligned with the relativistic jets, which have been shown to drive the outflow of fast moving knots \citep{2010ApJ...725.2038D}, their location suggests that cold SN dust grains have been ejected along the north-east and south-west jets of Cas\,A. If so, it is interesting to note that the dust in the outflow along the jets is not destroyed with the same efficiency as elsewhere in the remnant. Either the grain composition might be different and less easily destroyed by the reverse shock, or the faster moving material along the jet is less prone to dust sputtering. The latter argument is supported by models of grain destruction in reverse shocks which show that the sputtering yield decreases with increasing energy for He atom/ion inertial sputtering at velocities $>$ 200 km/s or equivalently T$_{\text{gas}}$ $>$ 10$^7$ K (e.g., \citealt{1978MNRAS.183..367B,1979ApJ...231...77D,1979ApJ...231..438D,1994ApJ...431..321T,2006ApJ...648..435N,2016A&A...587A.157B}).

\section{Dust masses and uncertainties}
\label{Discussion}

\subsection{SN dust masses and uncertainties}
We derive SN dust masses of 0.5\,M$_{\odot}$ for MgSiO$_{3}$ dust or 0.3\,M$_{\odot}$ for Mg$_{2.4}$SiO$_{4.4}$ grains with a silicate-type dust composition. We can obtain similarly good SED fits with 0.5-0.6\,M$_{\odot}$ of carbonaceous grains. Based on elemental yield predictions from nucleosynthesis models for type II and type IIb core-collapse supernova, we can rule out that Al$_{2}$O$_{3}$, Mg$_{0.7}$SiO$_{2.7}$ or CaAl$_{12}$O$_{19}$ grain species dominate the cold dust reservoir in Cas\,A. Although some carbon dust might form in specific ejecta layers (e.g., \citealt{2013ApJ...776..107S}), a dominant amorphous carbon dust component seems unlikely for this very oxygen-rich SNR. For the latter results, we have assumed an ISM model with $G$=0.6\,G$_{\text{0}}$ which was independently constrained based on SED modelling of the ISM dust surrounding Cas\,A and PDR modelling of the interstellar [C~{\sc{i}}] and CO emission along the sightline of Cas\,A. In Appendix \ref{Discuss_IS}, we investigated what effect this assumption has on the modelled SN dust masses. We find that a lower ISRF (or G=0.3\,G$_{\text{0}}$) results in higher residual SN dust emission at IR-submm wavelengths and thus higher SN silicate dust masses (0.9-1.4\,M$_{\odot}$). In the same way, we derive lower SN dust fluxes and thus lower SN dust masses (0.1-0.2\,M$_{\odot}$) for silicate-type grains for a higher ISRF (G=1.0\,G$_{\text{0}}$). Based on the estimated metal production by nucleosynthesis models up to 1.4\,M$_{\odot}$ (MgSiO$_{3}$) or 0.9\,M$_{\odot}$ (Mg$_{2.4}$SiO$_{4.4}$) for silicate-type grains\footnote{These maximum dust masses correspond to a supernova type II event with a 30\,M$_{\odot}$ progenitor. Although dust masses are lower for a type IIb event with a 18\,M$_{\odot}$ progenitor, we lack constraints on the element production in a type IIb event for a more massive progenitor.}, the ISM models with G=0.3\,G$_{\text{0}}$ would require all metals to be locked up into dust grains to fit the SED which is incompatible with the detection of metals in Cas\,A in the gas phase (e.g., \citealt{2008ApJ...673..271R,2014ApJ...786...55A}). While our two independent methods favour the ISM model with G=0.6\,G$_{\odot}$, we cannot rule out that the local ISRF near Cas\,A resembles the conditions in the solar neighbourhood (G=1.0\,G$_{\odot}$) and G=0.6\,G$_{\odot}$ merely represents the average ISRF along the line of sight to Cas\,A. In that case, the SN dust mass that is able to reproduce the SED would be lower and our models suggest the condensation of 0.2\,M$_{\odot}$ or 0.1\,M$_{\odot}$ of MgSiO$_{3}$ or Mg$_{2.4}$SiO$_{4.4}$ dust, respectively. In both cases, those values would be sufficient to account for the production of dust at high redshifts by supernovae (if Cas\,A can be considered to be representative for the dust condensation efficiency in other SNRs).

\subsection{Comparison with previous results for Cas\,A}
\label{CompareBefore.sec}
Our preferred SN dust masses derived here for silicate-type grains (0.3-0.5 M$_{\odot}$ with a lower limit of $M_{\text{d}}$ $\geq$ 0.1-0.2\,M$_{\text{0}}$) are larger than the dust masses previously derived from observational SED studies of Cas\,A ($<$0.1\,M$_{\odot}$, \citealt{2008ApJ...673..271R,2010A&A...518L.138B,2010ApJ...719.1553S,2014ApJ...786...55A}), but consistent with the SN dust mass derived based on SCUBA\,850\,$\mu$m polarimetric observations (\citealt{2009MNRAS.394.1307D}; but see Section \ref{Discuss_850mu}). The extension of our study to wavelengths longwards of 200\,$\mu$m, where the cold dust component's emission peaks, and the careful removal of the emission of other components has allowed us to measure cold dust emitting at temperatures of $T_{\text{d}}$ $\sim$ 30\,K within Cas\,A. The mass of SN dust derived from our resolved study of \textit{Spitzer} and \textit{Herschel} observations is, furthermore, in agreement with the dust mass derived by fitting the nebular line profile asymmetries in the integrated spectra of Cas\,A using a Monte Carlo dust radiative transfer model ($\sim$1.1\,M$_{\odot}$, \citealt{Bevan2016b}). These values are also consistent with the dust condensation models of \citet{2007MNRAS.378..973B} who predicted $\sim$0.6 M$_{\odot}$ of dust to form in the ejecta of supernovae having progenitor masses between 22 and 30 M$_{\odot}$, but significantly higher compared to the dust evolution models in \citet{2010ApJ...713..356N}, who predicted 0.08\,M$_{\odot}$ of new grain material of which 0.072\,M$_{\odot}$ was predicted to reside in the inner remnant regions unaffected by the reverse shock. 

To understand the difference in the derived SN dust masses based on the newly reduced \textit{Herschel} data set presented here and the analysis of \citet{2010A&A...518L.138B} (who find 0.075\,M$_{\odot}$ of T$_{\text{d}}$=35\,K dust), we apply the same fitting technique to their PACS and SPIRE flux densities (after subtraction of their non-thermal emission model) using our multi-component SN+ISM model. For an ISM model with ISRF of $G$~=~0.6\,$G_{\text{0}}$ and a Mg$_{0.7}$SiO$_{2.7}$ and MgSiO$_{3}$ grain composition for the hot and warm+cold SN dust components, we derive a global SN dust mass of 0.23$\pm$0.06\,M$_{\odot}$ with a dust temperature of $T_{\text{d}}$\,=\,33\,K for the most massive, cold SN dust component. This newly derived dust mass is a factor of 3 higher compared to the dust mass at a similar temperature of $T_{\text{d}}$$\sim$35\,K derived by \citet{2010A&A...518L.138B} based on the same set of photometry measurements. The dust mass absorption coefficient $\kappa_{\text{160}}$\,=\,12.7\,cm$^{2}$ g$^{-1}$ used in this paper is only a factor of 1.3 larger than $\kappa_{\text{160}}$\,=\,9.8\,cm$^{2}$ g$^{-1}$ in \citet{2010A&A...518L.138B} and cannot explain the difference in SN dust masses. We therefore argue that the difference between the two studies can be attributed to the assumed model for the ISM emission, which dominates at wavelengths $\geq$\,160\,$\mu$m and strongly depends on the radiation field that illuminates the ISM dust (see Appendix \ref{Discuss_IS}).

\subsection{Shock destruction of SN dust}
Since most of the cold dust mass in Cas\,A appears to be located in the unshocked ejecta, the mass of dust that eventually will mix with the surrounding ISM material may decrease after the passage of the reserve shock. Based on a comparison of the dust mass surface density in- and outside the reverse shock and a simple geometrical model, we would estimate a dust destruction efficiency of $\sim$70$\%$ (see Section \ref{DustSED}). Several theoretical models have also estimated the current dust mass content in Cas\,A and have modelled the dust evolution after the reverse shock has reached the centre of the remnant. In general, the sputtering rate for a grain depends on the pre-shock gas density, the magnetic field strength, shock velocity and the grain radius. For the reverse shock sputtering models of \citet{2010ApJ...713..356N} the largest grain radii considered were $<$10$^{-2}$\,$\mu$m, which were found to have lifetimes of 500-1000 yrs. Grains of radius a=1\,$\mu$m would have lifetimes more than 100 times greater, longer than the time to cross the reverse shock. Using a set of hydrodynamical simulations, \citet{2010ApJ...715.1575S} calculated the grain sputtering efficiency for a variety of grain species and size distributions. Their simulations of planar reverse shocks impacting clumps of newly formed ejecta indicated that grains with radii $<$ 0.1\,$\mu$m would be sputtered to smaller grain sizes and often completely destroyed, while larger grains would be able to survive the reverse shock due to less efficient coupling to the shocked gas which might leave them undisturbed after the passage of the reverse shock (e.g. \citealt{2004ApJ...614..796S}). The destruction efficiency, furthermore, seems to depend on grain composition, with average destroyed fractions of 20, 38, 80 and 100\,$\%$ grains for Fe, C, SiO$_{2}$ and Al$_{2}$O$_{3}$ grains \citep{2010ApJ...715.1575S}, respectively. \citet{2016A&A...589A.132B} found that non-thermal sputtering in dust clumps in Cas\,A plays an important role in dust destruction and predicted that it would reduce the current SN dust mass by 40 to 80\,$\%$.  The largest grain radii that they considered for the Cas\,A case were $<$ 0.02\,$\mu$m. \citet{2016arXiv160202754M} adopted an interstellar MRN \citep{1977ApJ...217..425M} power-law grain size distribution n(a) $\propto$ a$^{-3.5}$ with $a_{\text{min}}$ = 0.005\,$\mu$m and $a_{\text{max}}$ = 0.25\,$\mu$m for Cas\,A, found that only the largest grains survived, with $\sim$12\,$\%$ and $\sim$16\,$\%$ of silicate and carbon grain mass surviving the passage of the reverse shock. Based on hydrodynamical simulations of the evolution of the remnant, \citet{2016A&A...587A.157B} predict a current dust mass of 0.83 M$_{\odot}$ of Cas\,A, in good agreement with the dust masses derived from our observational study presented in this paper. Most of their modelled SN dust was made up of Mg$_{2}$SiO$_{4}$ grains (0.4 M$_{\odot}$) with smaller contributions from SiO$_{2}$ (0.16 M$_{\odot}$), Fe$_{3}$O$_{4}$ (0.13 M$_{\odot}$), amorphous carbon (0.12 M$_{\odot}$), Al$_{2}$O$_{3}$ (2.14$\times$10$^{-2}$ M$_{\odot}$) and MgSiO$_{3}$ (2.4$\times$10$^{-5}$ M$_{\odot}$) grains. According to their simulations, only 1.07$\times$10$^{-2}$ M$_{\odot}$ of dust (1.30\,$\%$ of the current dust mass) would subsequently be able to survive the reverse shock. Their models treated grain size distributions with radii less then 0.01\,$\mu$m, but they did find that grains of radius 0.1\,$\mu$m would survive the passage of the reverse shock. While the above theoretical models predict different dust survival fraction, and adopted small grain radii typical of the ISM (while larger grains might be expected in SNRs), the conclusion from these studies is that somewhere between 1 to 80\,$\%$ of the current dust mass in Cas\,A will be able to survive and be mixed with the interstellar medium. In order to explain the high dust mass content in the early Universe it would require that $>$ 20-30$\%$ of the current SN-formed dust mass (0.3-0.5\,M$_{\odot}$) survives passage through the SN reverse shock.

\subsection{IR-submm emission components}
By carefully determining the temperature and emission levels of the Galactic ISM foreground material, we have been able to model the ISM and SN dust emission at every wavelength based on a multi-component SED fitting approach. In Appendix \ref{Discuss_comparison}, we present a detailed comparison of our best fitting model with observations, and conclude that our multi-component ISM+SN model is adequate to reproduce the observed IR-submm emission towards Cas\,A within the uncertainties. While our model is consistent with the observational IR-submm constraints up to 500\,$\mu$m, our best fitting combined model of all emission components has a flux deficit of 9.5\,Jy at 850\,$\mu$m (although within the limits of the 850\,$\mu$m uncertainties). In Appendix \ref{Discuss_850mu}, we present a detailed comparison between the predicted 850\,$\mu$m model flux of various components (synchrotron, ISM and SN dust emission) and observations presented by \citet{2003Natur.424..285D} and \citet{2009MNRAS.394.1307D}. We show that the 850\,$\mu$m deficit can be accounted for using a more shallow synchrotron spectrum ($\alpha$=-0.54) without significantly affecting the synchrotron contribution at wavelengths $\lambda$ $\leq$\,500\,$\mu$m and thus the resulting SN dust masses determined from SED fitting. We, furthermore, show that the emission of ISM dust (8.6\,Jy) and SN dust (0.4\,Jy) in our model at 850\,$\mu$m is difficult to reconcile with the high degree of polarisation attributed to SN dust by \citet{2009MNRAS.394.1307D}, which would require a SN dust flux at 850\,$\mu$m of at least 6\,Jy. 

\begin{figure*}
	\includegraphics[width=16cm]{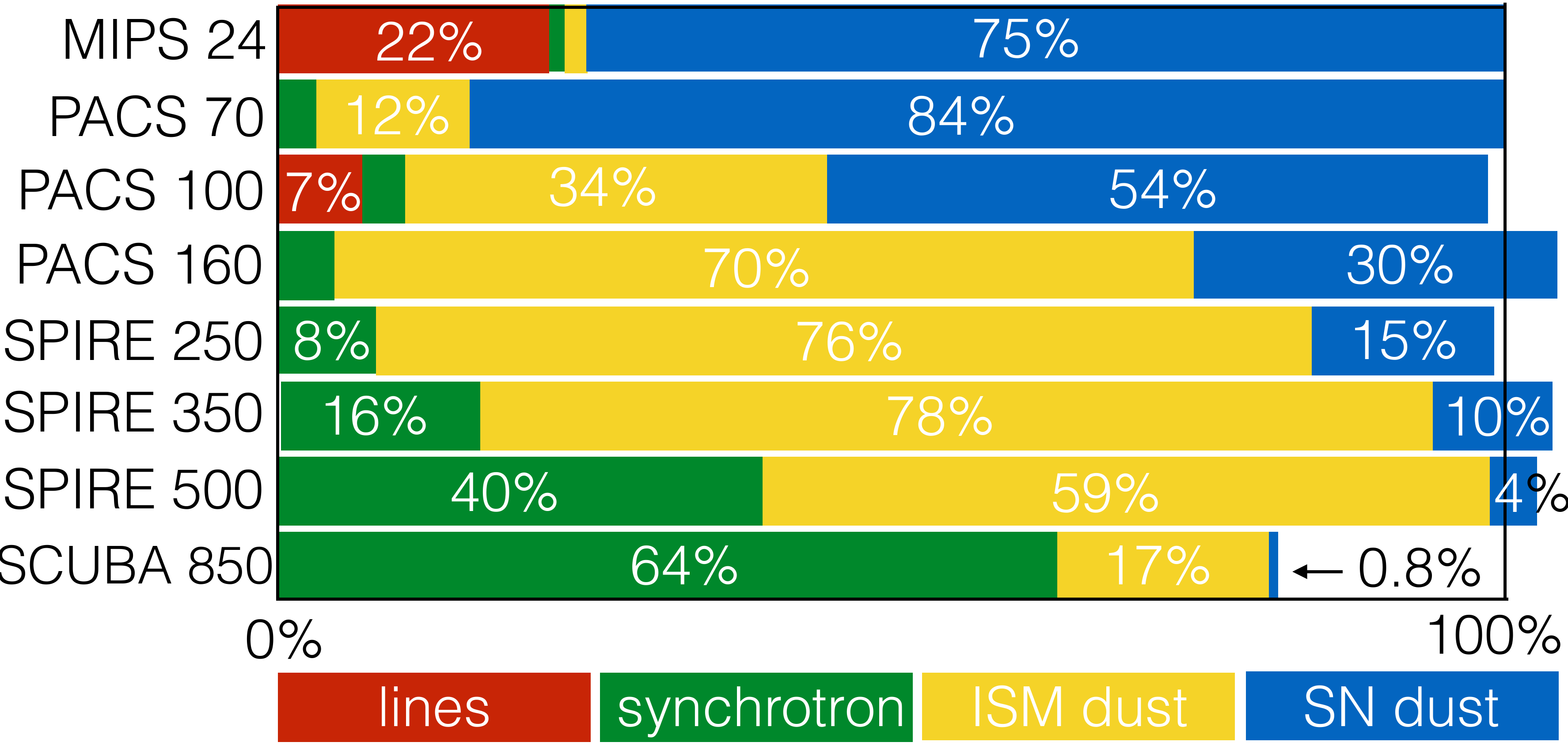}
    \caption{Schematic overview of the relative flux contributions of line, synchrotron, ISM and SN dust emission for Cas\,A in the MIPS\,24\,$\mu$m, PACS\,70, 100 and 160\,$\mu$m, and SPIRE\,250, 350 and 500\,$\mu$m and SCUBA\,850\,$\mu$m wavebands. The vertical black line on the right indicates the 100$\%$ level. The emission of various components does not always perfectly add up to 100$\%$ due to the flux uncertainties that have been taken into account in the SED fitting. The model deficit at 850\,$\mu$m is discussed in Appendix \ref{Discuss_850mu}.}
    \label{Ima_CasA_schematic}
\end{figure*}

Figure \ref{Ima_CasA_schematic} provides a schematic overview of the contributions of the different components to the total IR/submm emission for Cas\,A based on models constructed for each of these components. With SN dust contributions of only 15$\%$, 10$\%$ and 4$\%$ to the SPIRE\,250, 350 and 500\,$\mu$m wavebands, respectively, an accurate determination of the SN dust mass in Cas\,A requires reliable models for the emission of other components contributing at those wavelengths. The contribution of various emission components at IR-submm wavelengths is discussed in more detail in Appendix \ref{Sect_CasA_aftercorr}. 

The ISM and SN dust mass maps derived from the multi-component SED modelling can furthermore be used to derive interstellar and SN visual extinction maps (see Appendix \ref{AVmodel}). The high column densities of interstellar material significantly attenuate the lines of sight towards Cas\,A with an average $A_{\text{V}}$ of 6-8\,mag and peaks up to 15\,mag. The condensation of elements in the central regions of Cas\,A provides an additional visual extinction of up to $A_{\text{V}}$=1.5. The combined attenuation by ISM and SN dust makes it very difficult to detect the supernova remnant and a possible binary companion at optical wavelengths but near-IR imaging could be more effective.

\section{Conclusions}
\label{Conclusions}
We have combined \textit{Spitzer} IRAC, MIPS and IRS, \textit{WISE}, \textit{Planck}, and \textit{Herschel} PACS and SPIRE datasets to study the spatial distribution of SN dust produced in Cas\,A at a resolution of 0.6\,pc in order to determine the overall mass of freshly formed dust in Cas\,A. By extending our analysis to submillimetre wavelengths, we are able to probe the coldest dust grains that have condensed in the ejecta of Cas\,A. Due to the high degree of ISM contamination at longer wavelengths, we require a careful removal of the emission of foreground and background interstellar material and other contaminants (line emission and synchrotron radiation). We have constructed physically-motivated models to subtract the line emission and synchrotron radiation from these IR/submm images. More specifically, we
\begin{itemize}
\item used \textit{Spitzer} IRS, \textit{Herschel} PACS IFU and SPIRE FTS spectra to carefully estimate and remove contaminating emission from bright lines (mainly [Ar~{\sc{iii}}] 8.99\,$\mu$m and [O~{\sc{iii}}] 88\,$\mu$m) from the broadband images,
\item updated the spectral index of the synchrotron spectrum based on exquisitely calibrated \textit{Planck} mm-wave data, and extrapolated the synchrotron radiation to IR/submm wavelengths by normalising the integrated \textit{Planck} fluxes to a 3.7\,mm VLA image,
\item applied the THEMIS dust model \citep{2013A&A...558A..62J,2014A&A...565L...9K} that has been calibrated based on observations of dust extinction and emission in the Milky Way, to model the emission of foreground and background interstellar dust. This ISM dust model was scaled relative to the SPIRE 500\,$\mu$m flux after subtraction of the synchrotron component. The scaling of the radiation field (G\,=\,0.6\,$G_{\text{0}}$) that illuminates the ISM dust was constrained through the modelling of the ISM dust SEDs in the surroundings of Cas\,A. A similar scaling factor, $G$\,=\,0.6$^{+0.4}_{-0.3}$\,$G_{\text{0}}$, was retrieved through PDR modelling based on the [C~{\sc{i}}] 1-0, 2-1 and CO 4-3 line emission originating from ISM material in the line of sight to Cas\,A (Appendix \ref{Sect_RF}).
\end{itemize}

After subtraction of line emission and synchrotron radiation components, we fitted the 17-500\,$\mu$m SED with a multi-component ISM+SN dust model. We required three SN dust components (hot, warm and cold) to reproduce the multi-waveband emission. The hot dust component was assumed to be composed of silicates with a low Mg/Si ratio, Mg$_{0.7}$SiO$_{2.7}$, following \citet{2008ApJ...673..271R} and \citet{2014ApJ...786...55A} in order to model the emission peak at 21\,$\mu$m. We ran different sets of SED models with varying dust composition (MgSiO$_{3}$, Mg$_{2.4}$SiO$_{4.4}$, Mg$_{0.7}$SiO$_{2.7}$, Al$_{2}$O$_{3}$, CaAl$_{12}$O$_{19}$, various types of amorphous carbon) for the warm and cold SN dust components. We find that

\begin{itemize}
\item the best fitting models suggest that 0.5 M$_{\odot}$ of MgSiO$_{3}$ grains or 0.3 M$_{\odot}$ of Mg$_{2.4}$SiO$_{4.4}$ grains are present in Cas\,A, with an average cold dust temperature $T_{\text{d}}$ $\sim$ 30\,K.
\item strict lower limits of 0.1-0.2\,M$_{\odot}$ on the cold SN dust mass are derived by assuming a higher scaling factor for the ISRF (G=1.0\,$G_{\text{0}}$) heating the ISM dust.
\item the multi-wavelength emission of Cas\,A can also be fitted by 0.5-0.6\,M$_{\odot}$ of carbon grains, even though silicates are expected to dominate the dust composition in the O-rich remnant Cas\,A.
\item a total SN dust mass between 0.4\,M$_{\odot}$ and 0.6\,M$_{\odot}$ is required to reproduce the dust continuum observations assuming a mixture of 50$\%$ of silicate-type grains and 50$\%$ of carbonaceous grains.
\item from the amounts required to match the observed SED, we can exclude Mg$_{0.7}$SiO$_{2.7}$, CaAl$_{12}$O$_{19}$ and Al$_{2}$O$_{3}$ as dominant dust species based on the elemental yields predicted for a core-collapse supernova with a 30\,M$_{\odot}$ progenitor. 
\item the cold SN dust component is mainly distributed interior to the reverse shock of Cas\,A, suggesting that some part of the newly formed dust has been destroyed by the reverse shock. 
\item the drop in dust mass behind the reverse shock implies that $\sim$70$\%$ by mass of the dust is destroyed as it passes through the reverse shock in Cas\,A. 
\item the ISM model predicts an average interstellar extinction of $A_{\text{V}}$ = 6-8\,mag with maximum values of $A_{\text{V}}$ = 15\,mag towards Cas\,A. 
\end{itemize}

Based on our revised dust mass of 0.3-0.5\,M$_{\odot}$ for silicate-type dust in Cas\,A\footnote{The latter SN dust masses result from our SED modelling procedure using the latest ISM models and IR/submm dust emissivities derived from the most recent set of laboratory studies to date. The characterisation of dust optical properties and ISM dust composition is a dynamic area of research and the SN dust masses published in this work may need to be updated to account for any changes in ISM and SN dust models in the foreseeable future.}, we conclude that, if produced by other remnants, this dust mass is sufficient to explain the large dust masses observed in dusty star-forming galaxies at high-redshifts, if a non-negligible part of the initially condensed dust reservoir ($\geq$20-30$\%$) is capable of surviving the reverse shock.

\section*{Acknowledgements}
This work is dedicated to the memory of Bruce Swinyard, co-holder of 
the STFC grant that funded this work. Bruce played a 
key role in the conception and development of a number of important 
spaceborne infrared and submillimeter instruments, including the SPIRE 
instrument that flew on-board {\em Herschel}. He will be greatly missed.

We would like to thank the referee, Prof. Anthony Jones, for his careful reading of the paper and his comments and suggestions that helped to improve the work presented in this paper. We would like to thank Dave Green, Dan Milisavljevic, Rob Ivison, Edward Polehampton and Oliver Krause for interesting discussions that have helped to improve this work. We also would like to thank Marco Bocchio and Alexandre Boucaud for their assistance in creating the kernels to convolve the \textit{Herschel} images to the appropriate resolution with the most recent \textit{Herschel} PSF models. 

IDL gratefully acknowledges the support of the Science and Technology Facilities Council (STFC). 
MJB acknowledges support from STFC grant ST/M001334/1 and, since June 1st 
2016, from European Research Council (ERC) Advanced Grant {\sc sndust} 694520.'
MM is supported by an STFC Ernest Rutherford fellowship.
HLG acknowledges support from the European Research Council (ERC) in the form of Consolidator Grant {\sc CosmicDust}.

PACS was developed by a consortium of institutes
led by MPE (Germany) and including UVIE
(Austria); KU Leuven, CSL, IMEC (Belgium);
CEA, LAM (France); MPIA (Germany); INAFIFSI/
OAA/OAP/OAT, LENS, SISSA (Italy);
IAC (Spain). This development has been supported
by the funding agencies BMVIT (Austria),
ESA-PRODEX (Belgium), CEA/CNES (France),
DLR (Germany), ASI/INAF (Italy), and CICYT/
MCYT (Spain). SPIRE was developed
by a consortium of institutes led by Cardiff
University (UK) and including Univ. Lethbridge
(Canada); NAOC (China); CEA, LAM
(France); IFSI, Univ. Padua (Italy); IAC (Spain);
Stockholm Observatory (Sweden); Imperial College
London, RAL, UCL-MSSL, UKATC, Univ.
Sussex (UK); and Caltech, JPL, NHSC, Univ.
Colorado (USA). This development has been
supported by national funding agencies: CSA
(Canada); NAOC (China); CEA, CNES, CNRS
(France); ASI (Italy); MCINN (Spain); SNSB
(Sweden); STFC and UKSA (UK); and NASA
(USA). 

This research has made use of the NASA/ IPAC Infrared Science Archive, which is operated by the Jet Propulsion Laboratory, California Institute of Technology, under contract with the National Aeronautics and Space Administration.

%%%%%%%%%%%%%%%%%%%%%%%%%%%%%%%%%%%%%%%%%%%%%%%%%%



%\bibliographystyle{mnras}

\begin{thebibliography}{99}
\bibitem[\protect\citeauthoryear{Anderson \& Rudnick}{1996}]{1996ApJ...456..234A} Anderson, M.~C., \& Rudnick, L.\ 1996, \apj, 456, 234 
\bibitem[\protect\citeauthoryear{Andrews et al.}{2016}]{2016MNRAS.457.3241A} Andrews, J.~E., Krafton, K.~M., Clayton, G.~C., et al.\ 2016, \mnras, 457, 3241 
\bibitem[\protect\citeauthoryear{Aniano et al.}{2011}]{2011PASP..123.1218A} Aniano, G., Draine, B.~T., Gordon, K.~D., \& Sandstrom, K.\ 2011, \pasp, 123, 1218 
\bibitem[\protect\citeauthoryear{Arendt}{1989}]{1989ApJS...70..181A} Arendt, R.~G.\ 1989, \apjs, 70, 181 
\bibitem[\protect\citeauthoryear{Arendt et al.}{1999}]{1999ApJ...521..234A} Arendt, R.~G., Dwek, E., \& Moseley, S.~H.\ 1999, \apj, 521, 234 
\bibitem[\protect\citeauthoryear{Arendt et al.}{2014}]{2014ApJ...786...55A} Arendt, R.~G., Dwek, E., Kober, G., Rho, J., \& Hwang, U.\ 2014, \apj, 786, 55 
\bibitem[\protect\citeauthoryear{Balog et al.}{2014}]{2014ExA....37..129B} Balog, Z., M{\"u}ller, T., Nielbock, M., et al.\ 2014, Experimental Astronomy, 37, 129 
\bibitem[\protect\citeauthoryear{Barlow}{1978}]{1978MNRAS.183..367B} Barlow, M.~J.\ 1978, \mnras, 183, 367 
\bibitem[\protect\citeauthoryear{Barlow et al.}{2010}]{2010A&A...518L.138B} Barlow, M.~J., Krause, O., Swinyard, B.~M., et al.\ 2010, \aap, 518, L138 
\bibitem[\protect\citeauthoryear{Begemann et al.}{1997}]{1997ApJ...476..199B} Begemann, B., Dorschner, J., Henning, T., et al.\ 1997, \apj, 476, 199 
\bibitem[\protect\citeauthoryear{Bendo et al.}{2013}]{2013MNRAS.433.3062B} Bendo, G.~J., Griffin, M.~J., Bock, J.~J., et al.\ 2013, \mnras, 433, 3062 
\bibitem[\protect\citeauthoryear{Bersanelli et al.}{2010}]{2010A&A...520A...4B} Bersanelli, M., Mandolesi, N., Butler, R.~C., et al.\ 2010, \aap, 520, A4 
\bibitem[\protect\citeauthoryear{Bertoldi et al.}{2003}]{2003A&A...406L..55B} Bertoldi, F., Carilli, C.~L., Cox, P., et al.\ 2003, \aap, 406, L55
\bibitem[\protect\citeauthoryear{Bevan \& Barlow}{2016}]{2016MNRAS.456.1269B} Bevan, A., \& Barlow, M.~J.\ 2016, \mnras, 456, 1269
\bibitem[\protect\citeauthoryear{Bevan et al.}{2017}]{Bevan2016b} Bevan A., Barlow M.J. \& Milisavljevic D., 2017, MNRAS in press (arXiv:1611.05006)
\bibitem[\protect\citeauthoryear{Bianchi \& Schneider}{2007}]{2007MNRAS.378..973B} Bianchi, S., \& Schneider, R.\ 2007, \mnras, 378, 973 
\bibitem[\protect\citeauthoryear{Biscaro \& Cherchneff}{2016}]{2016A&A...589A.132B} Biscaro, C., \& Cherchneff, I.\ 2016, \aap, 589, A132
\bibitem[\protect\citeauthoryear{Bocchio et al.}{2016a}]{2016A&A...587A.157B} Bocchio, M., Marassi, S., Schneider, R., et al.\ 2016, \aap, 587, A157
\bibitem[\protect\citeauthoryear{Bocchio et al.}{2016b}]{2016A&A...591A.117B} Bocchio, M., Bianchi, S., \& Abergel, A.\ 2016, \aap, 591, A117 
\bibitem[\protect\citeauthoryear{Braun}{1987}]{1987A&A...171..233B} Braun, R.\ 1987, \aap, 171, 233 
\bibitem[\protect\citeauthoryear{Chevalier \& Kirshner}{1979}]{1979ApJ...233..154C} Chevalier, R.~A., \& Kirshner, R.~P.\ 1979, \apj, 233, 154 
\bibitem[\protect\citeauthoryear{Chevalier \& Oishi}{2003}]{2003ApJ...593L..23C} Chevalier, R.~A., \& Oishi, J.\ 2003, \apjl, 593, L23 
\bibitem[\protect\citeauthoryear{Compi{\`e}gne et al.}{2011}]{2011A&A...525A.103C} Compi{\`e}gne, M., Verstraete, L., Jones, A., et al.\ 2011, \aap, 525, A103 
\bibitem[\protect\citeauthoryear{DeLaney et al.}{2010}]{2010ApJ...725.2038D} DeLaney, T., Rudnick, L., Stage, M.~D., et al.\ 2010, \apj, 725, 2038 
\bibitem[\protect\citeauthoryear{DeLaney et al.}{2014}]{2014ApJ...785....7D} DeLaney, T., Kassim, N.~E., Rudnick, L., \& Perley, R.~A.\ 2014, \apj, 785, 7 
\bibitem[\protect\citeauthoryear{Docenko \& Sunyaev}{2010}]{2010A&A...509A..59D} Docenko, D., \& Sunyaev, R.~A.\ 2010, \aap, 509, A59
\bibitem[\protect\citeauthoryear{Dorschner et al.}{1995}]{1995A&A...300..503D} Dorschner, J., Begemann, B., Henning, T., Jaeger, C., \& Mutschke, H.\ 1995, \aap, 300, 503 
\bibitem[\protect\citeauthoryear{Douvion et al.}{2001}]{2001A&A...369..589D} Douvion, T., Lagage, P.~O., \& Pantin, E.\ 2001, \aap, 369, 589 
\bibitem[\protect\citeauthoryear{Draine}{1978}]{1978ApJS...36..595D} Draine, B.~T.\ 1978, \apjs, 36, 595   
\bibitem[\protect\citeauthoryear{Draine \& Salpeter}{1979a}]{1979ApJ...231...77D} Draine, B.~T., \& Salpeter, E.~E.\ 1979, \apj, 231, 77 
\bibitem[\protect\citeauthoryear{Draine \& Salpeter}{1979b}]{1979ApJ...231..438D} Draine, B.~T., \& Salpeter, E.~E.\ 1979, \apj, 231, 438 
%\bibitem[\protect\citeauthoryear{Draine \& Lee}{1984}]{1984ApJ...285...89D} Draine, B.~T., \& Lee, H.~M.\ 1984, \apj, 285, 89 
\bibitem[\protect\citeauthoryear{Dunne et al.}{2003}]{2003Natur.424..285D} Dunne, L., Eales, S., Ivison, R., Morgan, H., \& Edmunds, M.\ 2003, \nat, 424, 285 
\bibitem[\protect\citeauthoryear{Dunne et al.}{2009}]{2009MNRAS.394.1307D} Dunne, L., Maddox, S.~J., Ivison, R.~J., et al.\ 2009, \mnras, 394, 1307 
\bibitem[\protect\citeauthoryear{Dwek et al.}{1987}]{1987ApJ...315..571D} Dwek, E., Hauser, M.~G., Dinerstein, H.~L., Gillett, F.~C., \& Rice, W.~L.\ 1987, \apj, 315, 571 
\bibitem[\protect\citeauthoryear{Dwek et al.}{2007}]{2007ApJ...662..927D} Dwek, E., Galliano, F., \& Jones, A.~P.\ 2007, \apj, 662, 927 
\bibitem[\protect\citeauthoryear{Engelbracht et al.}{2007}]{2007PASP..119..994E} Engelbracht, C.~W., Blaylock, M., Su, K.~Y.~L., et al.\ 2007, \pasp, 119, 994 
\bibitem[\protect\citeauthoryear{Ennis et al.}{2006}]{2006ApJ...652..376E} Ennis, J.~A., Rudnick, L., Reach, W.~T., et al.\ 2006, \apj, 652, 376 
\bibitem[\protect\citeauthoryear{Eriksen et al.}{2009}]{2009ApJ...697...29E} Eriksen, K.~A., Arnett, D., McCarthy, D.~W., \& Young, P.\ 2009, \apj, 697, 29 
\bibitem[\protect\citeauthoryear{Fabbri et al.}{2011}]{2011MNRAS.418.1285F} Fabbri, J., Otsuka, M., Barlow, M.~J., et al.\ 2011, \mnras, 418, 1285 
\bibitem[\protect\citeauthoryear{Fazio et al.}{2004}]{2004ApJS..154...10F} Fazio, G.~G., Hora, J.~L., Allen, L.~E., et al.\ 2004, \apjs, 154, 10 
\bibitem[\protect\citeauthoryear{Fesen}{2001}]{2001ApJS..133..161F} Fesen, R.~A.\ 2001, \apjs, 133, 161 
\bibitem[\protect\citeauthoryear{Fesen et al.}{2001}]{2001AJ....122.2644F} Fesen, R.~A., Morse, J.~A., Chevalier, R.~A., et al.\ 2001, \aj, 122, 2644  
\bibitem[\protect\citeauthoryear{Fesen et al.}{2006}]{2006ApJ...645..283F} Fesen, R.~A., Hammell, M.~C., Morse, J., et al.\ 2006, \apj, 645, 283 
%\bibitem[\protect\citeauthoryear{Gail \& Sedlmayr}{2014}]{2014pccd.book.....G} Gail, H.-P., \& Sedlmayr, E.\ 2014, Physics and Chemistry of Circumstellar Dust Shells, by Hans-Peter Gail, Erwin Sedlmayr, Cambridge, UK: Cambridge University Press, 2014
%\bibitem[\protect\citeauthoryear{Galametz et al.}{2012}]{2012MNRAS.425..763G} Galametz, M., Kennicutt, R.~C., Albrecht, M., et al.\ 2012, \mnras, 425, 763 
\bibitem[\protect\citeauthoryear{Gall et al.}{2014}]{2014Natur.511..326G} Gall, C., Hjorth, J., Watson, D., et al.\ 2014, \nat, 511, 326 
%\bibitem[\protect\citeauthoryear{Galliano et al.}{2011}]{2011A&A...536A..88G} Galliano, F., Hony, S., Bernard, J.-P., et al.\ 2011, \aap, 536, A88 
\bibitem[\protect\citeauthoryear{Gomez et al.}{2012}]{2012ApJ...760...96G} Gomez, H.~L., Krause, O., Barlow, M.~J., et al.\ 2012, \apj, 760, 96
\bibitem[\protect\citeauthoryear{Gotthelf et al.}{2001}]{2001ApJ...552L..39G} Gotthelf, E.~V., Koralesky, B., Rudnick, L., et al.\ 2001, \apjl, 552, L39 
\bibitem[\protect\citeauthoryear{Griffin et al.}{2010}]{2010A&A...518L...3G} Griffin, M.~J., Abergel, A., Abreu, A., et al.\ 2010, \aap, 518, L3 
\bibitem[\protect\citeauthoryear{Groenewegen et al.}{2011}]{2011A&A...526A.162G} Groenewegen, M.~A.~T., Waelkens, C., Barlow, M.~J., et al.\ 2011, \aap, 526, A162 
\bibitem[\protect\citeauthoryear{Habing}{1968}]{1968BAN....19..421H} Habing, H.~J.\ 1968, \bain, 19, 421 
\bibitem[\protect\citeauthoryear{Hammer et al.}{2010}]{2010ApJ...714.1371H} Hammer, N.~J., Janka, H.-T., \& M{\"u}ller, E.\ 2010, \apj, 714, 1371 
\bibitem[\protect\citeauthoryear{Hines et al.}{2004}]{2004ApJS..154..290H} Hines, D.~C., Rieke, G.~H., Gordon, K.~D., et al.\ 2004, \apjs, 154, 290 
\bibitem[\protect\citeauthoryear{Houck et al.}{2004}]{2004SPIE.5487...62H} Houck, J.~R., Roellig, T.~L., Van Cleve, J., et al.\ 2004, \procspie, 5487, 62 
\bibitem[\protect\citeauthoryear{Hurford \& Fesen}{1996}]{1996ApJ...469..246H} Hurford, A.~P., \& Fesen, R.~A.\ 1996, \apj, 469, 246 
\bibitem[\protect\citeauthoryear{Indebetouw et al.}{2014}]{2014ApJ...782L...2I} Indebetouw, R., Matsuura, M., Dwek, E., et al.\ 2014, \apjl, 782, L2 
\bibitem[\protect\citeauthoryear{J{\"a}ger et al.}{2003}]{2003A&A...408..193J} J{\"a}ger, C., Dorschner, J., Mutschke, H., Posch, T., \& Henning, T.\ 2003, \aap, 408, 193 
%\bibitem[\protect\citeauthoryear{Jarrett et al.}{2011}]{2011ApJ...735..112J} Jarrett, T.~H., Cohen, M., Masci, F., et al.\ 2011, \apj, 735, 112 
\bibitem[\protect\citeauthoryear{Jarrett et al.}{2013}]{2013AJ....145....6J} Jarrett, T.~H., Masci, F., Tsai, C.~W., et al.\ 2013, \aj, 145, 6 
%\bibitem[\protect\citeauthoryear{Jones et al.}{2003}]{2003ApJ...587..227J} Jones, T.~J., Rudnick, L., DeLaney, T., \& Bowden, J.\ 2003, \apj, 587, 227
\bibitem[\protect\citeauthoryear{Jones}{2012a}]{2012A&A...540A...1J} Jones, A.~P.\ 2012, \aap, 540, A1 
\bibitem[\protect\citeauthoryear{Jones}{2012b}]{2012A&A...540A...2J} Jones, A.~P.\ 2012, \aap, 540, A2 
\bibitem[\protect\citeauthoryear{Jones}{2012c}]{2012A&A...542A..98J} Jones, A.~P.\ 2012, \aap, 542, A98   
\bibitem[\protect\citeauthoryear{Jones et al.}{2013}]{2013A&A...558A..62J} Jones, A.~P., Fanciullo, L., K{\"o}hler, M., et al.\ 2013, \aap, 558, A62 
\bibitem[\protect\citeauthoryear{Kennicutt et al.}{2003}]{2003PASP..115..928K} Kennicutt, R.~C., Jr., Armus, L., Bendo, G., et al.\ 2003, \pasp, 115, 928 
\bibitem[\protect\citeauthoryear{Kifonidis et al.}{2001}]{2001NuPhA.688..168K} Kifonidis, K., M{\"u}ller, E., \& Plewa, T.\ 2001, Nuclear Physics A, 688, 168 
\bibitem[\protect\citeauthoryear{Kilpatrick et al.}{2014}]{2014ApJ...796..144K} Kilpatrick, C.~D., Bieging, J.~H., \& Rieke, G.~H.\ 2014, \apj, 796, 144 
\bibitem[\protect\citeauthoryear{K{\"o}hler et al.}{2014}]{2014A&A...565L...9K} K{\"o}hler, M., Jones, A., \& Ysard, N.\ 2014, \aap, 565, L9 
\bibitem[\protect\citeauthoryear{Kotak et al.}{2009}]{2009ApJ...704..306K} Kotak, R., Meikle, W.~P.~S., Farrah, D., et al.\ 2009, \apj, 704, 306 
\bibitem[\protect\citeauthoryear{Kozasa et al.}{1991}]{1991A&A...249..474K} Kozasa, T., Hasegawa, H., \& Nomoto, K.\ 1991, \aap, 249, 474 
\bibitem[\protect\citeauthoryear{Krause et al.}{2004}]{2004Natur.432..596K} Krause, O., Birkmann, S.~M., Rieke, G.~H., et al.\ 2004, \nat, 432, 596 
%\bibitem[\protect\citeauthoryear{Krause et al.}{2005}]{2005Sci...308.1604K} Krause, O., Rieke, G.~H., Birkmann, S.~M., et al.\ 2005, Science, 308, 1604 
\bibitem[\protect\citeauthoryear{Krause et al.}{2008}]{2008Sci...320.1195K} Krause, O., Birkmann, S.~M., Usuda, T., et al.\ 2008, Science, 320, 1195 
\bibitem[\protect\citeauthoryear{Lamarre et al.}{2010}]{2010A&A...520A...9L} Lamarre, J.-M., Puget, J.-L., Ade, P.~A.~R., et al.\ 2010, \aap, 520, A9 
\bibitem[\protect\citeauthoryear{Laming \& Hwang}{2003}]{2003ApJ...597..347L} Laming, J.~M., \& Hwang, U.\ 2003, \apj, 597, 347 
%\bibitem[\protect\citeauthoryear{Li \& Draine}{2001}]{2001ApJ...554..778L} Li, A., \& Draine, B.~T.\ 2001, \apj, 554, 778 
\bibitem[\protect\citeauthoryear{Makiwa et al.}{2013}]{2013ApOpt..52.3864M} Makiwa, G., Naylor, D.~A., Ferlet, M., et al.\ 2013, \ao, 52, 3864 
%\bibitem[\protect\citeauthoryear{Makiwa et al.}{2016}]{2016MNRAS.458.2150M} Makiwa, G., Naylor, D.~A., van der Wiel, M.~H.~D., et al.\ 2016, \mnras, 458, 2150 
\bibitem[\protect\citeauthoryear{Mathis et al.}{1977}]{1977ApJ...217..425M} Mathis, J.~S., Rumpl, W., \& Nordsieck, K.~H.\ 1977, \apj, 217, 425 
\bibitem[\protect\citeauthoryear{Mathis et al.}{1983}]{1983A&A...128..212M} Mathis, J.~S., Mezger, P.~G., \& Panagia, N.\ 1983, \aap, 128, 212 
\bibitem[\protect\citeauthoryear{Matsuura et al.}{2011}]{2011Sci...333.1258M} Matsuura, M., Dwek, E., Meixner, M., et al.\ 2011, Science, 333, 1258 
\bibitem[\protect\citeauthoryear{Matsuura et al.}{2015}]{2015ApJ...800...50M} Matsuura, M., Dwek, E., Barlow, M.~J., et al.\ 2015, \apj, 800, 50 
\bibitem[\protect\citeauthoryear{Meikle et al.}{2007}]{2007ApJ...665..608M} Meikle, W.~P.~S., Mattila, S., Pastorello, A., et al.\ 2007, \apj, 665, 608 
\bibitem[\protect\citeauthoryear{Micelotta et al.}{2016}]{2016arXiv160202754M} Micelotta, E.~R., Dwek, E., \& Slavin, J.~D.\ 2016, MNRAS in press (arXiv:1602.02754)
%\bibitem[\protect\citeauthoryear{Micha{\l}owski}{2015}]{2015A&A...577A..80M} Micha{\l}owski, M.~J.\ 2015, \aap, 577, A80 
\bibitem[\protect\citeauthoryear{Mie}{1908}]{1908AnP...330..377M} Mie, G.\ 1908, Annalen der Physik, 330, 377
\bibitem[\protect\citeauthoryear{Milisavljevic \& Fesen}{2015}]{2015Sci...347..526M} Milisavljevic, D., \& Fesen, R.~A.\ 2015, Science, 347, 526 
\bibitem[\protect\citeauthoryear{Mookerjea et al.}{2006}]{2006MNRAS.371..761M} Mookerjea, B., Kantharia, N.~G., Roshi, D.~A., \& Masur, M.\ 2006, \mnras, 371, 761 
\bibitem[\protect\citeauthoryear{Morgan \& Edmunds}{2003}]{2003MNRAS.343..427M} Morgan, H.~L., \& Edmunds, M.~G.\ 2003, \mnras, 343, 427 
\bibitem[\protect\citeauthoryear{Mutschke et al.}{2002}]{2002A&A...392.1047M} Mutschke, H., Posch, T., Fabian, D., \& Dorschner, J.\ 2002, \aap, 392, 1047 
\bibitem[\protect\citeauthoryear{Naylor et al.}{2014}]{2014SPIE.9143E..2DN} Naylor, D.~A., Baluteau, J.-P., Bendo, G.~J., et al.\ 2014, \procspie, 9143, 91432D
\bibitem[\protect\citeauthoryear{Nozawa et al.}{2006}]{2006ApJ...648..435N} Nozawa, T., Kozasa, T., \& Habe, A.\ 2006, \apj, 648, 435
\bibitem[\protect\citeauthoryear{Nozawa et al.}{2010}]{2010ApJ...713..356N} Nozawa, T., Kozasa, T., Tominaga, N., et al.\ 2010, \apj, 713, 356 
\bibitem[\protect\citeauthoryear{Oni{\'c} \& Uro{\v s}evi{\'c}}{2015}]{2015ApJ...805..119O} Oni{\'c}, D., \& Uro{\v s}evi{\'c}, D.\ 2015, \apj, 805, 119 
\bibitem[\protect\citeauthoryear{Orlando et al.}{2016}]{2016ApJ...822...22O} Orlando, S., Miceli, M., Pumo, M.~L., \& Bocchino, F.\ 2016, \apj, 822, 22 
\bibitem[\protect\citeauthoryear{Ott}{2010}]{2010ASPC..434..139O} Ott, S.\ 2010, Astronomical Data Analysis Software and Systems XIX, 434, 139 
\bibitem[\protect\citeauthoryear{Owen \& Barlow}{2015}]{2015ApJ...801..141O} Owen, P.~J., \& Barlow, M.~J.\ 2015, \apj, 801, 141 
\bibitem[\protect\citeauthoryear{Owen}{2015}]{2015Owen} Owen, P.~J., 2015, Phd thesis, University College London
\bibitem[\protect\citeauthoryear{Papadopoulos et al.}{2004}]{2004MNRAS.351..147P} Papadopoulos, P.~P., Thi, W.-F., \& Viti, S.\ 2004, \mnras, 351, 147 
\bibitem[\protect\citeauthoryear{Paradis et al.}{2012}]{2012A&A...543A.103P} Paradis, D., Dobashi, K., Shimoikura, T., et al.\ 2012, \aap, 543, A103 
\bibitem[\protect\citeauthoryear{Pavlov et al.}{2000}]{2000ApJ...531L..53P} Pavlov, G.~G., Zavlin, V.~E., Aschenbach, B., Tr{\"u}mper, J., \& Sanwal, D.\ 2000, \apjl, 531, L53 
\bibitem[\protect\citeauthoryear{P{\'e}rez-Rend{\'o}n et al.}{2002}]{2002RMxAC..12...94P} P{\'e}rez-Rend{\'o}n, B., Garc{\'{\i}}a-Segura, G., \& Langer, N.\ 2002, Revista Mexicana de Astronomia y Astrofisica Conference Series, 12, 94 
\bibitem[\protect\citeauthoryear{P{\'e}rez-Rend{\'o}n et al.}{2009}]{2009A&A...506.1249P} P{\'e}rez-Rend{\'o}n, B., Garc{\'{\i}}a-Segura, G., \& Langer, N.\ 2009, \aap, 506, 1249  
\bibitem[\protect\citeauthoryear{Pilbratt et al.}{2010}]{2010A&A...518L...1P} Pilbratt, G.~L., Riedinger, J.~R., Passvogel, T., et al.\ 2010, \aap, 518, L1 
\bibitem[\protect\citeauthoryear{Planck Collaboration I}{2011}]{2011A&A...536A...1P} Planck Collaboration, Ade, P.~A.~R., Aghanim, N., et al.\ 2011, \aap, 536, A1 
%\bibitem[\protect\citeauthoryear{Planck Collaboration XI}{2014}]{2014A&A...571A..11P} Planck Collaboration, Abergel, A., Ade, P.~A.~R., et al.\ 2014, \aap, 571, A11 
\bibitem[\protect\citeauthoryear{Planck Collaboration XLIV}{2016a}]{2016arXiv160401029P} Planck Collaboration, Aghanim, N., Alves, M.~I.~R., et al.\ 2016, submitted to \aap (to arXiv:1604.01029) 
\bibitem[\protect\citeauthoryear{Planck Collaboration XXVI}{2016b}]{2015arXiv150702058P} Planck Collaboration, Ade, P.~A.~R., Aghanim, N., et al.\ 2016, \aap in press (arXiv:1507.02058) 
\bibitem[\protect\citeauthoryear{Planck Collaboration XXXI}{2016c}]{2016A&A...586A.134P} Planck Collaboration, Arnaud, M., Ashdown, M., et al.\ 2016, \aap, 586, A134 
\bibitem[\protect\citeauthoryear{Poglitsch et al.}{2010}]{2010A&A...518L...2P} Poglitsch, A., Waelkens, C., Geis, N., et al.\ 2010, \aap, 518, L2 
\bibitem[\protect\citeauthoryear{Pound \& Wolfire}{2008}]{2008ASPC..394..654P} Pound, M.~W., \& Wolfire, M.~G.\ 2008, Astronomical Data Analysis Software and Systems XVII, 394, 654 
\bibitem[\protect\citeauthoryear{Priddey et al.}{2003}]{2003MNRAS.344L..74P} Priddey, R.~S., Isaak, K.~G., McMahon, R.~G., Robson, E.~I., \& Pearson, C.~P.\ 2003, \mnras, 344, L74 
\bibitem[\protect\citeauthoryear{Reed et al.}{1995}]{1995ApJ...440..706R} Reed, J.~E., Hester, J.~J., Fabian, A.~C., \& Winkler, P.~F.\ 1995, \apj, 440, 706
\bibitem[\protect\citeauthoryear{Reynoso \& Goss}{2002}]{2002ApJ...575..871R} Reynoso, E.~M., \& Goss, W.~M.\ 2002, \apj, 575, 871 
\bibitem[\protect\citeauthoryear{Rho et al.}{2003}]{2003ApJ...592..299R} Rho, J., Reynolds, S.~P., Reach, W.~T., et al.\ 2003, \apj, 592, 299 
\bibitem[\protect\citeauthoryear{Rho et al.}{2008}]{2008ApJ...673..271R} Rho, J., Kozasa, T., Reach, W.~T., et al.\ 2008, \apj, 673, 271-282 
\bibitem[\protect\citeauthoryear{Rieke et al.}{2004}]{2004ApJS..154...25R} Rieke, G.~H., Young, E.~T., Engelbracht, C.~W., et al.\ 2004, \apjs, 154, 25 
\bibitem[\protect\citeauthoryear{Rouleau \& Martin}{1991}]{1991ApJ...377..526R} Rouleau, F., \& Martin, P.~G.\ 1991, \apj, 377, 526 
\bibitem[\protect\citeauthoryear{Rowlands et al.}{2014}]{2014MNRAS.441.1040R} Rowlands, K., Gomez, H.~L., Dunne, L., et al.\ 2014, \mnras, 441, 1040 
\bibitem[\protect\citeauthoryear{Sarangi \& Cherchneff}{2013}]{2013ApJ...776..107S} Sarangi, A., \& Cherchneff, I.\ 2013, \apj, 776, 107 
%\bibitem[\protect\citeauthoryear{Semenov et al.}{2003}]{2003A&A...410..611S} Semenov, D., Henning, T., Helling, C., Ilgner, M., \& Sedlmayr, E.\ 2003, \aap, 410, 611 
\bibitem[\protect\citeauthoryear{Sibthorpe et al.}{2010}]{2010ApJ...719.1553S} Sibthorpe, B., Ade, P.~A.~R., Bock, J.~J., et al.\ 2010, \apj, 719, 1553 
\bibitem[\protect\citeauthoryear{Silvia et al.}{2010}]{2010ApJ...715.1575S} Silvia, D.~W., Smith, B.~D., \& Shull, J.~M.\ 2010, \apj, 715, 1575 
\bibitem[\protect\citeauthoryear{Slavin et al.}{2004}]{2004ApJ...614..796S} Slavin, J.~D., Jones, A.~P., \& Tielens, A.~G.~G.~M.\ 2004, \apj, 614, 796 
\bibitem[\protect\citeauthoryear{Smith et al.}{2007}]{2007PASP..119.1133S} Smith, J.~D.~T., Armus, L., Dale, D.~A., et al.\ 2007, \pasp, 119, 1133 
%\bibitem[\protect\citeauthoryear{Smith et al.}{2009}]{2009ApJ...693..713S} Smith, J.~D.~T., Rudnick, L., Delaney, T., et al.\ 2009, \apj, 693, 713 
\bibitem[\protect\citeauthoryear{Sugerman et al.}{2006}]{2006Sci...313..196S} Sugerman, B.~E.~K., Ercolano, B., Barlow, M.~J., et al.\ 2006, Science, 313, 196 
\bibitem[\protect\citeauthoryear{Swinyard et al.}{2010}]{2010A&A...518L...4S} Swinyard, B.~M., Ade, P., Baluteau, J.-P., et al.\ 2010, \aap, 518, L4 
\bibitem[\protect\citeauthoryear{Swinyard et al.}{2014}]{2014MNRAS.440.3658S} Swinyard, B.~M., Polehampton, E.~T., Hopwood, R., et al.\ 2014, \mnras, 440, 3658 
\bibitem[\protect\citeauthoryear{Tielens et al.}{1994}]{1994ApJ...431..321T} Tielens, A.~G.~G.~M., McKee, C.~F., Seab, C.~G., \& Hollenbach, D.~J.\ 1994, \apj, 431, 321 
\bibitem[\protect\citeauthoryear{Tielens}{2005}]{2005pcim.book.....T} Tielens, A.~G.~G.~M.\ 2005, The Physics and Chemistry of the Interstellar Medium, by A.~G.~G.~M.~Tielens, pp.~.~ISBN 0521826349.~Cambridge, UK: Cambridge University Press,  2005., 
\bibitem[\protect\citeauthoryear{Todini \& Ferrara}{2001}]{2001MNRAS.325..726T} Todini, P., \& Ferrara, A.\ 2001, \mnras, 325, 726 
\bibitem[\protect\citeauthoryear{Vinyaikin}{2014}]{2014ARep...58..626V} Vinyaikin, E.~N.\ 2014, Astronomy Reports, 58, 626 
\bibitem[\protect\citeauthoryear{Wallstr{\"o}m et al.}{2013}]{2013A&A...558L...2W} Wallstr{\"o}m, S.~H.~J., Biscaro, C., Salgado, F., et al.\ 2013, \aap, 558, L2 
\bibitem[\protect\citeauthoryear{Watson et al.}{2015}]{2015Natur.519..327W} Watson, D., Christensen, L., Knudsen, K.~K., et al.\ 2015, \nat, 519, 327 
\bibitem[\protect\citeauthoryear{Werner et al.}{2004}]{2004ApJS..154....1W} Werner, M.~W., Roellig, T.~L., Low, F.~J., et al.\ 2004, \apjs, 154, 1 
\bibitem[\protect\citeauthoryear{Wesson et al.}{2015}]{2015MNRAS.446.2089W} Wesson, R., Barlow, M.~J., Matsuura, M., \& Ercolano, B.\ 2015, \mnras, 446, 2089 
\bibitem[\protect\citeauthoryear{Willingale et al.}{2002}]{2002A&A...381.1039W} Willingale, R., Bleeker, J.~A.~M., van der Heyden, K.~J., Kaastra, J.~S., \& Vink, J.\ 2002, \aap, 381, 1039 
\bibitem[\protect\citeauthoryear{Willingale et al.}{2003}]{2003A&A...398.1021W} Willingale, R., Bleeker, J.~A.~M., van der Heyden, K.~J., \& Kaastra, J.~S.\ 2003, \aap, 398, 1021 
\bibitem[\protect\citeauthoryear{Woosley \& Weaver}{1995}]{1995ApJS..101..181W} Woosley, S.~E., \& Weaver, T.~A.\ 1995, \apjs, 101, 181 
\bibitem[\protect\citeauthoryear{Wright et al.}{1999}]{1999ApJ...518..284W} Wright, M., Dickel, J., Koralesky, B., \& Rudnick, L.\ 1999, \apj, 518, 284 
\bibitem[\protect\citeauthoryear{Wright et al.}{2010}]{2010AJ....140.1868W} Wright, E.~L., Eisenhardt, P.~R.~M., Mainzer, A.~K., et al.\ 2010, \aj, 140, 1868-1881 
%\bibitem[\protect\citeauthoryear{Wu et al.}{2013}]{2013A&A...556A.116W} Wu, R., Polehampton, E.~T., Etxaluze, M., et al.\ 2013, \aap, 556, A116 
\bibitem[\protect\citeauthoryear{Young et al.}{2006}]{2006ApJ...640..891Y} Young, P.~A., Fryer, C.~L., Hungerford, A., et al.\ 2006, \apj, 640, 891 
%\bibitem[\protect\citeauthoryear{Zubko et al.}{2004}]{2004ApJS..152..211Z} Zubko, V., Dwek, E., \& Arendt, R.~G.\ 2004, \apjs, 152, 211
\end{thebibliography}

% This method is tedious and prone to error if you have lots of references; 

%%%%%%%%%%%%%%%%%%%%%%%%%%%%%%%%%%%%%%%%%%%%%%%%%%

%%%%%%%%%%%%%%%%% APPENDICES %%%%%%%%%%%%%%%%%%%%%

\appendix

\section{Planck flux measurements}
\label{Planck_flux}
In this section, we compare various \textit{Planck} measurements and explain why we opted to use the customised aperture photometry results from \citet{2016A&A...586A.134P}. The Second \textit{Planck} Catalogue of Compact Sources (PCCS2, \citealt{2015arXiv150702058P}) reports four different flux measurements for Cas A (see Table \ref{Table_Planckflux}) based on different source extraction techniques. The DETFLUX relies on the assumption of a point-like source, and filters out any extended emission. The other flux measurements are determined from the all-sky maps at the known position of Cas A. The APERFLUX corresponds to the flux measurement within a circular aperture centred at the position of the source, with radius equal to the FWHM of the effective beam at that frequency. The PSFFLUX is derived by fitting a PSF model to the flux observed at the source position. The GAUFLUX is obtained by fitting a two-dimensional Gaussian model with variable amplitude, size and shape to the emission at the source position. The PCCS2 catalogue reports the fluxes for the LFI filters, while the HFI fluxes are retrieved from the PCCS2E catalogue \citep{2015arXiv150702058P}. The latter catalogue does not contain a flux measurement for Cas A at 857\,GHz, which is not surprising given the high degree of confusion with ISM dust at that frequency. 

While the different flux measurements for the low-frequency filters are compatible, we find huge variations between the different flux extraction techniques at higher frequencies ($\nu$ $\geq$ 100\,GHz). The GAUFLUX measurement at 545 GHz is several times higher compared to the other flux measurements, which can be attributed to a larger estimated source size due to the dominant contribution of interstellar dust emission, which has a more extended morphology compared to Cas\,A at those wavelengths. The DETFLUX measurements underestimate the fluxes because of the better resolution in those bands which starts to resolve the emission of Cas\,A and, thus, misses some of the flux due to its point source assumption. Based on a comparison of the various \textit{Planck} fluxes with published fluxes at 87 GHz (96.7\,Jy, \citealt{1999ApJ...518..284W}), 353 GHz (50.8 Jy, \citealt{2003Natur.424..285D}), and our \textit{Herschel} SPIRE 500\,$\mu$m flux density (59.2\,Jy, see Table \ref{Table_Fluxfraction}), we conclude that the different flux measurements from the PCCS2 catalog all underestimate the fluxes at higher frequencies, where Cas\,A starts to be resolved with respect to the \textit{Planck} beam. We therefore use the customised flux measurements from \citet{2016A&A...586A.134P} where the source size and beam size were considered in order to optimise the extraction area for aperture photometry (see Table \ref{Table_Planckflux}, last row). Their flux uncertainties only account for observational uncertainties, while the measurements are undoubtedly affected by the chosen aperture for source extraction and background measurement. After experimenting with different source apertures and background regions in the \textit{Herschel} maps, we assume more conservative uncertainties on the \textit{Planck} photometry measurements which are a factor of two higher than reported by \citet{2016A&A...586A.134P}. 

% Table 1
\begin{table*}
	\centering
	\caption{Overview of the different flux measurements $F_{\nu}$ in units of Jy for Cas\,A from the Second \textit{Planck} Catalogue of Compact Sources (PCCS2). The FWHM of the \textit{Planck} beam in the different LFI and HFI filters is reported in the second row. The last row presents the fluxes reported by \citet{2016A&A...586A.134P}, that were extracted within an aperture optimised to the intrinsic source size and \textit{Planck} beam at every frequency. \citet{2016A&A...586A.134P} does not report any flux measurement at 545\,GHz due to the high level of confusion with the ISM dust emission. The uncertainties are the original errors reported by \citet{2016A&A...586A.134P}, while we assume uncertainties that are a factor of two larger due to the dependence of the flux measurement on the choice of the aperture for source extraction and background determination.}
	\label{Table_Planckflux}
	\begin{tabular}{lcccccccc} % four columns, alignment for each
		\hline
               Flux type & $F_{\text{545}}$ & $F_{\text{353}}$ & $F_{\text{217}}$ & $F_{\text{143}}$ & $F_{\text{100}}$ & $F_{\text{70}}$ & $F_{\text{44}}$ & $F_{\text{30}}$ \\
		\hline
                FWHM [$\arcmin$] & 5.0 & 5.0 & 5.5 & 7.1 & 10 & 14 & 24 & 33 \\
		\hline
                DETFLUX & 23.7$\pm$1.8 &  25.2$\pm$0.6 & 32.6$\pm$0.4 & 48.2$\pm$0.4 &  72.1$\pm$0.6 & 100.0$\pm$0.2 & 135.2$\pm$0.4 & 182.2$\pm$0.5 \\
                APERFLUX & 50.6$\pm$20.6 &  40.5$\pm$6.1 & 46.7$\pm$2.2 &  61.8$\pm$0.8 &  81.0$\pm$0.7 & 105.0$\pm$0.6 & 141.5$\pm$1.9 & 189.5$\pm$2.1 \\
                PSFFLUX &  55.3 $\pm$32.4 & 40.9$\pm$10.7 & 42.7$\pm$3.6 & 57.3$\pm$1.5 &  77.7$\pm$0.6 & 100.6$\pm$1.1 & 138.3$\pm$1.9 & 188.6$\pm$2.3 \\
                GAUFLUX & 129.0$\pm$0.9 & 38.0$\pm$4.9  & 45.4$\pm$2.6 &  60.2$\pm$2.5 & 79.1$\pm$2.9 & 97.9$\pm$0.3 & 133.6$\pm$0.3 & 189.5$\pm$0.4 \\
		\hline
                \textbf{Planck XXXI\,2016} & - & \textbf{52$\pm$7} & \textbf{62$\pm$4} & \textbf{73$\pm$4} & \textbf{94$\pm$5} & \textbf{109$\pm$6} & \textbf{140$\pm$7} & \textbf{228$\pm$11} \\
		\hline
	\end{tabular}
\end{table*}

\section{Line emission}
\label{Sect_line}

Several of the continuum bands contain a substantial contribution from bright emission lines contributing to the radiative cooling budget of the supernova remnant. We used the \textit{Spitzer} IRS spectra, PACS IFU spectra and SPIRE FTS spectra to determine the contribution of line emission to the IRAC, \textit{WISE}, PACS and SPIRE photometric wavebands.

In the IRAC\,8\,$\mu$m and \textit{WISE}\,12\,$\mu$m bands, there are potential contributions from the [Ar~{\sc{ii}}] 6.9\,$\mu$m, [Ar~{\sc{iii}}] 8.99\,$\mu$m, [S~{\sc{iv}}] 10.51\,$\mu$m, [Ne~{\sc{ii}}] 12.81\,$\mu$m and [Ne~{\sc{iii}}] 15.56\,$\mu$m nebular lines, while the \textit{WISE}\,22\,$\mu$m and MIPS\,24\,$\mu$m bands include possible emission from the [S~{\sc{iii}}] 18.71\,$\mu$m, [Fe~{\sc{iii}}] 22.93\,$\mu$m, [O~{\sc{iv}}] 25.89\,$\mu$m and [Fe~{\sc{ii}}] 25.99\,$\mu$m lines. The \textit{Spitzer} IRS spectra were used to estimate the contribution of these nebular lines in Cas\,A. To derive line emission maps, we fitted the continuum with a first order polynomial, subtracted it from each of the line channels and integrated over the continuum-subtracted SED to determine the line flux in every pixel. After fitting the line emission maps, we convolved the line maps from their original resolution to the SPIRE\,500\,$\mu$m FWHM resolution of $\sim$36.3$\arcsec$. Taking into account the transmission of the IRAC\,8\,$\mu$m, \textit{WISE}\,12 and 22\,$\mu$m, MIPS\,24\,$\mu$m at the wavelengths of these lines, we derived non-negligible contributions from the [Ar~{\sc{ii}}] and [Ar~{\sc{iii}}] lines to the IRAC\,8\,$\mu$m image (41.6\,$\%$) and from the [S~{\sc{iv}}] and [Ne~{\sc{ii}}] lines to the \textit{WISE}\,12\,$\mu$m image (17.0$\%$). The latter values are consistent with the line contributions of 28\,$\%$ and 18\,$\%$ to the 8.28\,$\mu$m and 12\,$\mu$m MSX bands estimated by \citet{2004ApJS..154..290H}. The [O~{\sc{iv}}] 25.89\,$\mu$m and [Fe~{\sc{ii}}] 25.99\,$\mu$m lines contribute 1.3\,$\%$ and 22.1\,$\%$ of the remnant's emission in the \textit{WISE}\,22\,$\mu$m and MIPS\,24\,$\mu$m images, respectively.

\begin{figure*}
	\includegraphics[width=5.8cm]{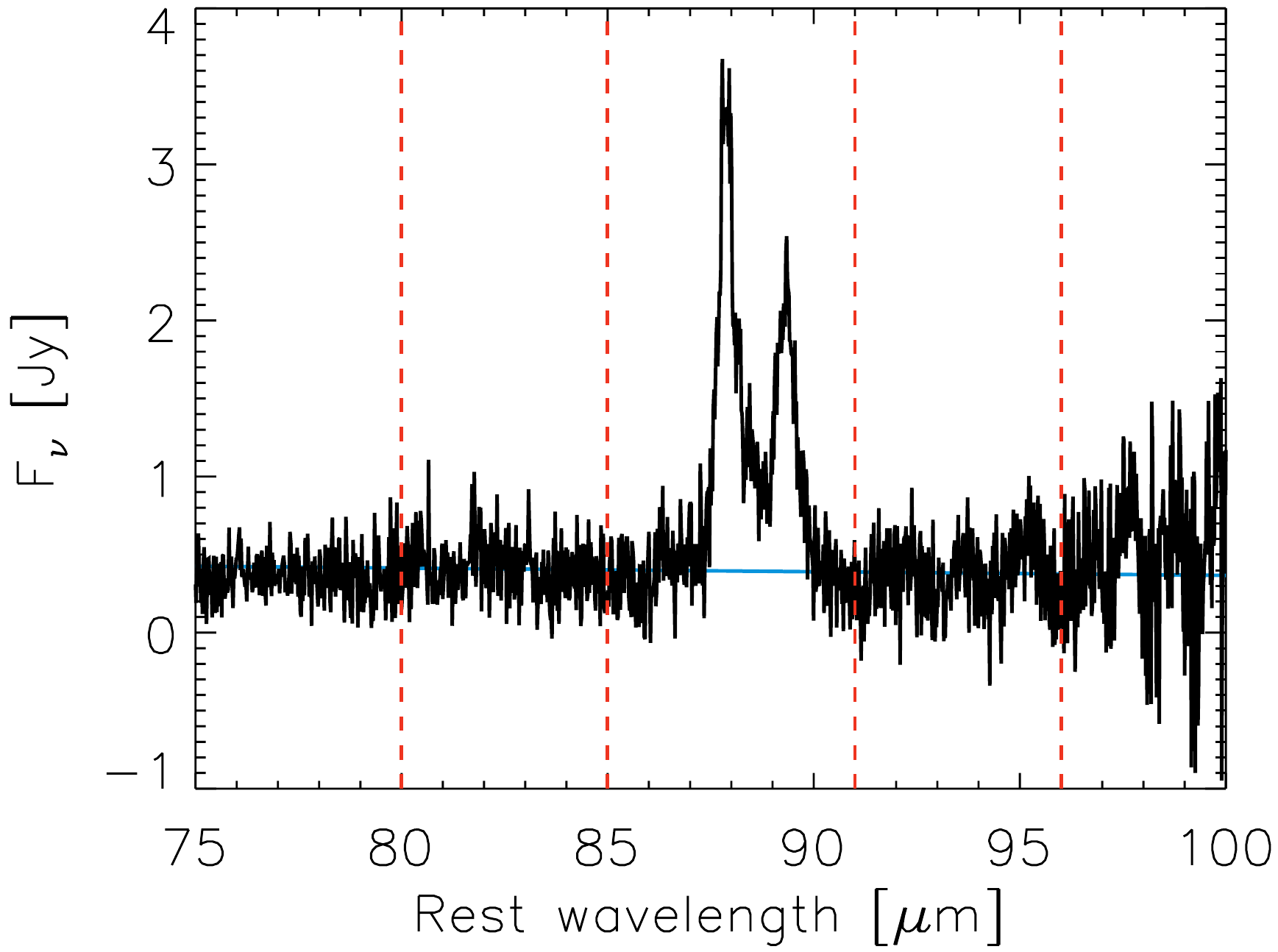} 
	\includegraphics[width=5.8cm]{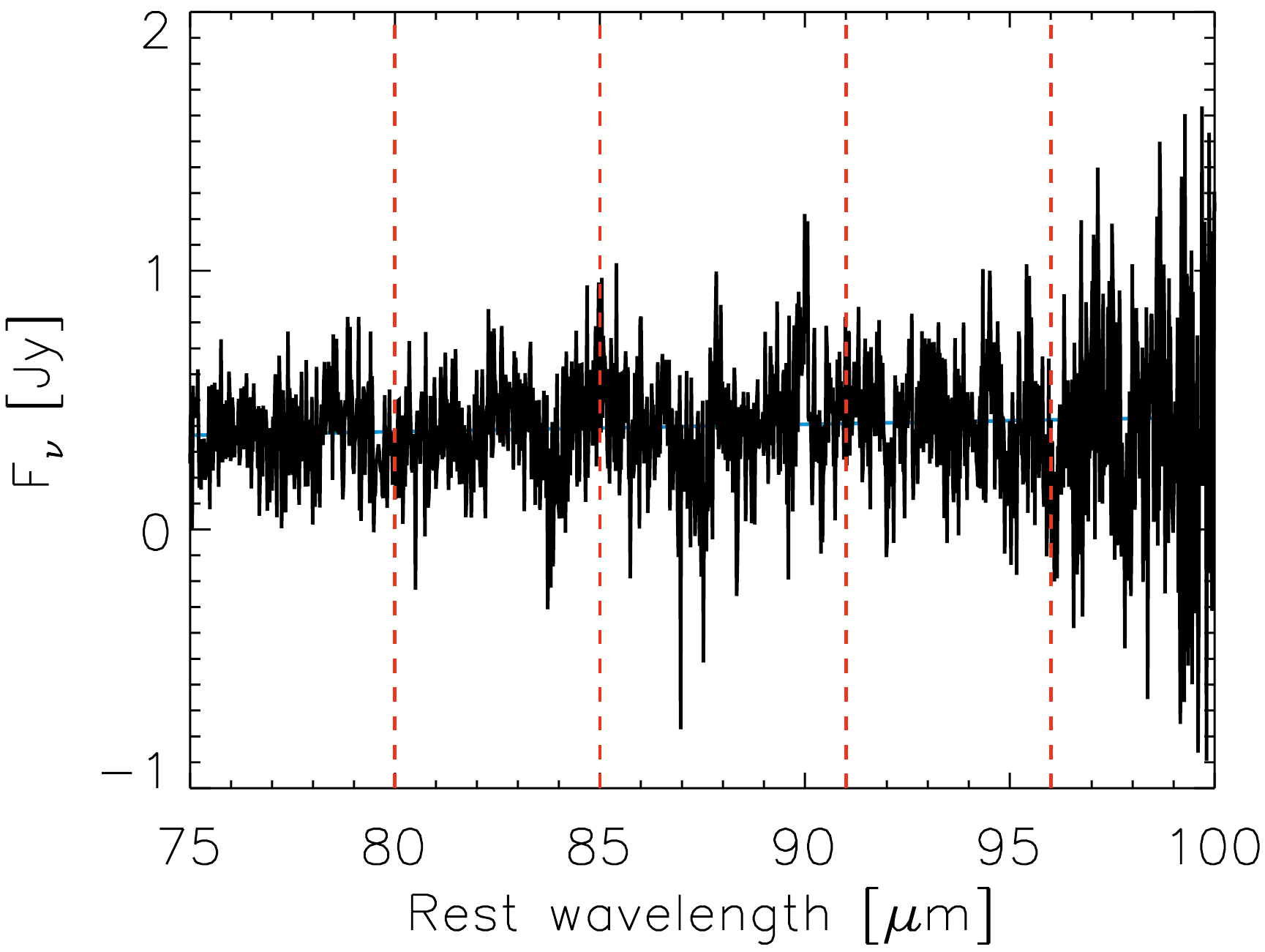} 
	\includegraphics[width=5.8cm]{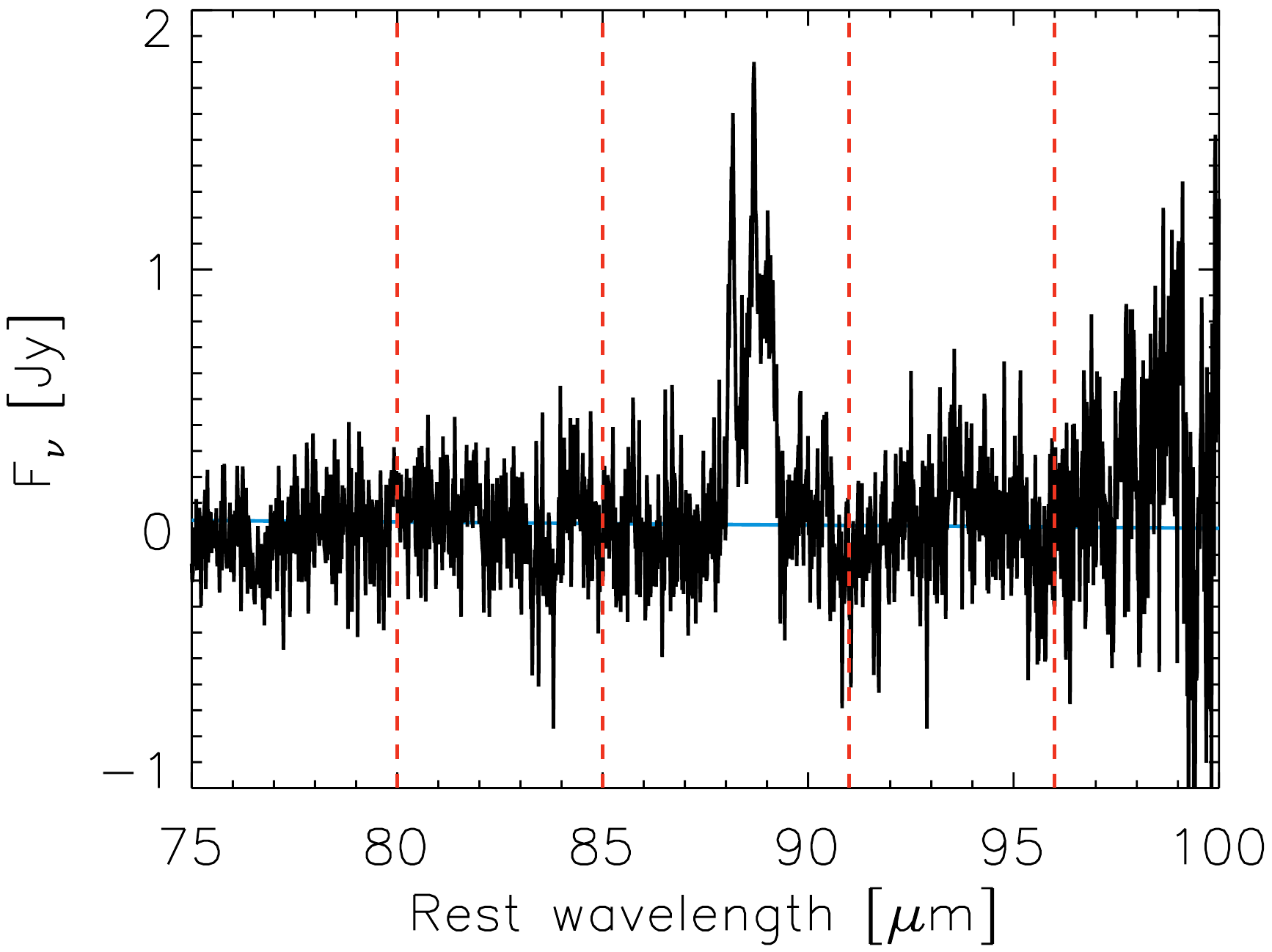} \\
	\includegraphics[width=5.8cm]{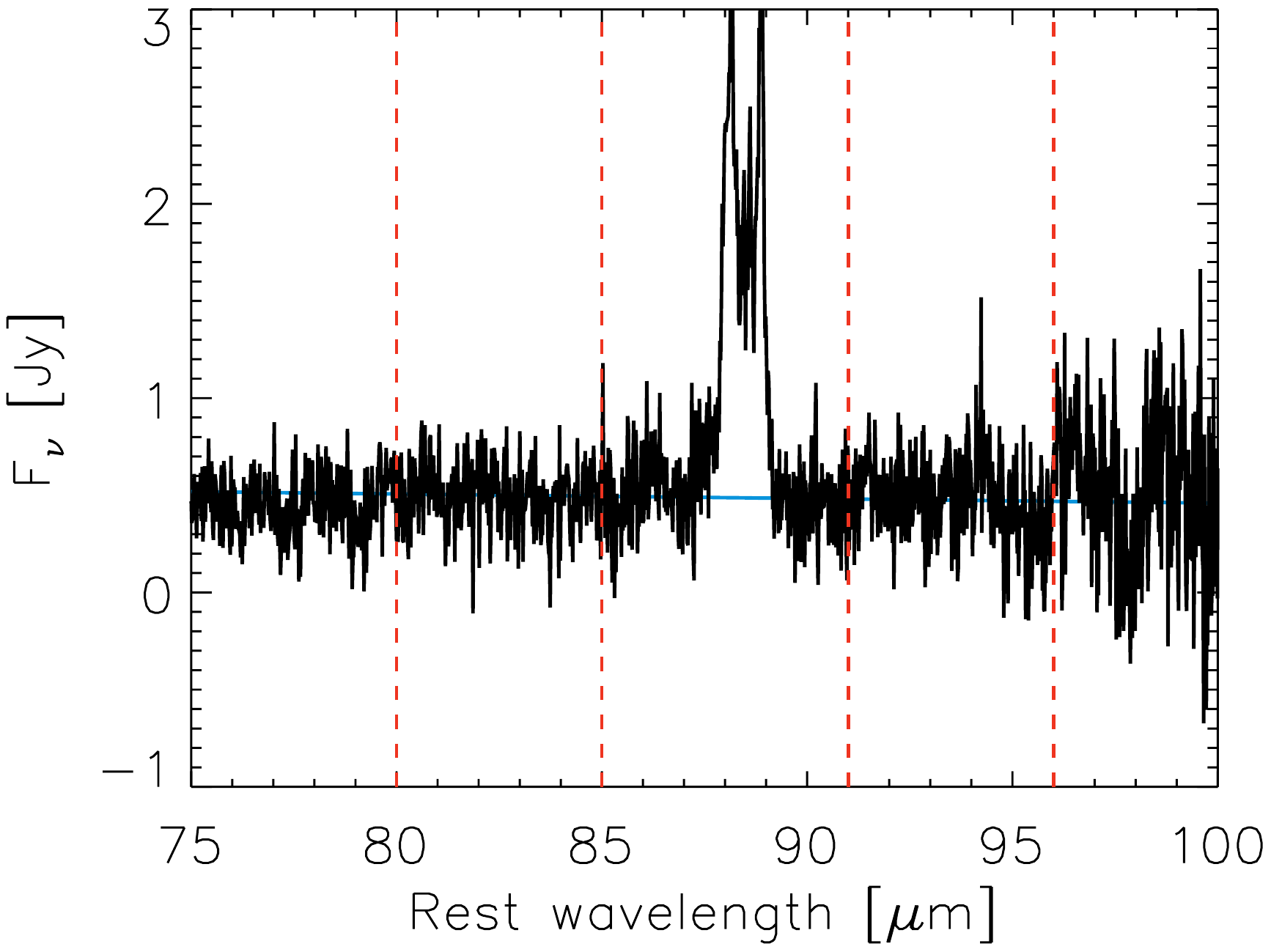} 
	\includegraphics[width=5.8cm]{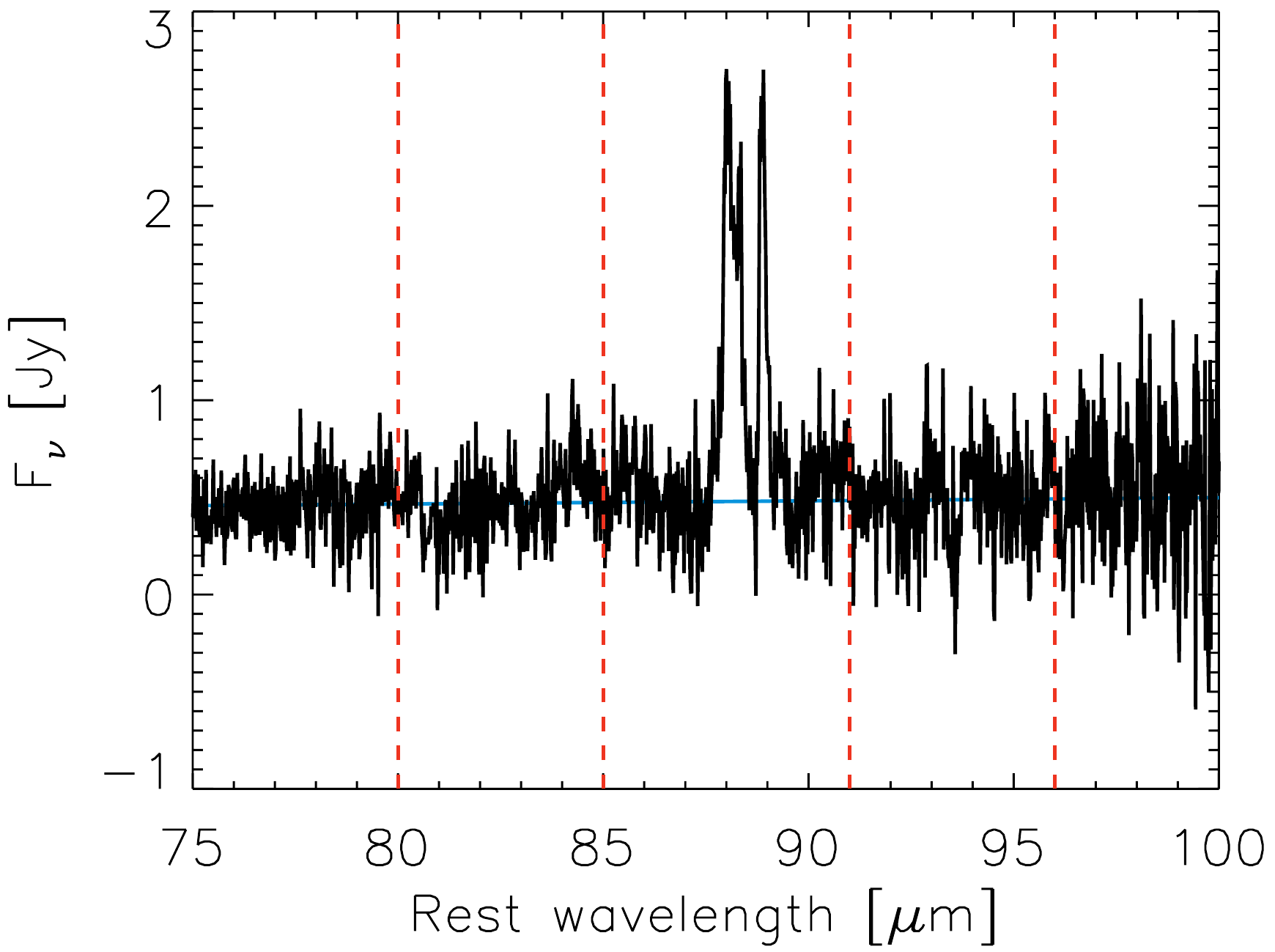} 
	\includegraphics[width=5.8cm]{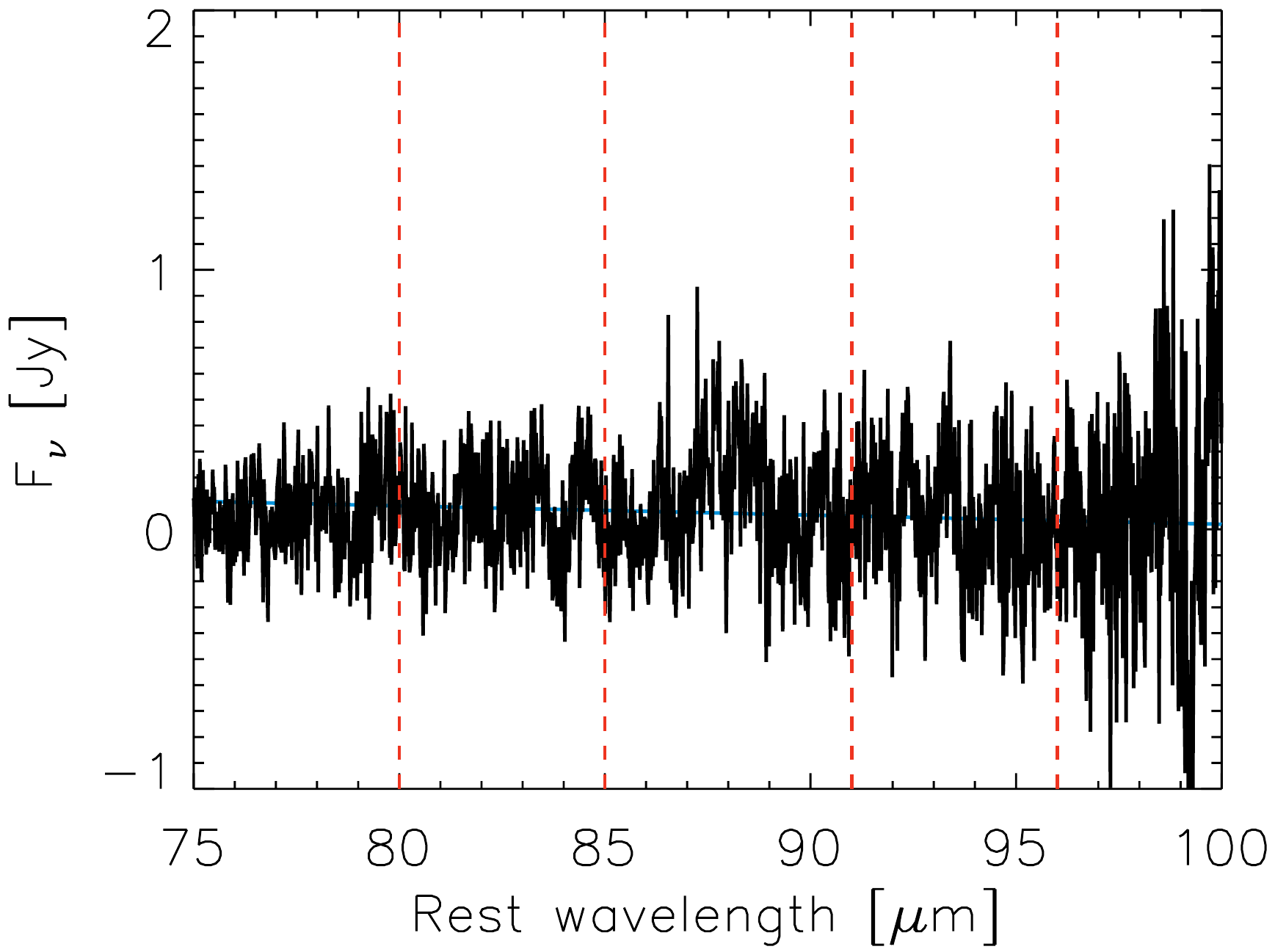} \\
	\includegraphics[width=5.8cm]{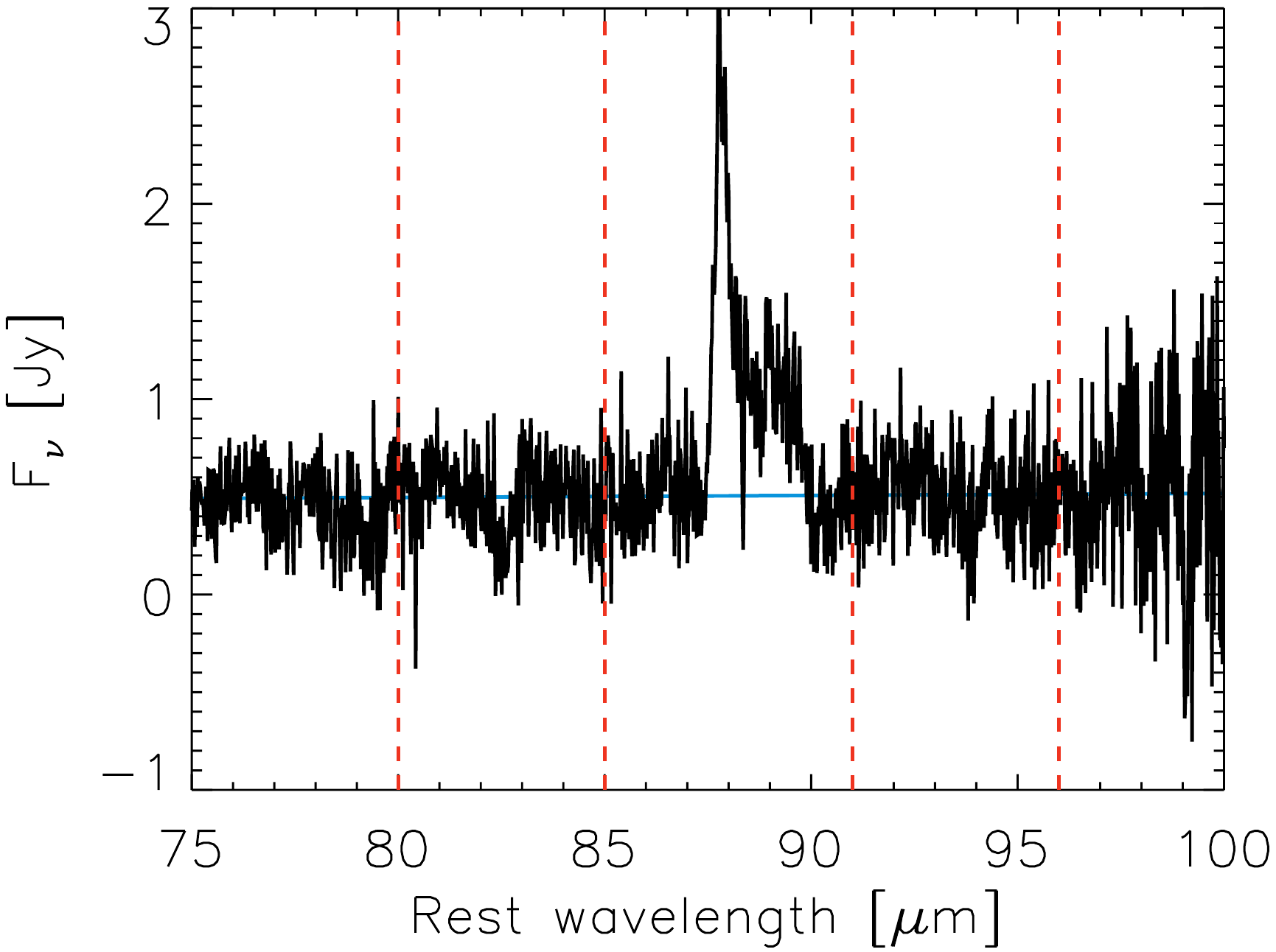} 
	\includegraphics[width=5.8cm]{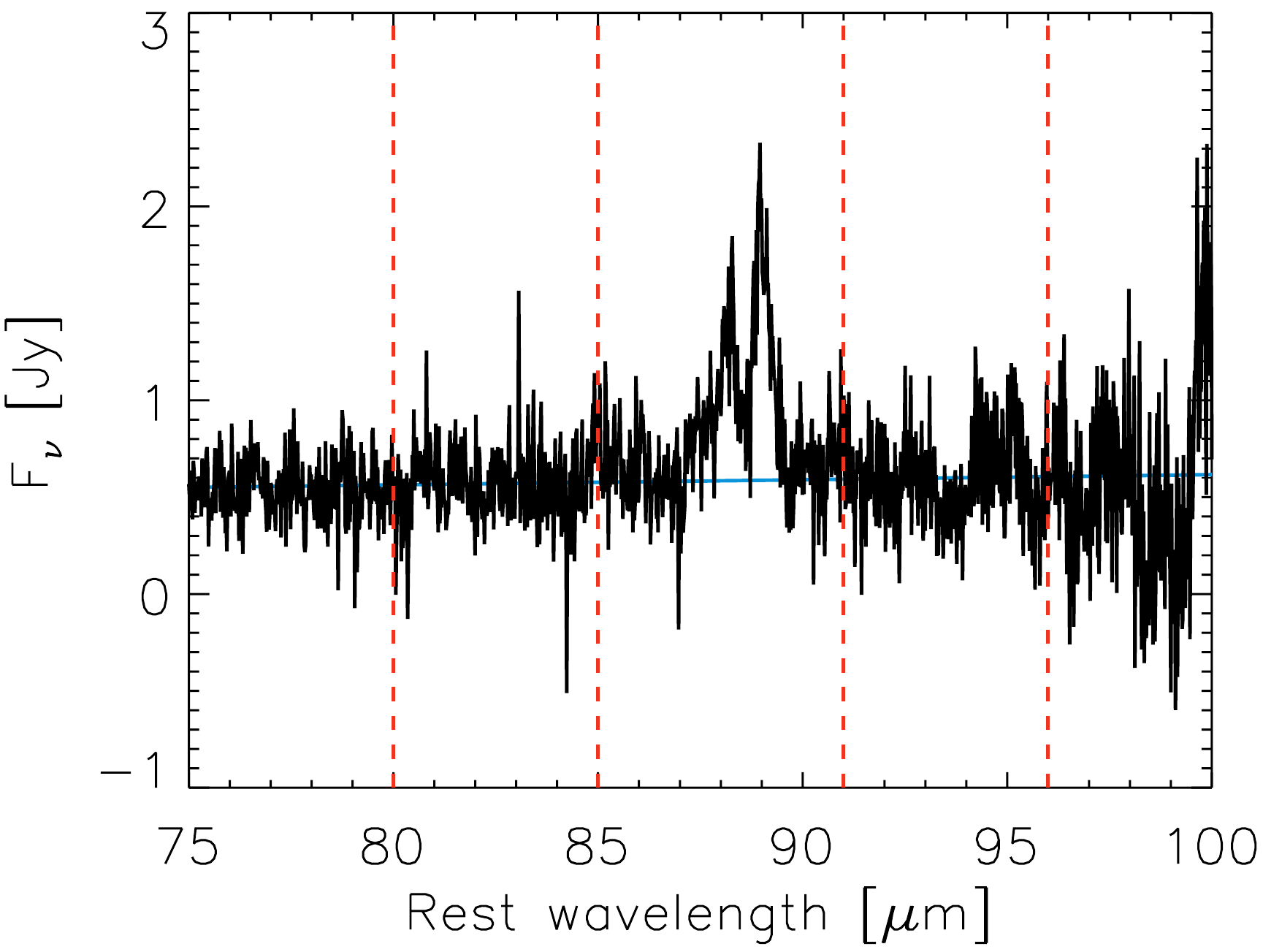} 
	\includegraphics[width=5.8cm]{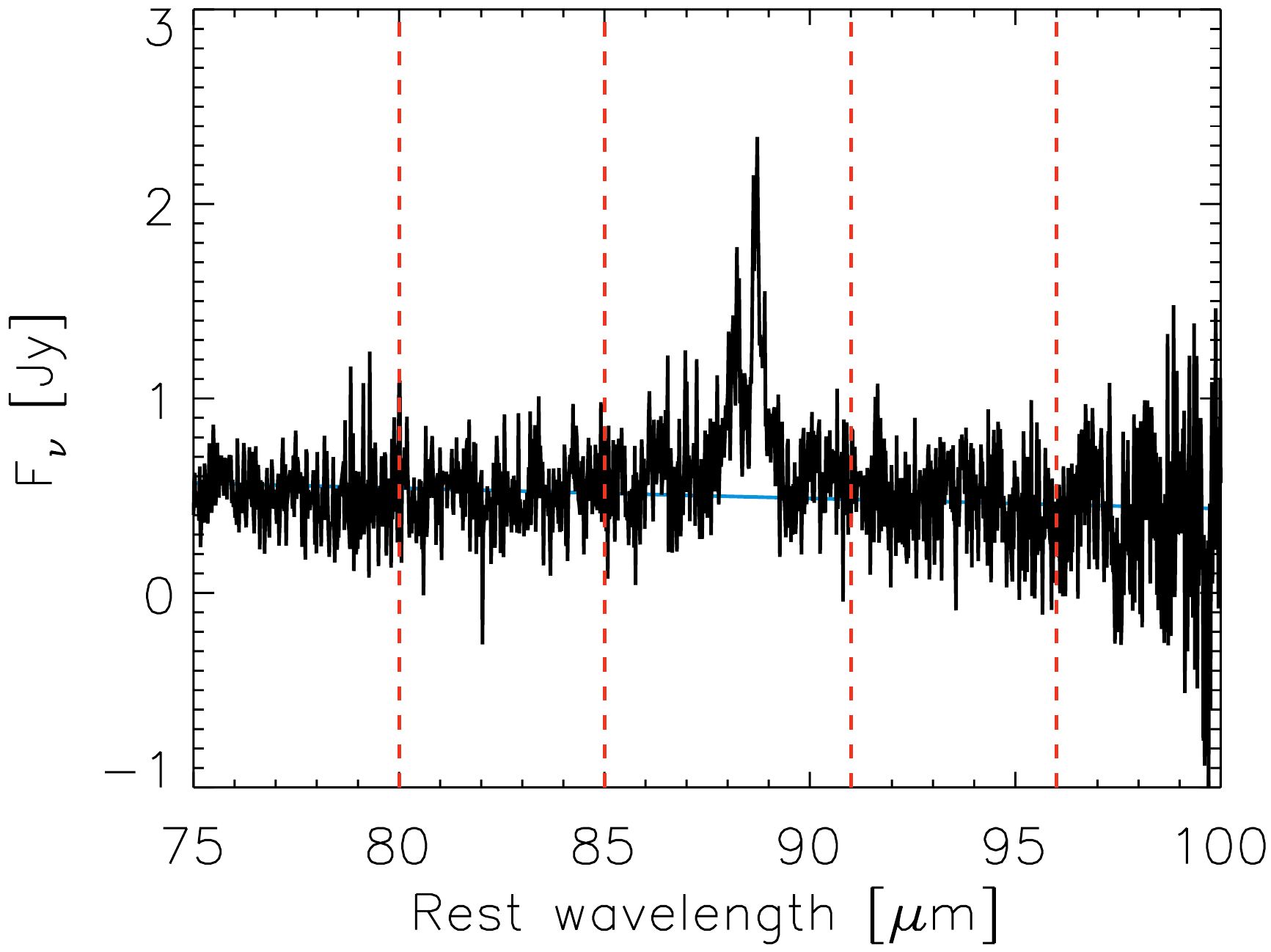} \\	
    \caption{The [O~{\sc{iii}}]\,88\,$\mu$m line spectra for the nine different pointings observed with the PACS IFU mode. The black line represents the observed spectrum, while the red dashed lines indicate the wavelength ranges on each side of the line profile that are used to fit the continuum. The solid blue lines represent the best fit to the continuum level.} 
    \label{Lineprofile_OIII}
\end{figure*}

The PACS IFU spectrometer observed the four wavelength ranges 51-72, 70-105, 102-146 and 140-210\,$\mu$m at each of the nine targeted positions in Cas\,A (see Fig. \ref{CasA_AOR_PACS_IFU}). At each position the four individual spectra were combined to a single spectrum by scaling them to match the fluxes in the overlapping wavelength ranges. To determine the [O~{\sc{i}}] 63\,$\mu$m, [O~{\sc{iii}}] 88\,$\mu$m, [O~{\sc{i}}] 145\,$\mu$m and [C~{\sc{ii}}] 158\,$\mu$m line fluxes, we fitted the continuum with a first degree polynomial, subtracted the continuum and summed the flux under the line profile. Figure \ref{Lineprofile_OIII} shows the [O~{\sc{iii}}] 88\,$\mu$m line profiles for each of the nine targeted fields. We did not use Gaussian models to fit the line emission since the lines are often very irregular and separated into blue- and red-shifted components. By comparing the line fluxes to the PACS broadband fluxes within the same regions (taking into account the transmission of the PACS filters at the central wavelengths of these lines), we were able to derive the contribution of the different lines to the PACS broadband maps. The contribution of the [O~{\sc{iii}}] 88\,$\mu$m line to the PACS\,70\,$\mu$m image is usually 1$\%$ or lower, and so is within the limits of the photometric uncertainties in this waveband. Similarly, the contributions of the [O~{\sc{i}}] 63\,$\mu$m and  [C~{\sc{ii}}] 158\,$\mu$m lines to the PACS\,70\,$\mu$m and 160\,$\mu$m wavebands, respectively, are less than 1\,$\%$ and thus within the few percent photometric uncertainty. The [O~{\sc{iii}}] 88\,$\mu$m line contribution to the PACS\,100\,$\mu$m waveband is more significant, with 2.7\,$\%$ to 5.3\,$\%$ of the PACS\,100\,$\mu$m emission attributable to line cooling from ionised gas. The largest contribution (5.3\,$\%$) comes from the central regions of Cas\,A, in the area internal to the reverse shock, with the second highest contributions (4-5\,$\%$) in the shocked ejecta in the south. Since we do not have a full map of the [O~{\sc{iii}}] 88\,$\mu$m line emission to correct the PACS\,100\,$\mu$m map for line contamination, we use the  [Si~{\sc{ii}}] 34.82\,$\mu$m line map derived from the \textit{Spitzer} IRS spectra. Even though the ionisation potential (IP) of S$^{++}$ (34.79\,eV) more closely resembles the IP of O$^{+}$ (35.12 eV), the critical density (for collisions with electrons) of the [S~{\sc{iv}}]\,10.51\,$\mu$m line ($n_{\text{crit}}$ = 5$\times$10$^{4}$ cm$^{-3}$) is two orders of magnitude higher compared to the critical density for [O~{\sc{iii}}]. The [S~{\sc{iv}}] 10.51\,$\mu$m line emission in Cas\,A is also significantly weaker than the [Si~{\sc{ii}}] 34.82\,$\mu$m line. 

With an ionisation potential of 8.15 eV to produce Si$^{+}$, the [Si~{\sc{ii}}] line can trace both the neutral and ionised gas phase. Because the gas in Cas\,A can be ionised due to shocks and/or the radiation field, we assume that most of the [Si~{\sc{ii}}] emission comes from the ionised gas phase. The latter assumption is supported by the absence of any detected [O~{\sc{i}}] 63\,$\mu$m or 145\,$\mu$m line emission from Cas\,A \citep{2010A&A...509A..59D,2015Owen}. With a critical density of 10$^{3}$ cm$^{-3}$ for collisions with electrons, and an excitation temperature for the upper level of 413\,K, the [Si~{\sc{ii}}] line is thought to trace a gas reservoir similar to the [O~{\sc{iii}}] 88\,$\mu$m line with a critical density $n_{\text{crit}}$=5$\times$10$^{2}$ cm$^{-3}$ and excitation temperature of 163\,K. We scale the [Si~{\sc{ii}}] map to the [O~{\sc{iii}}] line contribution of 16\,Jy at PACS\,100\,$\mu$m determined from the \textit{ISO} LWS data for Cas\,A (covering most of the nebula) by \citet{2010A&A...518L.138B}, and subtracted it from the PACS\,100\,$\mu$m map. We do not apply any corrections to other wavebands because the [O~{\sc{i}}] 63\,$\mu$m and [C~{\sc{ii}}] 158\,$\mu$m line contributions are small and within the photometric waveband uncertainties.

We inspected the SPIRE FTS spectra to see whether there is a possible contribution from the CO(10-9) 260.4\,$\mu$m, CO(8-7) 325.5\,$\mu$m and CO(5-4) 520.6\,$\mu$m lines to the SPIRE\,250, 350 and 500\,$\mu$m wavebands, respectively, but do not find a line contribution that could be considered to be significant compared to the photometric uncertainties.  

\section{Characterisation of the ISRF using submm emission lines}
\label{Sect_RF}
\begin{figure*}
	\includegraphics[width=8.2cm]{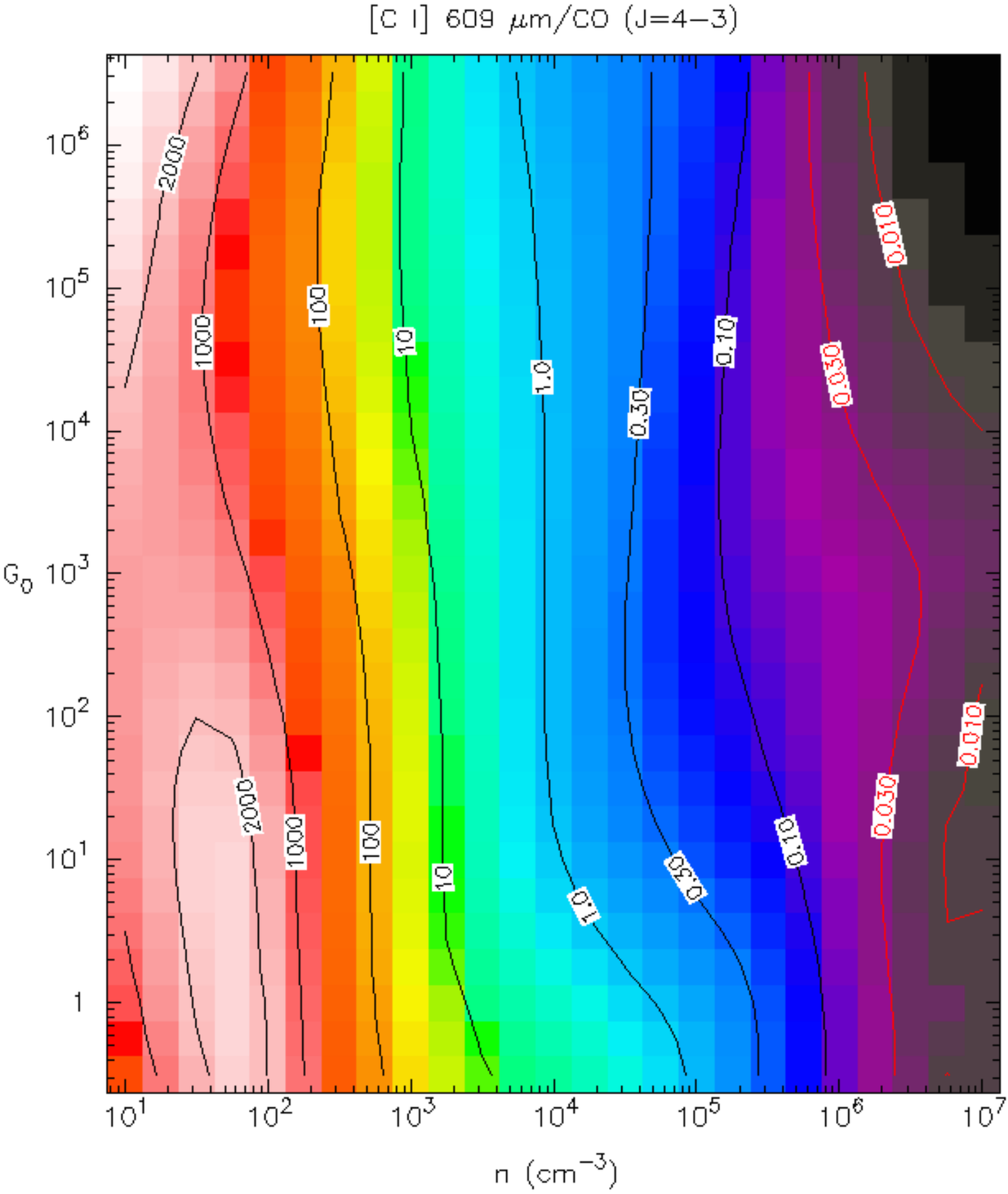}
	\includegraphics[width=8.2cm]{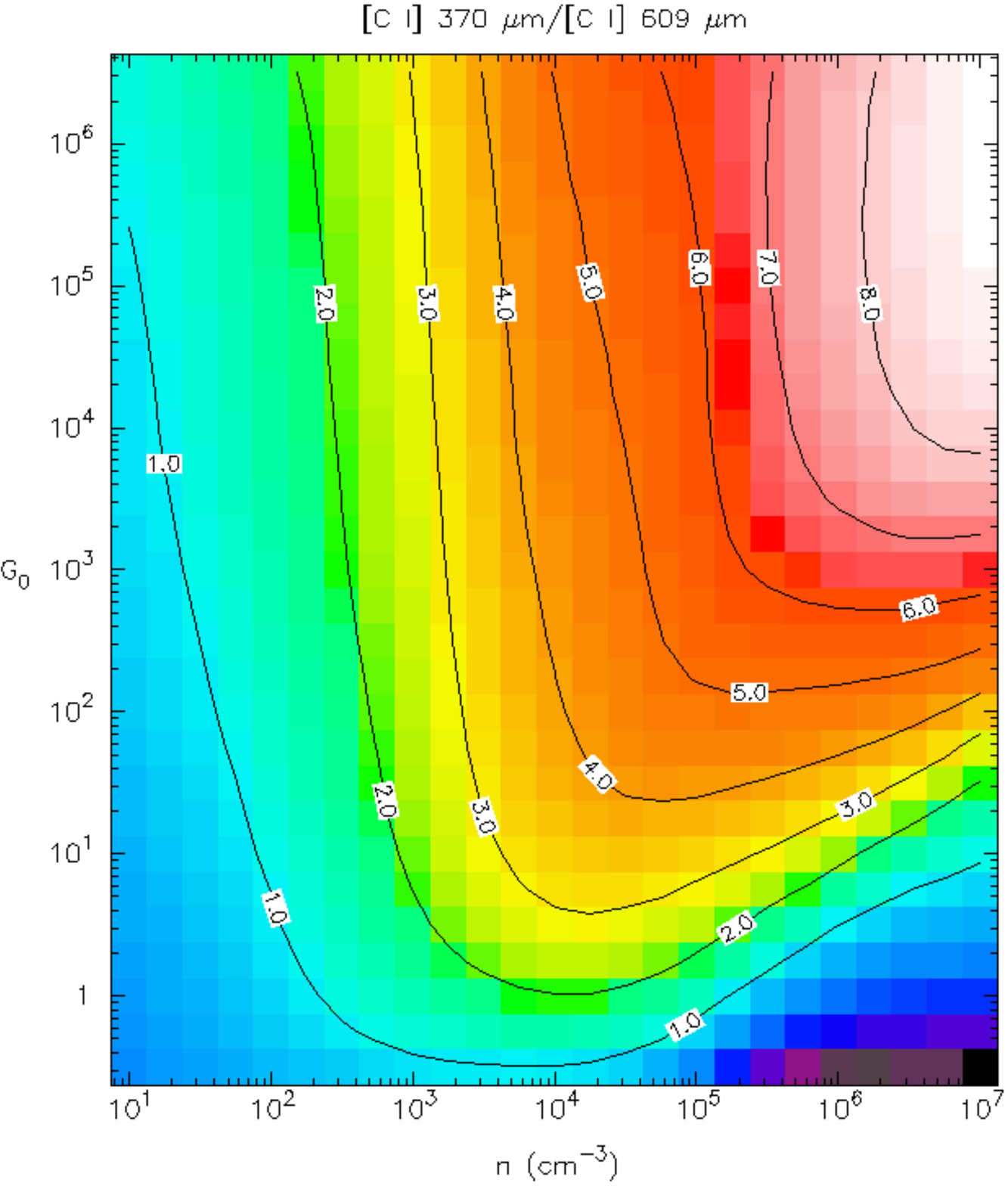} \\
    \caption{Diagnostic plots from the PDR toolbox \citep{2008ASPC..394..654P} showing the variation of the line ratios [C~{\sc{i}}]1-0/CO(4-3) (left) and [C~{\sc{i}}]2-1/ [C~{\sc{i}}]1-0 (right) as a function of the radiation field, $G$ and H-nucleus density, $n_{\text{H}}$. }
    \label{Ima_PDR_diagnostics}
\end{figure*}

In Section \ref{Sect_ISdust}, we have analysed the variation of the ISRF scaling factor, G, in regions that are dominated by ISM dust emission surrounding Cas\,A. Here, we present an alternative method to derive the strength of the UV radiation field along the line of sight to Cas\,A. Specifically, we relied on the observed submm line emission to constrain the parameters of suitable PDR models. Several submm lines are observed with the SPIRE FTS spectrometer, but we focus here on the CO(4-3), [C~{\sc{i}}] $^{3}P_{2}$ $\rightarrow$ $^{3}P_{1}$ and [C~{\sc{i}}] $^{3}P_{1}$ $\rightarrow$ $^{3}P_{0}$ lines at 461\,GHz, 809\,GHz and 492\,GHz, respectively, which are all tracers of the molecular gas in the ISM. Table \ref{Table_PDRlines} provides an overview of the line parameters (line frequency and detector FWHM, the energy of the upper level, and its critical density). Due to the high critical density of the CO(4-3) line transition (see Table \ref{Table_PDRlines}), the [C~{\sc{i}}]\,492/CO(4-3) line ratio is a good tracer of the density of the interstellar gas (see Fig. \ref{Ima_PDR_diagnostics}, left panel). With similar critical densities for both of the [C~{\sc{i}}] lines, the [C~{\sc{i}}] 809/[C~{\sc{i}}] 492 line ratio is a diagnostic of the ISRF scaling factor, via their temperature dependence at low values of $G$ ($G$ $<$ 10\,$G_{\text{0}}$, see Fig. \ref{Ima_PDR_diagnostics}, right panel). We neglect higher CO transitions in our analysis due to the lower detection significance for higher transitions and the possible association of higher CO transitions to dense knots inside Cas\,A (see \citealt{2013A&A...558L...2W}).

The three emission lines of interest were fit simultaneously, using a third order polynomial for the continuum while the line profile is fit with a sinc function. We assume the line profiles have a width that corresponds to the spectral resolution of the instrument (ranging from 280 to 970 km s$^{-1}$ towards longer wavelengths). Based on the typical line widths ($\leq$ 6 km s$^{-1}$) for the [C~{\sc{i}}]\,492\,GHz, $^{12}$CO(2-1) and $^{13}$CO(2-1) line emission observed from interstellar clouds in the Perseus arm where Cas A is situated with ground-based telescopes \citep{2006MNRAS.371..761M,2014ApJ...796..144K}, we are confident that this assumption holds. Even if line broadening occurred due to the interaction of the interstellar clouds with the expanding supernova remnant, we do not expect to find line widths broader than a few 100 km s$^{-1}$ (a careful check of the line widths in the SPIRE FTS spectra confirms this hypothesis).

To diagnose the ISM conditions based on these three lines, we compared the [C~{\sc{i}}]\,492/CO(4-3) and [C~{\sc{i}}]\,809/[C~{\sc{i}}]\,492 line ratios to PDR models. More specifically, we used the online PDR Toolbox (PDRT, \citealt{2008ASPC..394..654P}) to determine the ISRF scaling factor, $G$, and the gas density, $n_{\text{H}}$, in the PDR. Even though the lines were extracted within different apertures due to varying beamsize with wavelength \citep{2013ApOpt..52.3864M}, the surface brightness units allow us to compare the line strength without applying any aperture corrections. The line ratios in the PDR Toolbox were calculated for a plane-parallel geometry with elemental abundances and grain properties similar to solar metallicity. While lower CO transitions might be optically thick, we assume that the CO(4-3) and the [C~{\sc{i}}] lines are optically thin and that we therefore do not need to apply any corrections for opacity to compare the observed line ratios with the PDR model predictions. A direct comparison of observed line ratios with the PDR models also relies on the presumption that UV radiation or collisions dominate the line excitation, and that excitation by shocks is negligible. \citet{2014ApJ...796..144K} found that towards regions in the south and west of Cas\,A, the interaction of the shock front and the interstellar clouds causes a broadening of CO lines (line widths of 8-10 km s$^{-1}$ have been observed as compared to the 1-3 km s$^{-1}$ line widths typical for Galactic molecular clouds), which suggests the importance of shock excitation in those regions. In the north, turbulence induced by an interaction with interstellar material results in a wide distribution of radial velocities for mid-infrared fine-structure lines in the \textit{Spitzer} IRS spectra \citep{2014ApJ...796..144K}. When modelling the fine-structure lines from the SPIRE FTS spectra, we should therefore be cautious in interpreting line ratios beyond the forward shock front, which might be excited by shocks. 

\begin{figure*}
	\includegraphics[width=17cm]{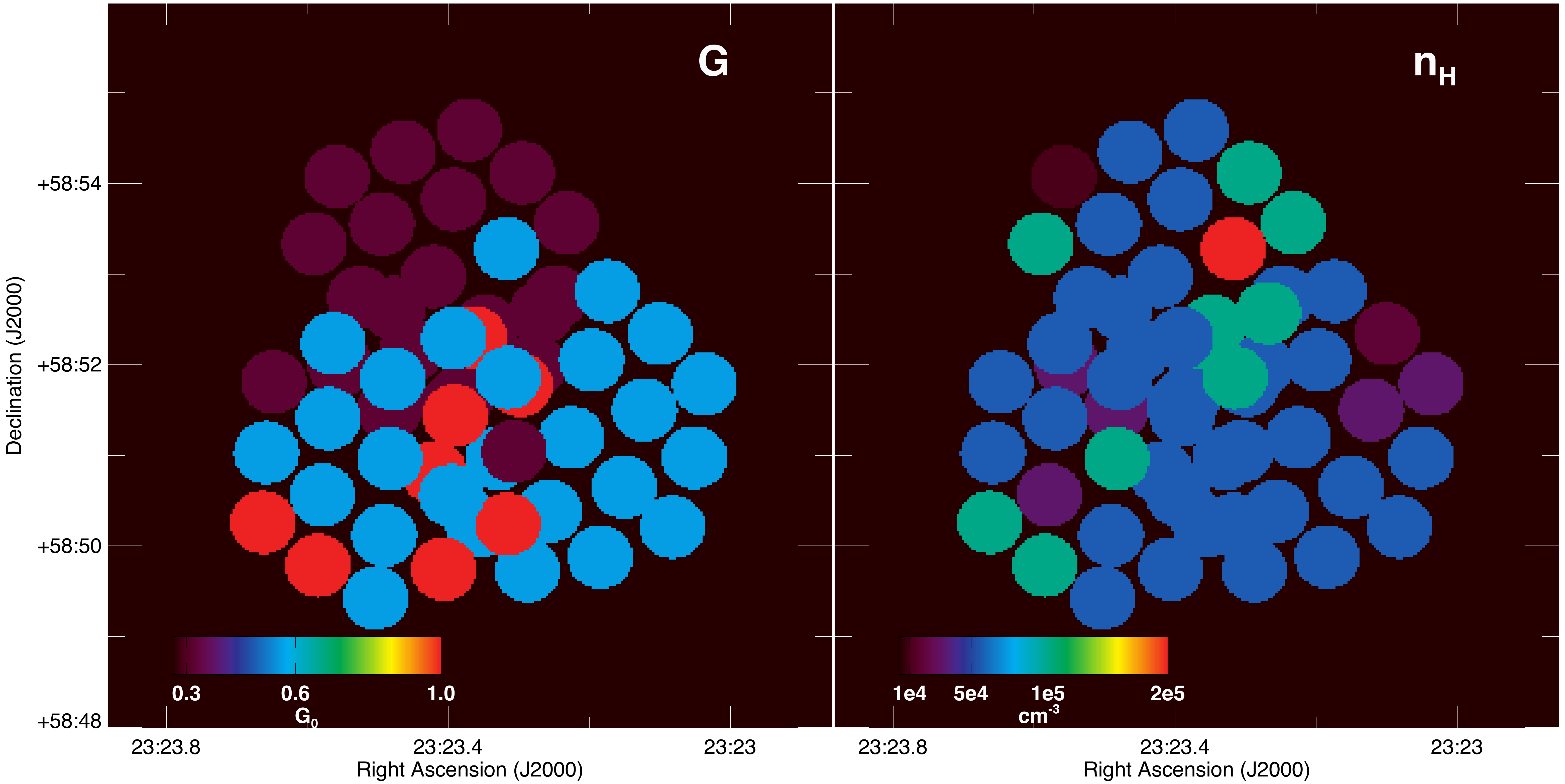}
    \caption{The best fitting PDR parameters, $G$ and $n_{\text{H}}$, derived based on the CO(4-3), [C~{\sc{i}}] 1-0, and [C~{\sc{i}}] 2-1 line emission measured by the SPIRE LWS bolometers. }
    \label{Ima_PDR_results}
\end{figure*}

Figure \ref{Ima_PDR_results} gives an overview of the best fitting parameters derived from our PDR modelling of the spectrum in every FTS bolometer. Only bolometers with line detections with a signal-to-noise level $\geq$ 2 for all three lines are modelled. The strength of the radiation field was found to range between  G = 0.3\,$G_{\text{0}}$ and 1.0\,$G_{\text{0}}$, with the majority of bolometers in the central and north-west pointings yielding results consistent with G = 0.6\,$G_{\text{0}}$\footnote{Due to the restricted sampling of the radiation field and gas density in the PDR models, no intermediate values are allowed for gas densities between 1, 1.8, 3.2, 5.6 and 10 $\times$ 10$^{4}$ cm$^{-3}$ and $G$ values between G~=~0.3$G_{\text{0}}$, 0.6$G_{\text{0}}$ and 1.0$G_{\text{0}}$.}. The latter scaling factor is consistent with the $G$ values derived from the ISM dust SED modelling of the regions surrounding Cas\,A (see Fig. \ref{Ima_CasA_interstellar}, top left panel). The hydrogen nucleus density was found to range between $n_{\text{H}}$ = 3.2 $\times$ 10$^{4}$ cm$^{-3}$ and 10$^{5}$ cm$^{-3}$. For the northern pointing, the line spectra from the majority of bolometers were found to require values of $G$ = 0.3\,$G_{\text{0}}$ and a H-nucleus density $n_{\text{H}}$ = 10$^{5}$ cm$^{-3}$, while the dust SED modelling suggests lower gas densities and stronger radiation fields ($G$ $\sim$1\,$G_{\text{0}}$) just north of Cas\,A. Possible shock excitation, caused by the interaction of the expanding shock front with interstellar clouds, could contribute to the line emission. \citet{2014ApJ...796..144K} have indeed identified high-velocity features in their MIR spectra due to turbulence governed by the interaction of the shock front with dense interstellar material. Alternatively, some of the line emission might be associated with dense clumps in the supernova remnant.

Compared to the PDR analysis of \citet{2006MNRAS.371..761M}, based on a [C~{\sc{i}}] 492\,GHz map, who found gas densities $n_{\text{H}}$ between 10$^{2}$ and 10$^{3}$ cm$^{-3}$ under the assumption of an illuminating radiation field with $G$ = 1.0\,$G_{\text{0}}$, the gas densities derived from our PDR analysis are almost two orders of magnitude higher. Due to the different sets of tracers used by \citet{2006MNRAS.371..761M} ([C~{\sc{i}}]\,492\,GHz, $^{12}$CO(2-1), $^{13}$CO(2-1) and in this work ([C~{\sc{i}}]\,492 and 809\,GHz, $^{12}$CO(4-3)), we might expect to probe regions deeper within the cloud, resulting in higher gas densities. The fixed radiation field of $G$ = 1.0\,$G_{\text{0}}$ used by \citet{2006MNRAS.371..761M} furthermore implies that a lower column of material would be required to reproduce the line emission compared to radiation fields of 0.3 $G_{\text{0}}$ or 0.6 $G_{\text{0}}$, which could bias their PDR analysis towards low gas densities, while our analysis is biased to higher gas densities due to using a different set of tracers.

Figure \ref{Ima_PDR_results} indicates that the strength of the radiation field that illuminates the ISM material along the sight lines to Cas\,A does not vary strongly and is consistent with an average value of 0.6 $G_{\text{0}}$. We therefore adopt an ISRF scaling factor of 0.6 $G_{\text{0}}$ to model the ISM dust emission along the line of sight of Cas\,A (see Section \ref{DustSED}). In Section \ref{Discuss_IS}, we explore the effects on the SN dust mass estimates of assuming lower (0.3 $G_{\text{0}}$) and higher (1.0 $G_{\text{0}}$) scaling factors. A radiation field G = 0.6 $G_{\text{0}}$ corresponds to an ISM dust temperature of $T_{\text{d}}$ $\sim$ 16.4\,K. This dust temperature is consistent with the ISM dust temperature near Cas\,A, $T_{\text{d}}$ $\sim$ 16.5\,K, derived by \citet{2010ApJ...719.1553S}. 

% Table 7
\begin{table}
	\centering
	\caption{Characteristics of the lines used to constrain PDR parameters along the line of sight to Cas\,A. Columns 2 and 3 specify the line rest frequency and the FWHM of the SPIRE FTS bolometer beams at those frequencies. The energy of the upper state and their critical densities (as calculated by \citealt{2004MNRAS.351..147P}) are presented in columns 4 and 5.}
	\label{Table_PDRlines}
	\begin{tabular}{lcccc} % four columns, alignment for each
	\hline
         Line & Frequency & FWHM & $E_{\text{upper}}$/k & $n_{\text{crit}}$ \\
         & [GHz] & [$\arcsec$] & [K] & [cm$^{-3}$] \\
	\hline
	~CO(4-3) & 461.041 & 41.7 & 55 & 2$\times$10$^{4}$ \\
	~[C~{\sc{i}}] $^{3}P_{2}$ $\rightarrow$ $^{3}P_{1}$ & 809.342 & 38.1 & 62.4 & 10$^{3}$ \\
	~[C~{\sc{i}}] $^{3}P_{1}$ $\rightarrow$ $^{3}P_{0}$ & 492.161 & 33.0 & 23.6 & 500 \\			
	\hline
	\end{tabular}
\end{table}

% Figure 8

% Figure 9

\section{Global SED fitting}
\label{Sect_globalfit}
We use the global fluxes (see Table \ref{Table_Fluxfraction}), after subtraction of synchrotron radiation and line contributions, to determine the total amount of emission from ISM dust projected on Cas\,A and SN dust in Cas\,A. Similar to the spatially resolved SED fitting procedure in Section \ref{DustSED}, the global emission spectrum is fitted with a four-component SED model with an ISM dust component and hot, warm and cold SN dust components. The ISM model is based on the THEMIS dust model from \citet{2013A&A...558A..62J} illuminated by a radiation field of $G$=0.6\,$G_{\text{0}}$ scaled to the dust mass that is derived from the SPIRE\,500\,$\mu$m flux and the relations in Eq. \ref{Eq1_label}. The hot SN dust component is assumed to have a fixed dust composition of silicates with a low Mg/Si ratio of 0.7 (i.e., Mg$_{0.7}$SiO$_{2.7}$) with dust temperatures in the range 100-200\,K. The SED modelling procedure is repeated for the dust species listed in Table \ref{Table_DustSpecies} under the assumption that it dominates the dust composition of the warm+cold SN dust components. We assume dust temperatures between 40-100\,K and 10-40\,K for the warm and cold SN dust components. Table \ref{Table_SEDfit_global} gives an overview of the best fitting parameters derived from the global SED fitting. 

% Table 5
\begin{table*}
	\centering
	\caption{The results of the \textbf{global SED fitting} procedure for a variety of SN dust species. We present dust masses and temperatures for the hot (columns 2 and 3), warm (columns 4 and 5) and cold SN dust components (columns 6 and 7) for three different cases with ISRF scaling factors G = 0.3\,$G_{\odot}$, 0.6\,$G_{\odot}$ and 1.0\,$G_{\odot}$, respectively. Columns 8 and 9 list the total dust mass (i.e., the sum of the hot, warm and cold dust masses) and the reduced $\chi^{2}$ values that correspond to the best fitting SED model. The uncertainties on the SN dust masses correspond to the 1$\sigma$ errors computed from the covariance matrix during the SED fitting based on the observational uncertainties of the flux in every waveband. Columns 10 and 11 present the lower limits on the SN dust temperature and mass. The lower limit on the total SN dust mass was calculated by scaling the ISM dust model to the extreme value of the SPIRE 500\,$\mu$m flux density (i.e., $F_{\text{500}}$ $+$ $\sigma_{500}$) to maximise the contribution from ISM dust emission to the global SED (i.e., leaving no room for a SN dust contribution at 500\,$\mu$m). The dust temperature corresponds to the dust temperature of the coldest SN dust component obtained from this lower limit fit. The numbers in boldface correspond to the results from SED fitting for an ISM model with G=0.6\,G$_{\text{0}}$ and a SN dust model with silicate-type grains (MgSiO$_{3}$, Mg$_{2.4}$SiO$_{4.4}$), and are believed to be representative of the SN dust conditions in Cas\,A.}
	\label{Table_SEDfit_global}
	\begin{tabular}{|lcccccccc|cc|} % four columns, alignment for each
	\hline 	
	Dust species & \multicolumn{2}{c}{Hot dust} & \multicolumn{2}{c}{Warm dust} & \multicolumn{2}{c}{Cold dust} & \multicolumn{2}{c}{Total dust} & \multicolumn{2}{|c|}{Total dust} \\
	                     & \multicolumn{2}{c}{Best fit} &\multicolumn{2}{c}{Best fit} &\multicolumn{2}{c}{Best fit} &\multicolumn{2}{c}{Best fit} & \multicolumn{2}{|c|}{Lower limit} \\
%	\hline 		
	 & $T_{\text{d}}$ & $M_{\text{d}}$ & $T_{\text{d}}$ & $M_{\text{d}}$ & $T_{\text{d}}$ & $M_{\text{d}}$ &  $M_{\text{d}}$ &  $\chi^{2}$ & $T_{\text{d}}$ & $M_{\text{d}}$(lower)  \\
	 & (K) & (10$^{-3}$ M$_{\odot}$) & (K) & (10$^{-2}$ M$_{\odot}$) & (K) & (M$_{\odot}$) & (M$_{\odot}$) & & (K) & (M$_{\odot}$)  \\
%	\hline 	 
	\hline 	
	\multicolumn{11}{c}{\textbf{$G$~=~0.3~$G_{\text{0}}$}} \\ 
%	\hline 
%	\hline 
	MgSiO$_{3}$ & 120 & 0.6$\pm$0.1 & 66 & 1.5$\pm$0.2 & 24 & 2.2$\pm$0.3 & 2.2$\pm$0.3 & 0.18 & 30 & 0.6$\pm$0.1   \\
	Mg$_{0.7}$SiO$_{2.7}$ & 190 & 0.04$\pm$0.01 & 56 & 26.9$\pm$1.8  & 25 & 13.7$\pm$1.9 & 14.0$\pm$1.9 & 0.81 & 26 & 8.3$\pm$1.2 \\
	Mg$_{2.4}$SiO$_{4.4}$ & 130 & 0.3$\pm$0.1 & 66 & 1.8$\pm$0.2 & 26 & 1.2$\pm$0.2 & 1.2$\pm$0.2 & 0.55 & 30 & 0.6$\pm$0.1 \\		
	Al$_{2}$O$_{3}$ (porous) & 115 & 0.7$\pm$0.1 & 72 & 0.7$\pm$0.1 & 24 & 0.6$\pm$0.1 & 0.6$\pm$0.1 & 0.19 & 33 & 0.12$\pm$0.02 \\	
	CaAl$_{12}$O$_{19}$ & 105 & 1.1$\pm$0.1 & 74 & 1.0$\pm$0.1 & 16 & 34.8$\pm$5.2 & 34.8$\pm$5.2 & 1.34 & 20 & 4.7$\pm$0.6 \\
	Am. carbon "AC1" & 115 & 0.7$\pm$0.1 & 85 & 0.8$\pm$0.1 & 27 & 0.6$\pm$0.1 & 0.6$\pm$0.1 & 0.50 & 40 & 0.08$\pm$0.02 \\	
	a-C ($E_{\text{g}}$=0.1eV) & 115 & 0.7$\pm$0.1 & 77 & 1.0$\pm$0.1 & 25 & 1.2$\pm$0.1 & 1.2$\pm$0.1 & 0.31 & 37 & 0.2$\pm$0.1 \\	
%	\hline 	 
%	\hline 	
	\multicolumn{11}{c}{\textbf{$G$~=~0.6~$G_{\text{0}}$}} \\ 
%	\hline 
%	\hline 
	MgSiO$_{3}$ & \textbf{120} & \textbf{0.6$\pm$0.1} & \textbf{68} & \textbf{1.2$\pm$0.2} & \textbf{28} & \textbf{0.6$\pm$0.1} & \textbf{0.6$\pm$0.2} & \textbf{0.20} & \textbf{38} & \textbf{0.13$\pm$0.02}   \\
	Mg$_{0.7}$SiO$_{2.7}$ & 200 & 0.03$\pm$0.01 & 55 & 33.8$\pm$1.9 & 21 & 26.1$\pm$8.4 & 26.4$\pm$8.4 & 0.58 & 29 & 0.6$\pm$0.8 \\
	Mg$_{2.4}$SiO$_{4.4}$ & \textbf{130} & \textbf{0.3$\pm$0.1} & \textbf{66} & \textbf{1.8$\pm$0.2} & \textbf{29} & \textbf{0.4$\pm$0.1} & \textbf{0.4$\pm$0.1} & \textbf{0.47} & \textbf{38} & \textbf{0.10$\pm$0.03} \\
	Al$_{2}$O$_{3}$ (porous) & 110 & 1.0$\pm$0.1 & 72 & 0.7$\pm$0.1 & 27 & 0.2$\pm$0.1 & 0.2$\pm$0.1 & 0.22 & 40 & 0.02$\pm$0.01 \\
	CaAl$_{12}$O$_{19}$ & 100 & 1.6$\pm$0.2 & 74 & 0.9$\pm$0.1 & 17 & 12.8$\pm$3.6 & 12.8$\pm$3.6 & 1.26 & 24 & 0.7$\pm$0.2 \\
	Am. carbon "AC1" & 115 & 0.7$\pm$0.1 & 84 & 0.8$\pm$0.1 & 29 & 0.3$\pm$0.1 & 0.3$\pm$0.1 & 0.29 & 79 & 0.013$\pm$0.001 \\ % 
	a-C ($E_{\text{g}}$=0.1eV) & 115 & 0.7$\pm$0.1 & 77 & 1.0$\pm$0.1 & 27 & 0.5$\pm$0.2 & 0.5$\pm$0.1 & 0.24 & 40 & 0.03$\pm$0.02 \\				
%	\hline 	 
%	\hline 	
	\multicolumn{11}{c}{\textbf{$G$~=~1.0~$G_{\text{0}}$}} \\ 
%	\hline 
%	\hline 
	MgSiO$_{3}$ & 115 & 0.7$\pm$0.1 & 68 & 1.1$\pm$0.2 & 39 & 0.06$\pm$0.02 & 0.07$\pm$0.02 & 0.34 & 63 & 0.021$\pm$0.002   \\
	Mg$_{0.7}$SiO$_{2.7}$ & 185 & 0.04$\pm$0.01 & 55 & 32.6$\pm$2.2 & 30 & 0.3$\pm$0.8 & 0.6$\pm$0.8 & 0.61 & 56 & 0.28$\pm$0.02 \\ 
	Mg$_{2.4}$SiO$_{4.4}$ & 125 & 0.4$\pm$0.1 & 63 & 2.5$\pm$0.2 & 33 & 0.01$\pm$0.06 & 0.04$\pm$0.06 & 0.45 & 64 & 0.022$\pm$0.002 \\	
	Al$_{2}$O$_{3}$ (porous) & 110 & 1.0$\pm$0.1 & 70 & 0.8$\pm$0.1 & 12 & 1.0$\pm$1.8 & 1.0$\pm$1.8 & 0.28 & 74 & 0.007$\pm$0.001 \\
	CaAl$_{12}$O$_{19}$ & 100 & 1.5$\pm$0.2 & 75 & 0.8$\pm$0.1 & 17 & 5.1$\pm$4.3 & 5.1$\pm$4.3 & 1.24 & 75 & 0.010$\pm$0.001 \\
	Am. carbon "AC1" & 115 & 0.7$\pm$0.1 & 82 & 1.0$\pm$0.1 & 23 & 0.4$\pm$0.7 & 0.4$\pm$0.7 & 0.28 & 86 & 0.008$\pm$0.001 \\ %	
	a-C ($E_{\text{g}}$=0.1eV) & 115 & 0.7$\pm$0.1 & 75 & 1.2$\pm$0.1 & 22 & 1.3$\pm$0.1 & 1.3$\pm$0.1 & 0.31 & 77 & 0.010$\pm$0.001 \\ %						
	\hline
	\end{tabular}
\end{table*}

Similarly to the resolved SED fitting procedure, the SED models with a Mg$_{0.7}$SiO$_{2.7}$, CaAl$_{12}$O$_{19}$ and Al$_{2}$O$_{3}$ dust composition for the warm and cold SN dust components result in SN dust masses that are unrealistically high given the constraints on metal production based on nucleosynthesis models for supernova type II and type IIb events for a 30\,M$_{\odot}$ and 18\,M$_{\odot}$ progenitor, respectively. These grains can therefore be ruled out as the dominant dust species that form in Cas\,A\footnote{Note that we can not rule out that some of these dust species formed within Cas\,A, but they will not dominate the mass fraction of freshly formed grains.}. For carbonaceous grains, the best fit would imply a dust mass of 0.3-0.5\,M$_{\odot}$ with a lower limit of 0.01-0.03\,M$_{\odot}$ of carbon grains. Besides 0.3-0.5\,M$_{\odot}$ of carbon locked up in grains being considered unrealistic due to the lack of sufficient carbon production by the progenitor of Cas\,A, the ejecta of Cas\,A have been shown to be predominantly oxygen-rich \citep{1979ApJ...233..154C,2010A&A...509A..59D}. Assuming that the newly formed grains are silicates (MgSiO$_{3}$ or Mg$_{2.4}$SiO$_{4.4}$), we would need 0.4 to 0.6\,M$_{\odot}$ of these oxygen-rich grains, with a lower limit of 0.1\,M$_{\odot}$, at a temperature of $T_{\text{d}}$$\sim$30\,K to reproduce the global emission spectrum of Cas\,A.

\begin{figure}
	\includegraphics[width=8.75cm]{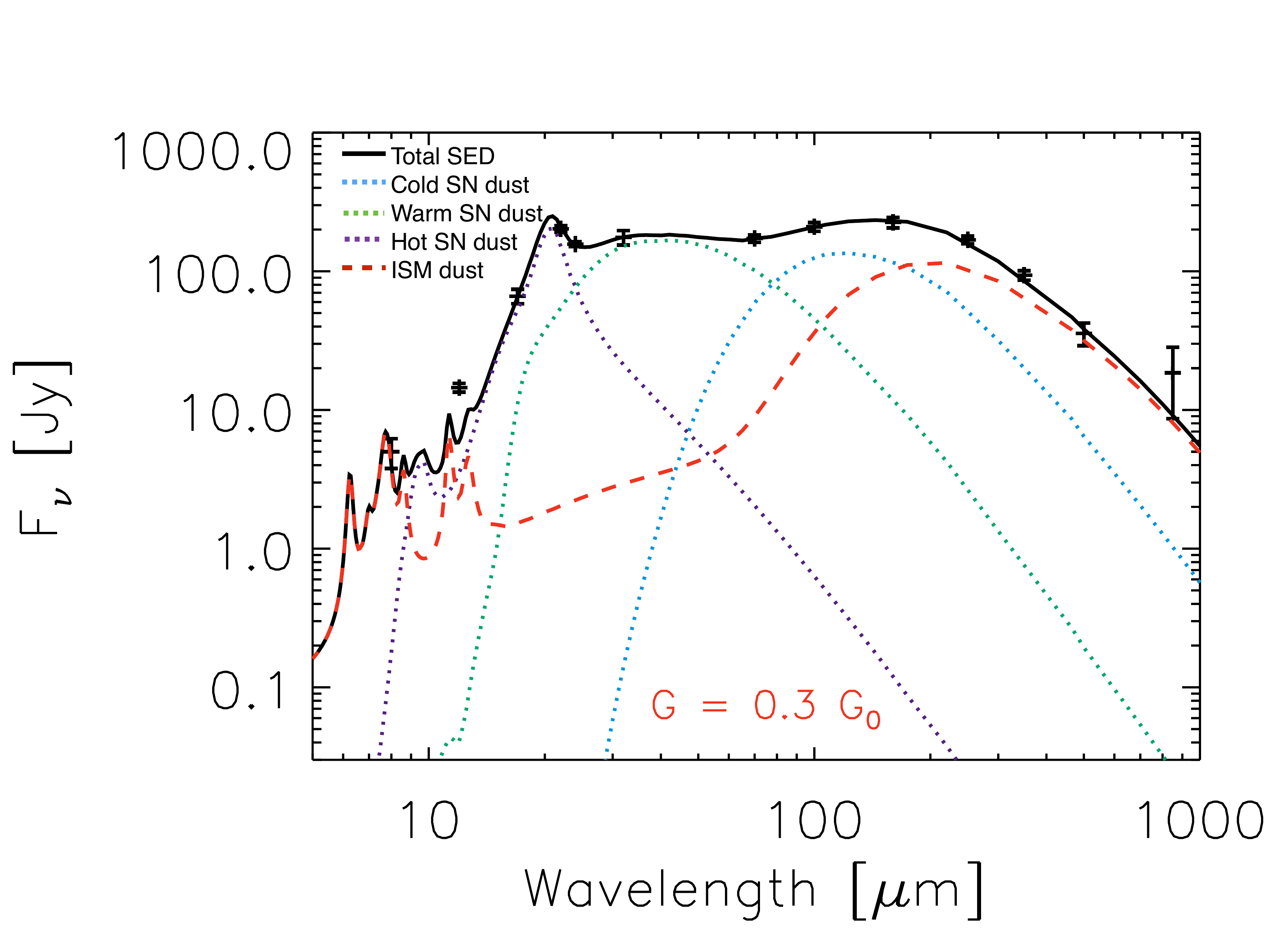} \\ %{CasA_SEDfit_global_G0_0_3_Mgproto_SNat500_label.pdf} \\
	\includegraphics[width=8.75cm]{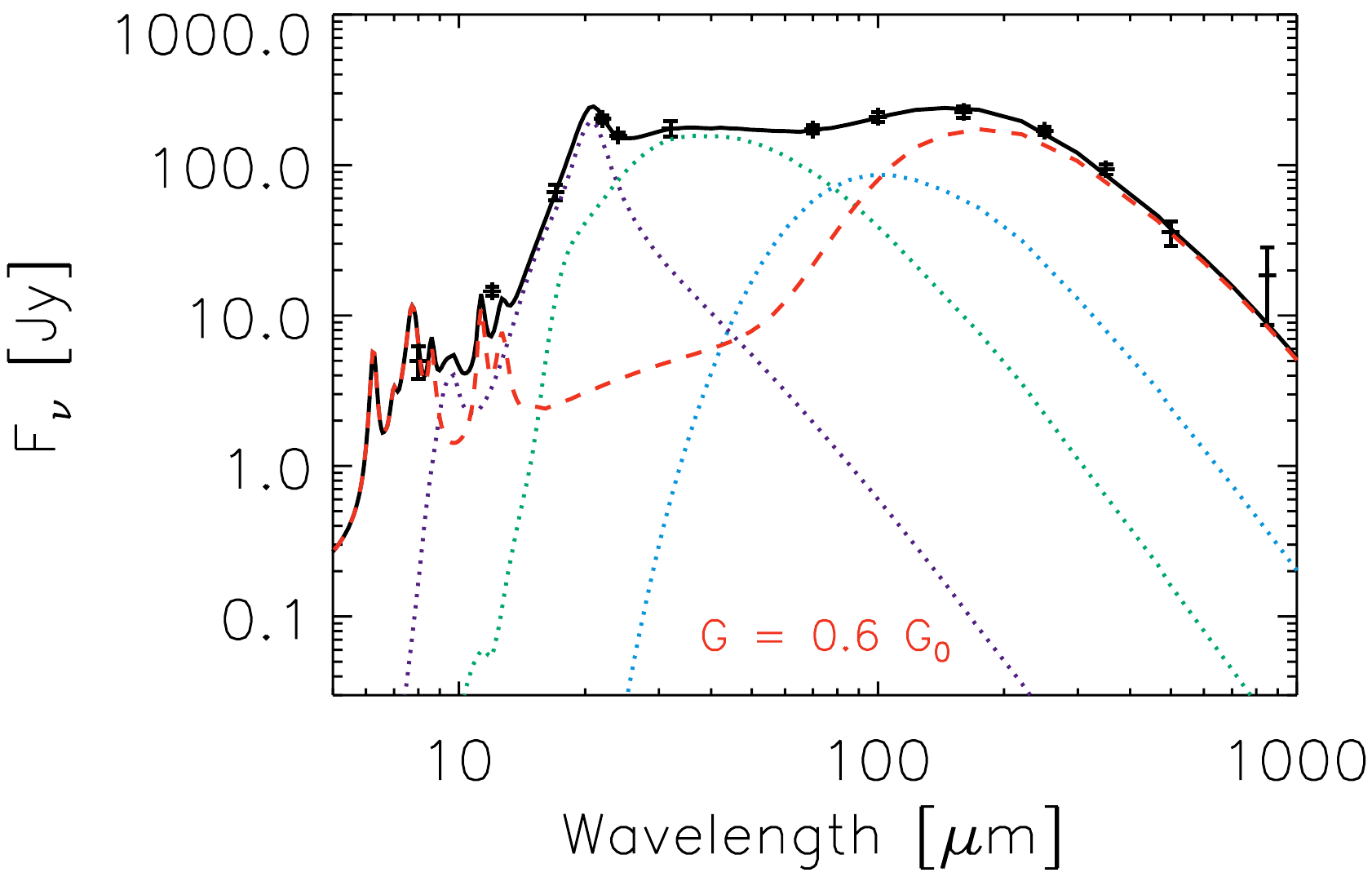} \\ %{CasA_SEDfit_global_G0_0_6_Mgproto_SNat500.pdf} \\	
	\includegraphics[width=8.75cm]{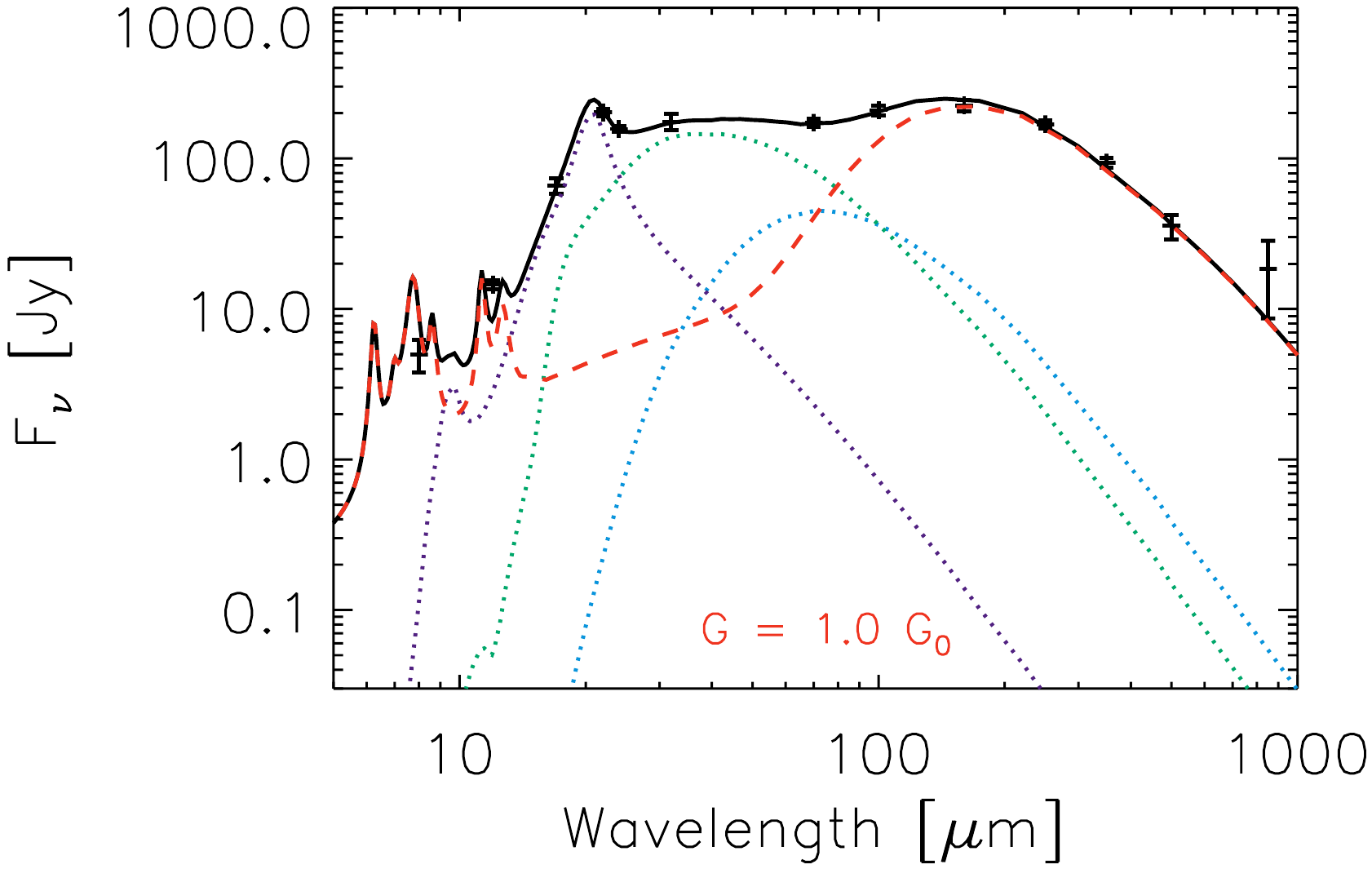} %{CasA_SEDfit_global_G0_1_0_Mgproto_SNat500.pdf}
    \caption{The best fitting global SED models fit to the 17-500\,$\mu$m photometry after subtraction of the contribution from line emission and synchrotron radiation, assuming a Galactic ISM dust model illuminated by a radiation field G=0.3\,$G_{\text{0}}$ (top), 0.6\,$G_{\text{0}}$ (middle) and 1.0\,$G_{\text{0}}$ (bottom). The legend in the top panel explains the line formatting and colours associated to the various SN and ISM dust components.} 
    \label{Ima_CasA_globalSED}
\end{figure}

Figure \ref{Ima_CasA_globalSED} shows the best fitting global SED model assuming an ISM dust SED model with radiation fields $G$=0.3\,$G_{\text{0}}$, 0.6\,$G_{\text{0}}$ and 1.0\,$G_{\text{0}}$. For these SED fits, we have assumed Mg$_{0.7}$SiO$_{2.7}$ dust for the hot SN dust component and MgSiO$_{3}$ dust for the warm and cold SN dust components. Even though the IRAC\,8\,$\mu$m and WISE\,12\,$\mu$m data points were not used to constrain the multi-component ISM+SN model, the best fitting model SED seems to reproduce the MIR emission in those wavebands within the error bars (with the exception of the  WISE\,12\,$\mu$m datapoint for the G\,=\,0.3\,$G_{\text{0}}$ model). The good agreement between model and observations seems to indicate that a combination of a warm SN dust component and the emission features originating from small hydrocarbons (or PAHs) in the diffuse ISM are able to account for the MIR emission observed towards Cas\,A. The SCUBA\,850\,$\mu$m data point is also omitted from the fitting procedure. We will discuss the SCUBA\,850\,$\mu$m observations in Appendix \ref{Discuss_850mu}. 

The SN dust masses derived from the spatially resolved SED fits (see Section \ref{DustSED}) are in overall agreement with the dust masses derived from the global SED fits within the limits of observational uncertainties. Since the variation in the SN dust emission spectrum and ISM dust contribution within an aperture of radius 165$\arcsec$ can introduce possible uncertainties in the global SED parameters (because local variations in ISM dust contribution and SN dust mass and temperature are averaged out in the global spectrum that includes all the emission within a R~=~165\,$\arcsec$ region),   we attach more weight to the results of the spatially resolved SED fitting analysis.
%Figure 13

%%%%%%%%%%%%%%%
%%%Model verification%%%
%%%%%%%%%%%%%%%

\section{Model verification}
\label{Sect_modelverification}
\subsection{Comparison between model and observations}
\label{Discuss_comparison}

\begin{figure}
	\includegraphics[width=9cm]{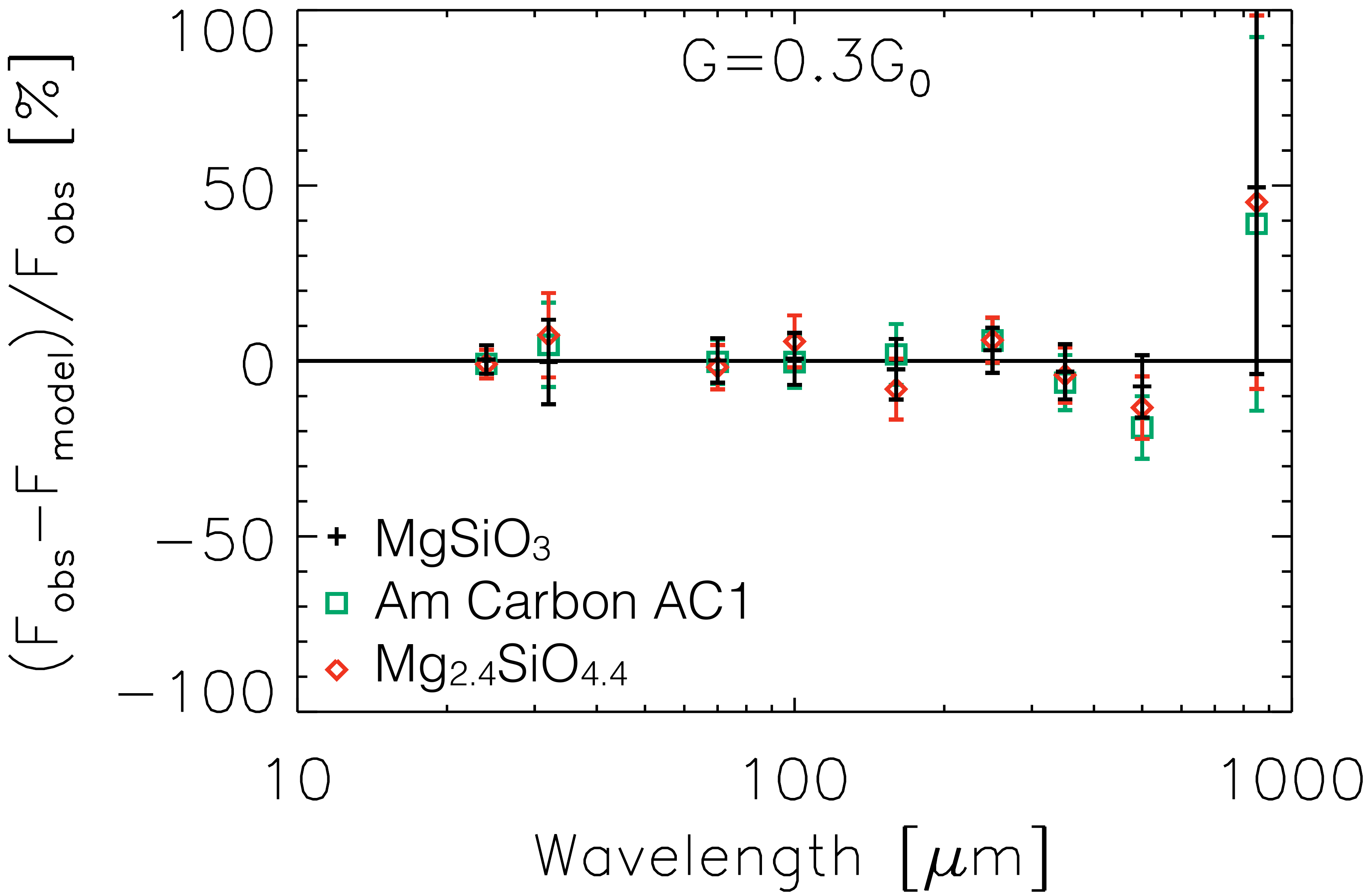} \\
	\includegraphics[width=9cm]{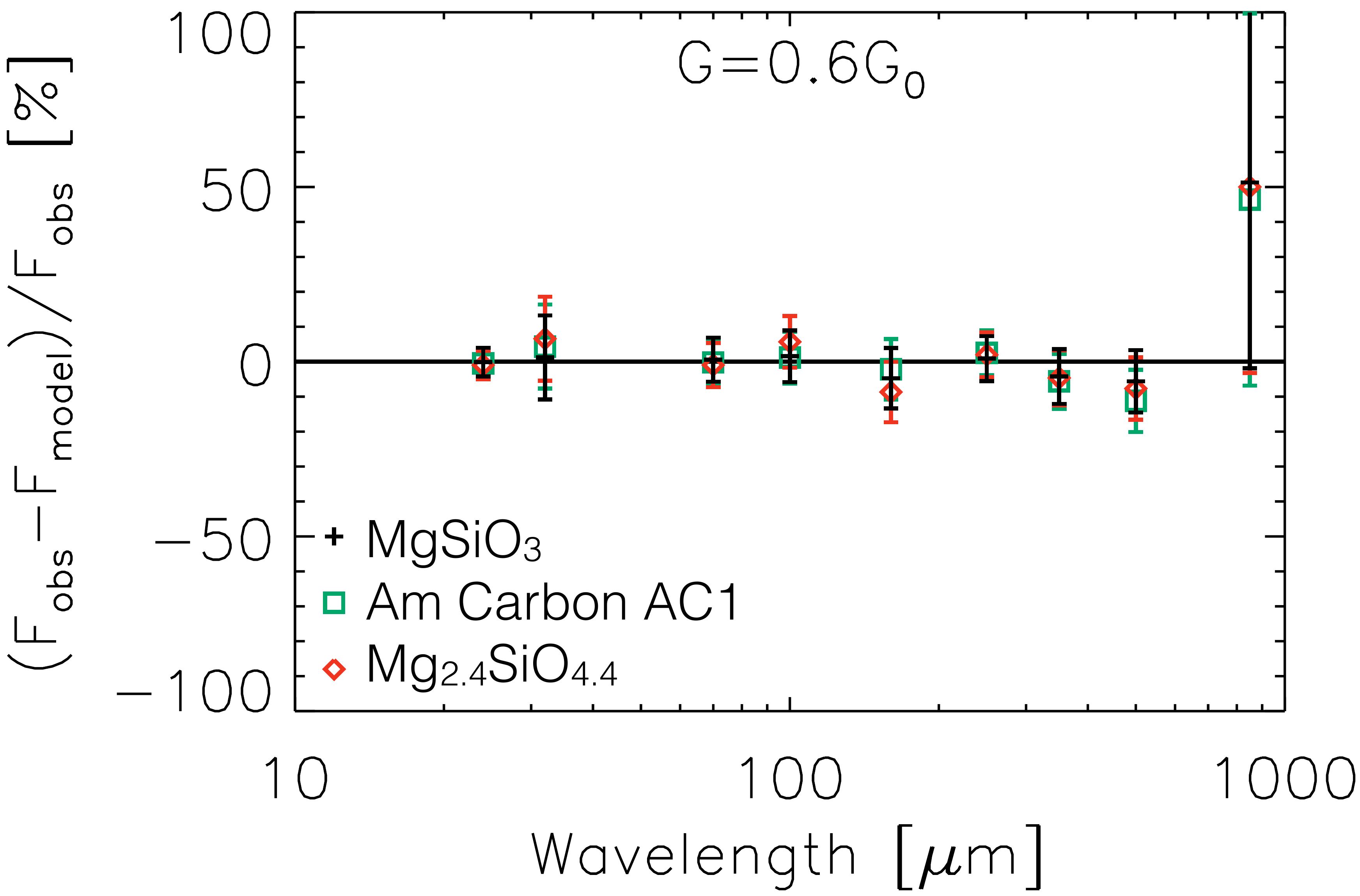} \\
	\includegraphics[width=9cm]{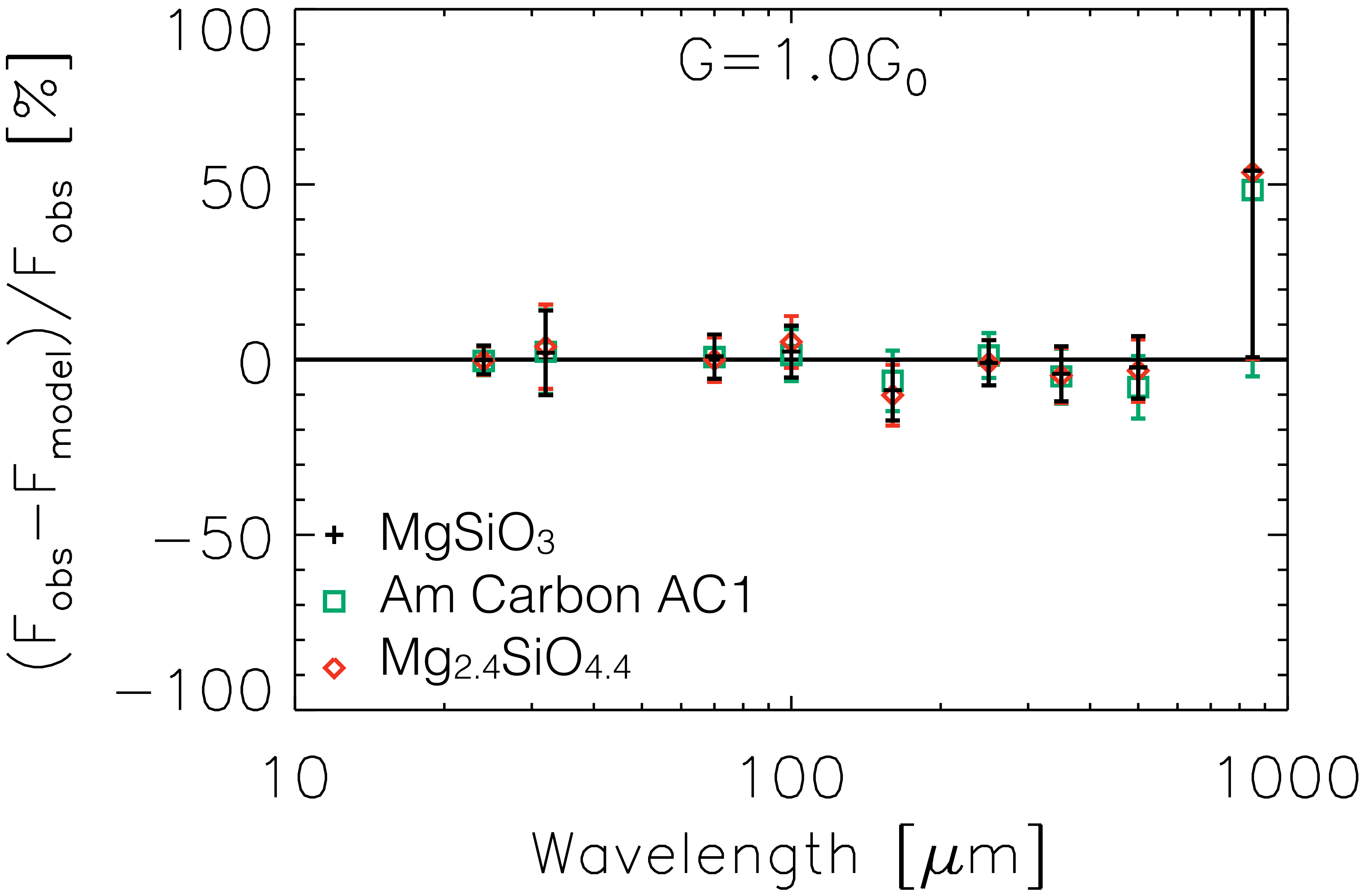}	
    \caption{Percentage offsets from the observations, $\frac{F_{\text{obs}}-F_{\text{model}}}{F_{\text{obs}}}$, for the global SED models. The top, middle and bottom panels show the model deviations for ISM dust models with G=0.3\,$G_{\odot}$, 0.6\,$G_{\odot}$ and 1.0\,$G_{\odot}$, respectively. The model offsets are shown for different SN dust compositions (MgSiO$_{3}$ in black, amorphous carbon in green and Mg$_{2.4}$SiO$_{4.4}$ in red) used for the warm and cold components. The rightmost datapoint corresponds to the SCUBA\,850\,$\mu$m flux reported by \citet{2009MNRAS.394.1307D}. Its offset from the best fitting models is discussed in Section \ref{Discuss_850mu}.}
    \label{Ima_CasA_globalSED_res}
\end{figure}

Here, we compare the modelled global dust SED and spatially resolved dust emission maps to observations to verify whether the multi-component ISM+SN model is capable of reproducing the observed IR-submm SEDs. The global relative offsets between observed and modelled fluxes (i.e., $F_{\text{obs}}$-$F_{\text{model}}$/$F_{\text{obs}}$) at various wavelengths are shown in Figure \ref{Ima_CasA_globalSED_res} for models with the three different ISM dust models (G~=~0.3, 0.6 and 1.0\,$G_{\odot}$) and various SN dust species. All dust SED models show small residuals, independent of the assumed dust composition or strength of the illuminating ISRF, and reproduce the MIPS\,24\,$\mu$m, IRS\,32\,$\mu$m continuum, PACS\,70, 100 and 160\,$\mu$m and SPIRE\,250, 350 and 500\,$\mu$m datapoints within the error bars. While the SCUBA\,850\,$\mu$m is not used to constrain the SED fitting procedure, all models are at the low end of the observed SCUBA 850\,$\mu$m data flux (see also Appendix \ref{Discuss_850mu}). 

%Figure 14

\begin{figure*}
	\includegraphics[width=12.35cm]{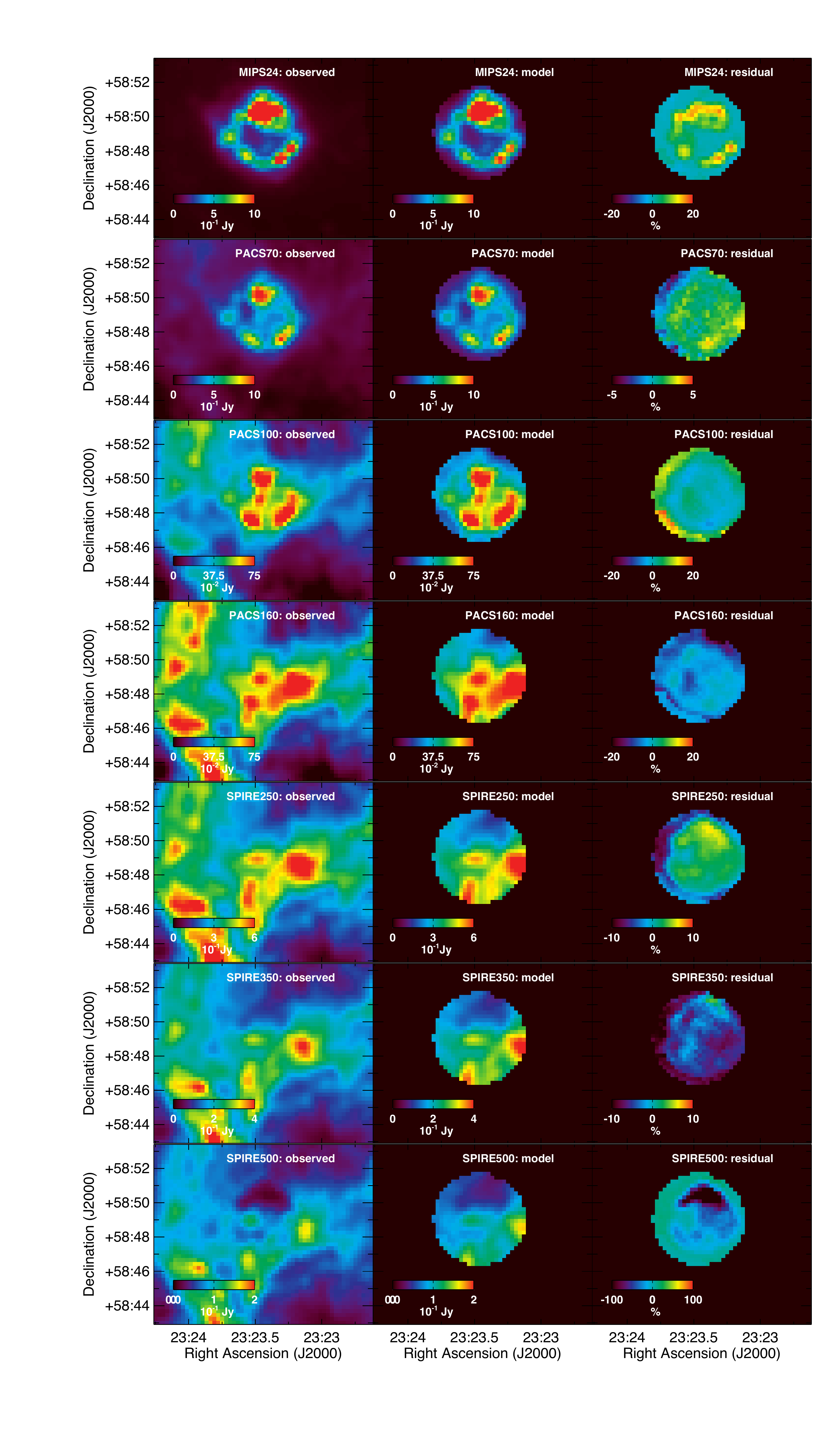} 
		\includegraphics[width=4.72cm]{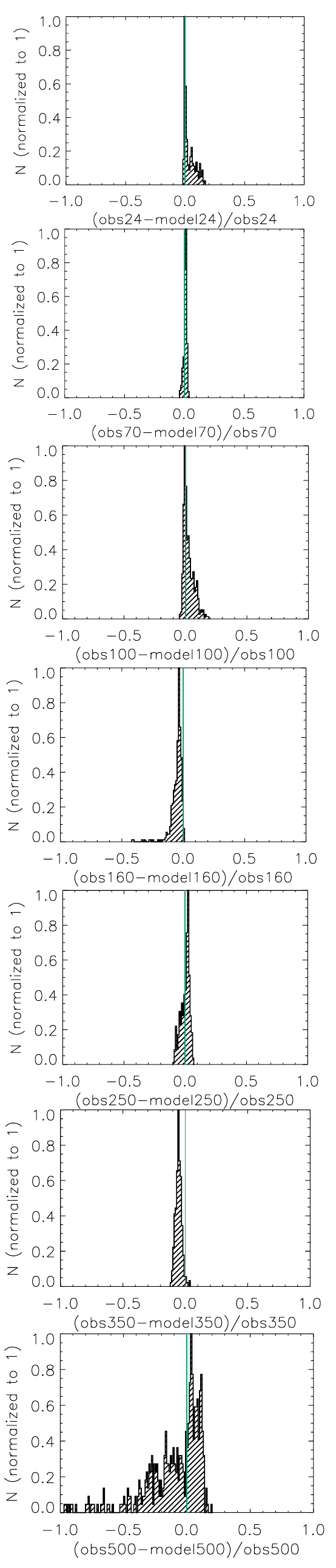} 		
    \caption{The observed (left column) and modelled (second column) MIPS\,24\,$\mu$m, PACS\,70, 100 and 160\,$\mu$m and SPIRE 250, 350 and 500\,$\mu$m images. The observed maps correspond to the \textit{Spitzer} and \textit{Herschel} maps after subtraction of the synchrotron radiation and line emission components (leaving a combination of ISM and SN dust emission). The modelled images result from a multi-component SED fit in each individual pixel. The third column shows the residual image, $\frac{F_{\text{obs}}-F_{\text{model}}}{F_{\text{obs}}}$, as percentage deviations to highlight deviations of the model from the observations. The scale of the residual images has been adjusted to cover the entire range of residuals. In the right-hand column, the histogram of residuals is shown for every waveband.}
    \label{Ima_CasA_resolvedSED_resmap}
\end{figure*}

To verify whether the multi-component model reproduces the observed spectrum in every pixel, in Figure \ref{Ima_CasA_resolvedSED_resmap} we compare the emission predicted by the SED model with observations. To highlight any model deviations from the observations, the residual images (i.e., observed flux - modelled flux / observed flux) and histogram of residuals are shown in the third and fourth columns of Figure \ref{Ima_CasA_resolvedSED_resmap}, respectively. The residuals between the model and observations show random offsets centred around 0  at every wavelength rather than a systematic offset, which suggests that the multi-component SN+ISM dust model is adequate to fit the IR/submm SED. At 500\,$\mu$m, several pixels show an offset between observations and model up to -50$\%$. Given that the offset arises in regions where the ISM contribution is low, and the model overestimates the ISM dust emission in those regions, we do not believe that this discrepancy will affect the determination of the SN dust mass in Cas\,A. 

%Model vs Obs comparison

\subsection{The effect of variations in the adopted radiation field}
\label{Discuss_IS}

In this Section we analyse the effects of variations in the strength of the ISRF illuminating the ISM dust along the sight line of Cas\,A on the determination of the SN dust mass. For the multi-component SED fitting procedures on resolved and global scales (see Section \ref{DustSED} and Appendix \ref{Sect_globalfit}, respectively), we have assumed a constant radiation field strength of $G$=0.6\,G$_{\text{0}}$. Based on local variations in $G$ in the areas surrounding Cas\,A (see Fig. \ref{Ima_CasA_interstellar}, left panel) it is however likely that regional variations occur, which might have an effect on the SN dust temperatures and dust masses derived for Cas\,A. The three sets of SED models that were run for $G$ = 0.3\,$G_{\text{0}}$, 0.6\,$G_{\text{0}}$ and 1.0\,$G_{\text{0}}$ give an idea of the influence of the variation in radiation field illuminating the ISM material on the SN dust mass and temperature estimates. We have not considered the results of ISM models with a stronger ISRF (i.e., $G$$>$1.0\,$G_{\text{0}}$) due to an overestimation of IR-submm fluxes for Cas\,A upon scaling these ISM dust model SEDs to the SPIRE\,500\,$\mu$m flux (after subtraction of the synchrotron radiation). The best fitting SN dust masses and temperatures derived for these three sets of SED models can be examined in Tables \ref{Table_SEDfit_resolved} and \ref{Table_SEDfit_global} for the spatially resolved and global SED fits, respectively.

\begin{figure*}
	\includegraphics[width=18cm]{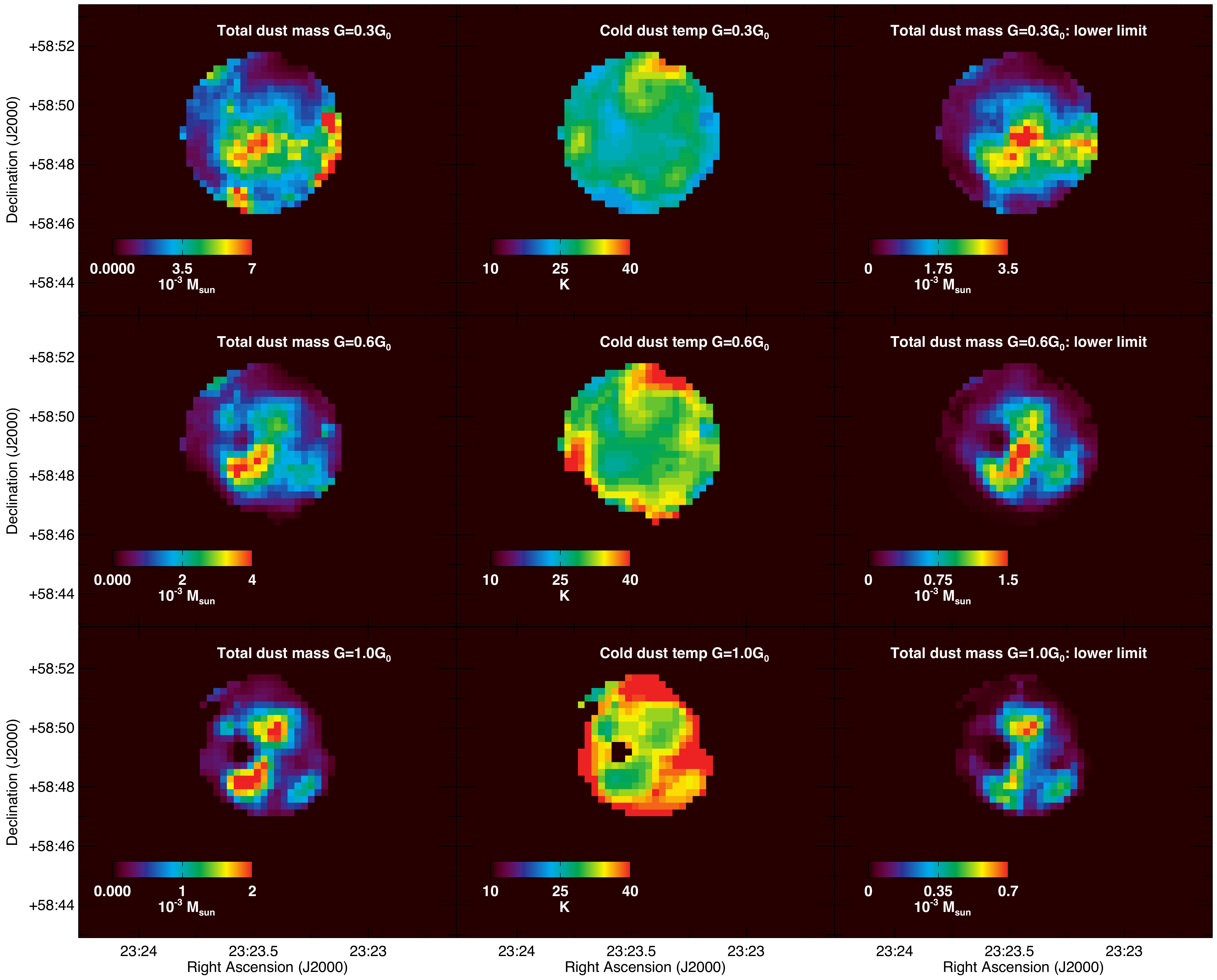}
    \caption{Total SN dust mass maps (first column) and cold SN dust temperature maps (second column) derived from spatially resolved SED fitting with a Mg$_{0.7}$SiO$_{2.7}$+MgSiO$_{3}$ SN dust composition and ISM dust models with radiation fields of $G$ = 0.3\,$G_{\text{0}}$, 0.6\,$G_{\text{0}}$ and 1.0\,$G_{\text{0}}$ (see top, middle and bottom row, respectively). The last column shows the lower limit to the total SN dust mass, derived by fixing the normalisation of the ISM dust model to reproduce the $F_{\text{500}}$+$\sigma_{\text{500}}$ flux density.}
    \label{Ima_CasA_resolvedSED_maps_G0}
\end{figure*}

Comparing the SED fitting results for different ISRFs, a general trend emerges that ISM models with lower dust temperatures (corresponding to lower $G$) yield larger SN dust masses, which trivially follows from the reduced IR/submm emission of ISM dust and is compensated for with an increase in the SN dust emission at those wavelengths. Figure \ref{Ima_CasA_resolvedSED_maps_G0} demonstrates how the derived SN dust temperatures, and consequently the SN dust masses, are sensitive to the radiation field that is assumed to illuminate the ISM dust. The overall distribution of SN dust temperatures within Cas\,A is similar for the three models regardless of the ISRF scaling factor. But with cold SN dust temperatures ranging from 20\,K to 40\,K (middle panels), the total dust mass can differ by factors of a few. More specifically, we derive dust masses of 1.4\,M$_{\odot}$, 0.5\,M$_{\odot}$ and 0.2\,M$_{\odot}$ from the resolved SED maps assuming that the warm and cold SN dust is composed of MgSiO$_{3}$ grains with ISM dust illuminated by a radiation field G=0.3\,G$_{\text{0}}$, 0.6\,G$_{\text{0}}$ and 1.0\,G$_{\text{0}}$, respectively. Similarly, the resolved SED fitting procedure for a Mg$_{2.4}$SiO$_{4.4}$ dust composition results in dust masses of 0.9\,M$_{\odot}$, 0.3\,M$_{\odot}$ and 0.1\,M$_{\odot}$ for different ISRFs, respectively. The SN dust masses derived for a G=0.3\,G$_{\text{0}}$ ISM model exceed the maximum dust masses (0.4-1.4\,M$_{\odot}$ and 0.3-0.9\,M$_{\odot}$ for MgSiO$_{3}$ and Mg$_{2.4}$SiO$_{4.4}$ dust) derived from the metal production predicted by nucleosynthesis models (under the assumption that all the metals are locked in dust grains), and are therefore considered unrealistic.  Given the uncertainties on the nucleosynthesis models \citep{1995ApJS..101..181W} and the mass of the progenitor, we argue that the condensation of 0.3-0.5\,M$_{\odot}$ (for a 0.6\,G$_{\text{0}}$ ISM model) or 0.1-0.2\,M$_{\odot}$ (for a 1.0\,G$_{\text{0}}$ ISM model) of dust in Cas\,A is reasonable. 

Based on a total ejecta mass estimate of 2-4\,M$_{\odot}$ derived both from X-ray observations of Cas\,A \citep{2002A&A...381.1039W} and kinetic arguments to determine the mass and explosion energy from optical observations \citep{2003ApJ...593L..23C,2003ApJ...597..347L}, these dust masses imply that the condensation of elements has been extremely efficient. Given that the progenitor has been predicted to have had a mass of 23-30 M$_{\odot}$ \citep{2001NuPhA.688..168K,2002RMxAC..12...94P,2006ApJ...640..891Y,2009A&A...506.1249P}, the total ejecta mass could be larger than 4 M$_{\odot}$, with the missing gas mass in the form of clumped neutral gas, which might also harbour the detected cold dust in the ejecta. Alternatively, most of the initial stellar mass might have been lost by the progenitor prior to explosion with estimated values of the swept-up mass of $\sim$\,8\,M$_{\odot}$ up to possibly 20\,M$_{\odot}$ \citep{2003A&A...398.1021W}.  

%Different G0 maps

We have shown that the ISM dust model with $G$ = 0.6\,$G_{\text{0}}$ most closely resembles the ISRF along the sight line of Cas\,A (based on SED fitting in the surrounding regions of Cas\,A, and a PDR analysis of submm emission lines; see Figures \ref{Ima_CasA_interstellar} and \ref{Ima_PDR_results}, respectively). We therefore attach more weight to the SN dust masses of 0.5\,M$_{\odot}$ (MgSiO$_{3}$), 0.3\,M$_{\odot}$ (Mg$_{2.4}$SiO$_{4.4}$) and 0.5-0.6\,M$_{\odot}$ (carbonaceous grains) derived from SED models with $G$ = 0.6\,$G_{\text{0}}$ ISM dust models. Due to possible variations in the ISRF along the line of sight, we can not rule out that the local radiation field near Cas\,A is closer to the solar ISRF which would imply SN dust masses of 0.1-0.2\,M$_{\odot}$. Being conservative, we can conclude that the mass of dust produced in Cas\,A is between 0.1 and 0.6\,M$_{\odot}$ and likely consists of a combination of mostly silicate-type (MgSiO$_{3}$ or Mg$_{2.4}$SiO$_{4.4}$) grains and some form of carbonaceous dust.

\section{Model predictions}
\label{Sec_modelprediction}
\subsection{Cas\,A: various IR-submm components}
\label{Sect_CasA_aftercorr}

\begin{figure*}
    \includegraphics[width=17.5cm]{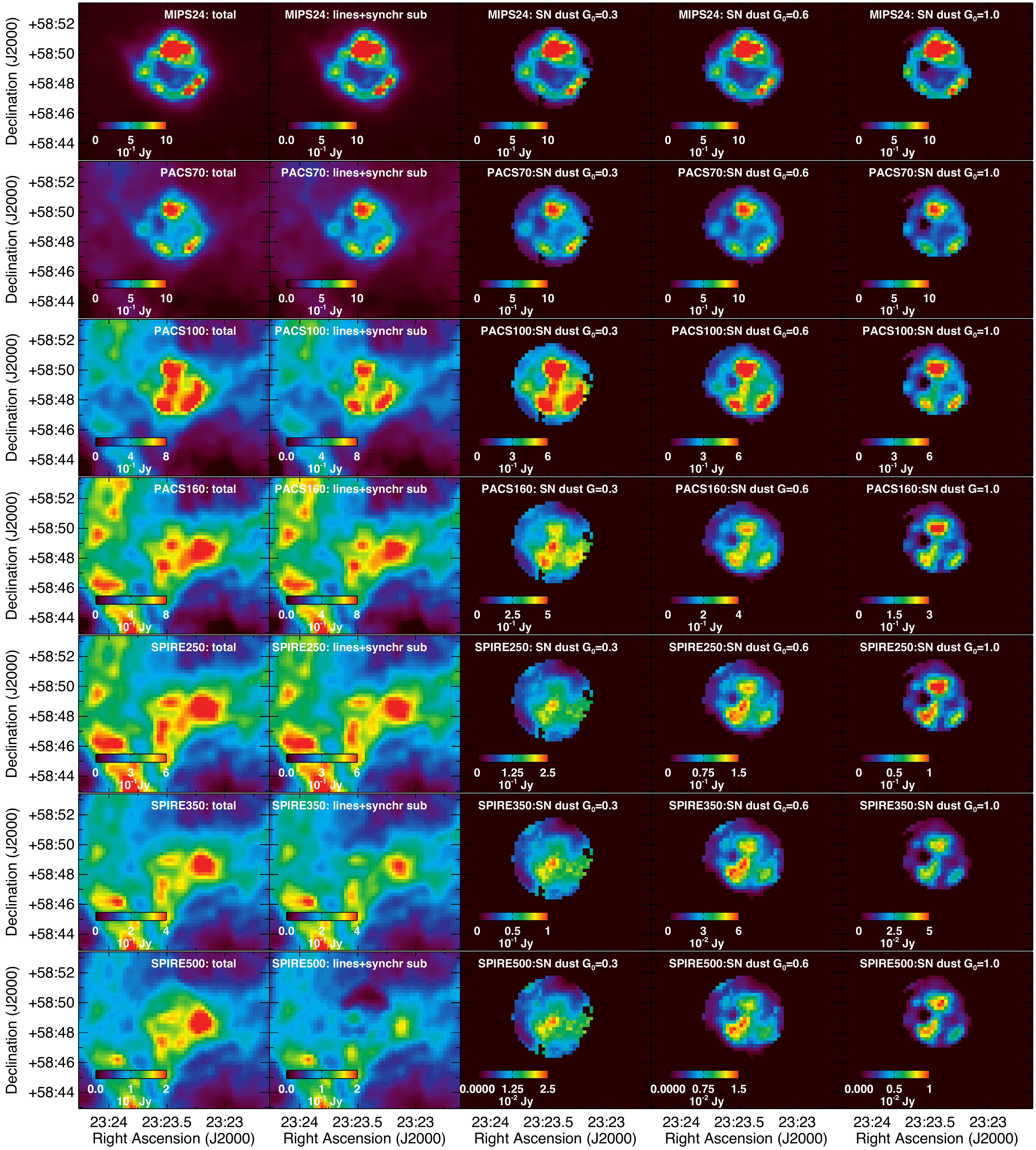}
    \caption{Infrared and sub-millimetre images of Cas\,A for different emission components. From top to bottom, MIPS\,24\,$\mu$m, PACS\,70, 100 and 160\,$\mu$m, SPIRE\,250, 350 and 500\,$\mu$m images are shown. The first column shows total emission maps, the second column shows the same maps after correcting for synchrotron radiation and line emission. The third, fourth and fifth columns show the dust emission intrinsic to the supernova remnant derived after subtracting best fitting models for ISM dust illuminated by radiation fields of G=0.3\,$G_{\text{0}}$, 0.6\,$G_{\text{0}}$ and 1.0\,$G_{\text{0}}$, respectively.} 
    \label{Ima_CasA_comp}
\end{figure*}

Based on the models constructed for each of the emission components, we can determine the relative contribution of various contaminants at every wavelength. Figure \ref{Ima_CasA_schematic} gives a schematic overview of the relative contributions at IR and submm wavelengths. Based on our models for line emission, synchrotron radiation and ISM dust emission, we can furthermore construct residual images with the remaining SN dust emission. The images in Figure \ref{Ima_CasA_comp} (first column) provide an overview of the total emission in the MIPS\,24\,$\mu$m, PACS\,70, 100 and 160\,$\mu$m, and SPIRE\,250, 350 and 500\,$\mu$m wavebands. The second column shows the same maps after subtraction of the contributions from line emission and synchrotron radiation, while the third, fourth and fifth columns show the emission intrinsic to the supernova remnant derived from modelling the residual SED with a multi-component SN+ISM dust model for ISRF scaling factors of 0.3\,$G_{\text{0}}$, 0.6\,$G_{\text{0}}$ and 1.0\,$G_{\text{0}}$, respectively. 

At mid-infrared wavelengths, the largest contributor to the broadband emission is from warm dust in the supernova remnant, with contributions of 92$\%$, 97$\%$, 75$\%$ and 94$\%$ in the IRS\,17\,$\mu$m, \textit{WISE}\,22\,$\mu$m, MIPS\,24\,$\mu$m and IRS\,32\,$\mu$m wavebands, respectively. In the PACS\,70\,$\mu$m waveband, the emission is still dominated by warm SN dust emission (84\,$\%$), while the PACS\,100\,$\mu$m waveband has a non-negligible contribution from ISM dust emission (34$\%$) in addition to a large SN dust contribution (54\,$\%$). With the emission of ISM dust peaking just longwards of 100\,$\mu$m (see Figure \ref{Ima_SED_Jones}), the contribution from SN dust emission drops significantly in the PACS\,160\,$\mu$m and SPIRE wavebands. For ISM dust illuminated by a radiation field with G=0.6\,$G_{\text{0}}$, the emission of dust intrinsic to Cas\,A contributes 30$\%$, 15$\%$, 10$\%$ and 4$\%$ of the total emission in the PACS\,160\,$\mu$m and SPIRE\,250, 350 and 500\,$\mu$m maps, respectively. In the PACS\,100 and 160\,$\mu$m maps, we can identify an emission peak in the inner regions of the remnant in addition to the three peaks of SN dust emission in the reverse shock regions. In the SPIRE wavebands, this central peak becomes more prominent compared to the reverse shock region. This SN dust component in the inner regions was already identified by \citet{2010A&A...518L.138B} and \citet{2014ApJ...786...55A}. The same emission pattern identified in the PACS wavebands  (i.e., one central peak and three secondary peaks in the reverse shock) is prominently present in the SN dust emission maps derived for an ISRF field G=0.6\,$G_{\text{0}}$. We argue that this supports the assumption that the G=0.6\,$G_{\text{0}}$ ISM model best represents the ISM conditions towards Cas\,A.

 \citet{2010A&A...518L.138B} performed a similar decomposition of the different emission components to the \textit{Herschel} PACS and SPIRE wavebands, using an earlier version of the \texttt{HIPE} pipeline to reduce the \textit{Herschel} data. Although their assumptions when modelling the synchrotron, line and interstellar dust emission were different\footnote{\citet{2010A&A...518L.138B} modelled the synchrotron emission based on a VLA 6\,cm image and a spectral index of -0.70. The ISM dust emission is estimated based on average PACS\,100/160 and PACS\,70/160 colours in bright regions of the ISM and used to scale the PACS\,160\,$\mu$m image to subtract the ISM dust contribution at 70\,$\mu$m, 100\,$\mu$m and submm wavelengths.}, the contributions of the different emission sources at various wavelengths are similar. We find a contribution from synchrotron radiation of 3.6$\%$, 3.5$\%$, 4.7$\%$, 7.9$\%$, 16.3$\%$, and 39.7$\%$ in the PACS\,70, 100, 160\,$\mu$m and SPIRE\,250, 350 and 500\,$\mu$m bands, while \citet{2010A&A...518L.138B} reported contributions of 3.7$\%$, 4.2$\%$, 6.8$\%$, 9.2$\%$, 21.1$\%$, and 47.9$\%$ in those wavebands, respectively. While our adopted spectral index (-0.644) is even shallower compared to their value of -0.70, their somewhat higher synchrotron contribution most likely arises from their extrapolation of the emission at 6 cm. The normalisation of our synchrotron component is based on recent \textit{Planck} measurements, which are closer in wavelength to the \textit{Herschel} data points. The ISM dust contributions of 10.7$\%$, 39.6$\%$, 74.1$\%$, 83.9$\%$, 75.0$\%$, and 52.9$\%$ in those wavebands estimated by \citet{2010A&A...518L.138B} are consistent within 5-10$\%$ to our estimated ISM contributions for a radiation field with scaling factor G=0.6\,$G_{\text{0}}$ (12.2$\%$, 34.4$\%$, 70.2$\%$, 76.4$\%$, 77.6$\%$, and 59.3$\%$ at 70, 100, 160, 250, 350 and 500\,$\mu$m, respectively). The contributions estimated by \citet{2010A&A...518L.138B} for the SN dust emission\footnote{We summed the flux from the warm and cool SN dust components of \citet{2010A&A...518L.138B} to derive their total SN dust emission for Cas\,A at each wavelength.} at 70\,$\mu$m (85.8$\%$), 100\,$\mu$m (47.9$\%$) and 160\,$\mu$m (19.3$\%$), and at 250\,$\mu$m (6.9$\%$), 350\,$\mu$m (3.9$\%$) and 500\,$\mu$m (0$\%$) are within the range of percentages derived here from our spatially resolved decomposition analysis (83.6, 53.9, 29.6, 14.9, 9.7, 4.4\,$\%$, respectively). Our supernova dust contributions are consistently higher by 4 to 10\,$\%$ in the PACS\,100 and 160\,$\mu$m and SPIRE\,250, 350 and 500\,$\mu$m wavebands, due to our lower estimates for the synchrotron and ISM dust emission at submm wavelengths. 

\subsection{A simulated 850\,$\mu$m map}
\label{Discuss_850mu}

%Figure 120 
\begin{figure*}
	\includegraphics[width=18cm]{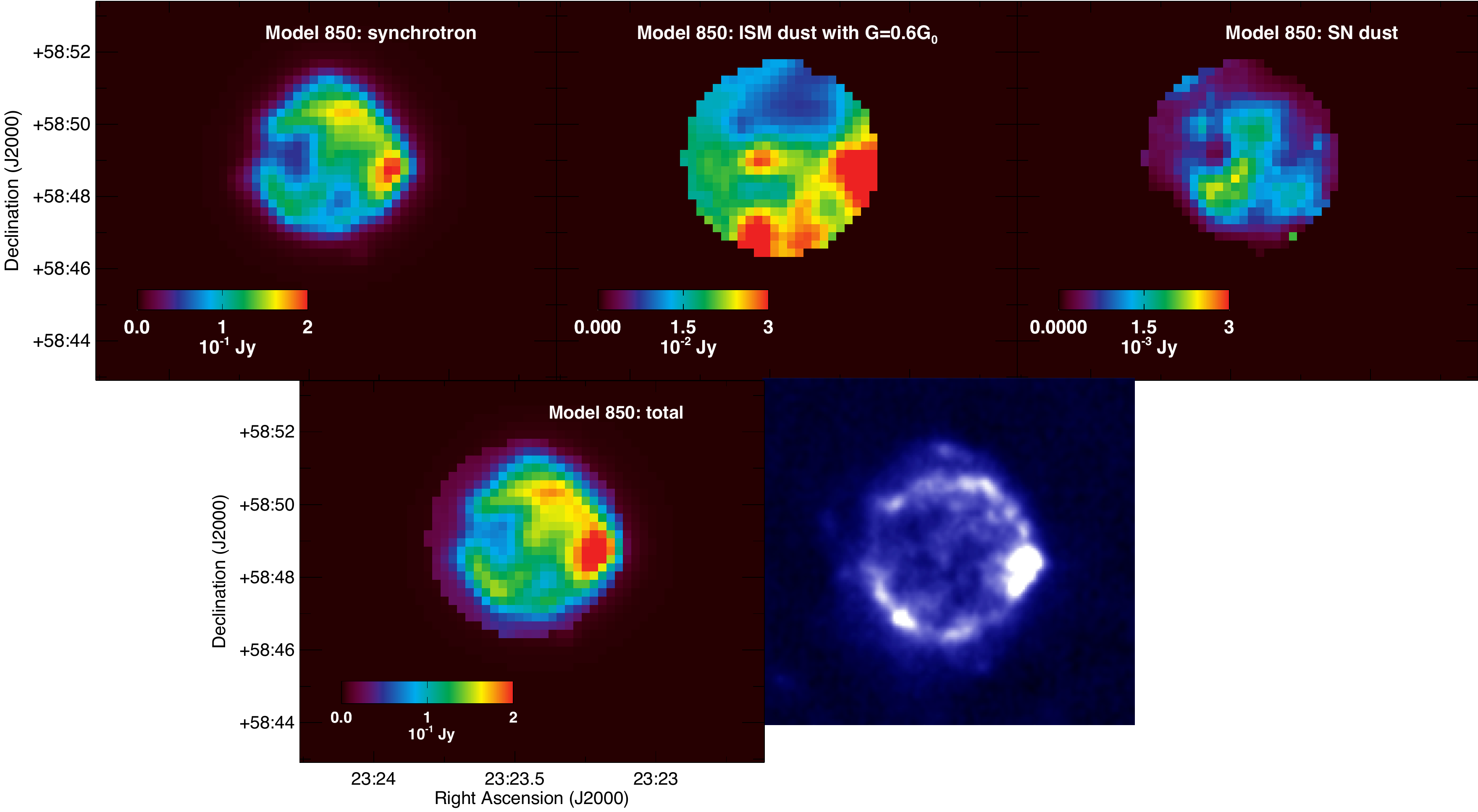}
    \caption{Top panels, from left to right: the modelled synchrotron radiation, the ISM emission and the SN dust emission at 850\,$\mu$m, based on our best fitting SED model for an ISM dust model illuminated by a radiation field $G$ = 0.6\,$G_{\text{0}}$ and a SN dust composition of Mg$_{0.7}$SiO$_{2.7}$ (hot component) + MgSiO$_{3}$ (warm and cold components). The bottom panels shows the total model emission at 850\,$\mu$m (left), compared to the SCUBA\,850\,$\mu$m image published by \citet{2003Natur.424..285D} (see their Figure 1).}
    \label{Fig_Simu_850mu}
\end{figure*}

By extrapolating the best fitting multi-component SED model to longer wavelengths and convolving the model with the SCUBA 850\,$\mu$m filter response curve, we are able to produce a modelled image of the ISM and SN dust emission at 850\,$\mu$m. To this image, we add the modelled synchrotron radiation (using the coefficients derived in Section \ref{Sect_synchr}) to simulate a predicted total 850\,$\mu$m image based on the models derived in this paper. Figure \ref{Fig_Simu_850mu} presents the modelled synchrotron radiation, ISM and SN dust emission at 850\,$\mu$m (top row). The bottom left panel shows the sum of all model contributions at 850\,$\mu$m, and can directly be compared to the observed SCUBA\,850\,$\mu$m image (bottom right panel). Given that our model resolution of 36$\arcsec$ is coarser compared to the resolution of the SCUBA 850\,$\mu$m image (15$\arcsec$), our model is not sensitive to all the structures detected in the SCUBA 850\,$\mu$m image. The overall emission features in our simulated SCUBA 850\,$\mu$m map compare well to the observed SCUBA 850\,$\mu$m map, with both showing a clear emission peak in the West and secondary peaks in the North-West and South-East.

Although the modelled and observed SCUBA 850\,$\mu$m images show a great resemblance, our global model flux at 850\,$\mu$m (measured within an aperture with radius of 165$\arcsec$), $F_{\text{model,850}}$ = 41.3$\pm$9.9 Jy\footnote{The estimated uncertainty on the total modelled 850\,$\mu$m emission is derived from the combined uncertainty on the synchrotron model flux and the errors on the relative contribution of ISM and SN dust emission at 850\,$\mu$m.} is lower compared to the SCUBA 850\,$\mu$m flux measurement $F_{\text{obs,850}}$ = 50.8$\pm$5.6 Jy \citep{2003Natur.424..285D}, but still consistent within the large error bars due to the uncertainty on the exact synchrotron contribution at 850\,$\mu$m. Our estimated contribution of synchrotron radiation (32.3$\pm$8.1 Jy) is similar to the values used by \citet{2003Natur.424..285D} (34.9 Jy) and \citet{2009MNRAS.394.1307D} (30.7 Jy). Our current model predicts that the ISM dust emission contributes about 8.6$\pm$5.7\,Jy, with only 0.4\,$\pm$0.1\,Jy originating from SN dust. The latter ISM and SN dust contributions are incompatible with the lower limit of 6.0\,Jy for the SN dust emission from Cas\,A (and thus upper limit for the 850\,$\mu$m ISM dust flux of 14.1\,Jy) derived by \citet{2009MNRAS.394.1307D} based on submm polarimetry. The high level of polarisation (30$\%$) of SN dust found by \citet{2009MNRAS.394.1307D} seems hard to explain in the framework of our model with a SN dust flux of only 0.4\,Jy. If the polarised emission would instead come from ISM dust, it surpasses the average intrinsic dust polarisation of 12$\%$ measured by \citet{2016arXiv160401029P}. The alignment of dust grains in the ISM surrounding Cas\,A might be stronger than usually found in the diffuse ISM, which might provide a possible explanation for the high degree of polarisation. The extrapolation of our model is consistent with the conclusion from \citet{2004Natur.432..596K} that most of the emission at 850\,$\mu$m arises from ISM dust emission after subtraction of the synchrotron component. 

Our current model for synchrotron+ISM+SN dust emission at 850\,$\mu$m does show a 9.5\,Jy deficit compared to the SCUBA 850\,$\mu$m flux (50.8$\pm$5.6\,Jy, \citealt{2003Natur.424..285D}) and \textit{Planck} 850\,$\mu$m flux (52$\pm$7\,Jy, \citealt{2016A&A...586A.134P}). Given that we already derive a large SN dust mass (0.4-0.7\,M$_{\odot}$) making a 0.4\,Jy flux contribution at 850\,$\mu$m, it seems highly unlikely that the model SN dust flux could be increased without invoking unrealistically high SN dust masses. We furthermore reject a scenario in which the offset at 850\,$\mu$m is caused by model deviations from the Galactic ISM dust emission, given that the small relative offsets between the best fitting ISM dust models and the observed PACS 100 and 160\,$\mu$m and SPIRE 250, 350 and 500\,$\mu$m data points are within the observational uncertainties. We thus argue that the missing emission at 850\,$\mu$m most likely arises from synchrotron radiation. Using a power law index of -0.54 determined based only on \textit{Planck} data (see Section \ref{Sect_synchr}) would be sufficient to increase the synchrotron radiation at 850\,$\mu$m to 37.6\,Jy. The updated model for a MgSiO$_{3}$ warm+cold SN dust composition and ISM model with $G$=0.6\,$G_{\text{0}}$ predicts 850\,$\mu$m contributions of 7.9\,Jy and 0.5\,Jy from ISM and SN dust emission, which would result in a total modelled flux at 850\,$\mu$m of 46\,Jy, consistent with the observed SCUBA\,850\,$\mu$m (50.6$\pm$5.6\,Jy) and \textit{Planck} 850\,$\mu$m (52$\pm$7\,Jy) fluxes. A small change in spectral index thus enables one to account for the 850\,$\mu$m deficit without affecting the main results of the paper.

\subsection{An interstellar extinction map}
\label{AVmodel}
Based on the maps of the total SN dust mass and ISM dust mass obtained from the spatially resolved SED fitting in Section \ref{DustSED}, we can infer the spatial variation in optical extinction towards Cas\,A. For a foreground dust screen model, the $V$ band attenuation is calculated based on the average extinction to gas column-density ratio in the solar neighbourhood, $A_{\text{V}}$/$N_{\text{H}}$ = 6.53$\times$10$^{-22}$ mag cm$^{2}$ \citep{2012A&A...543A.103P}, and a dust-to-hydrogen ratio $\Sigma M_{\text{d}}$/($N_{\text{H}}$$m_{\text{H}}$) = 0.0056 (derived from the \citealt{2013A&A...558A..62J} ISM dust model), which results in 
\begin{equation}
A_{\text{V}}~=~1.46 \frac{\Sigma M_{\text{d}}}{10^{5} M_{\odot} {\rm kpc}^{-2}} {\rm mag}.
\end{equation}

\begin{figure*}
	\includegraphics[width=18cm]{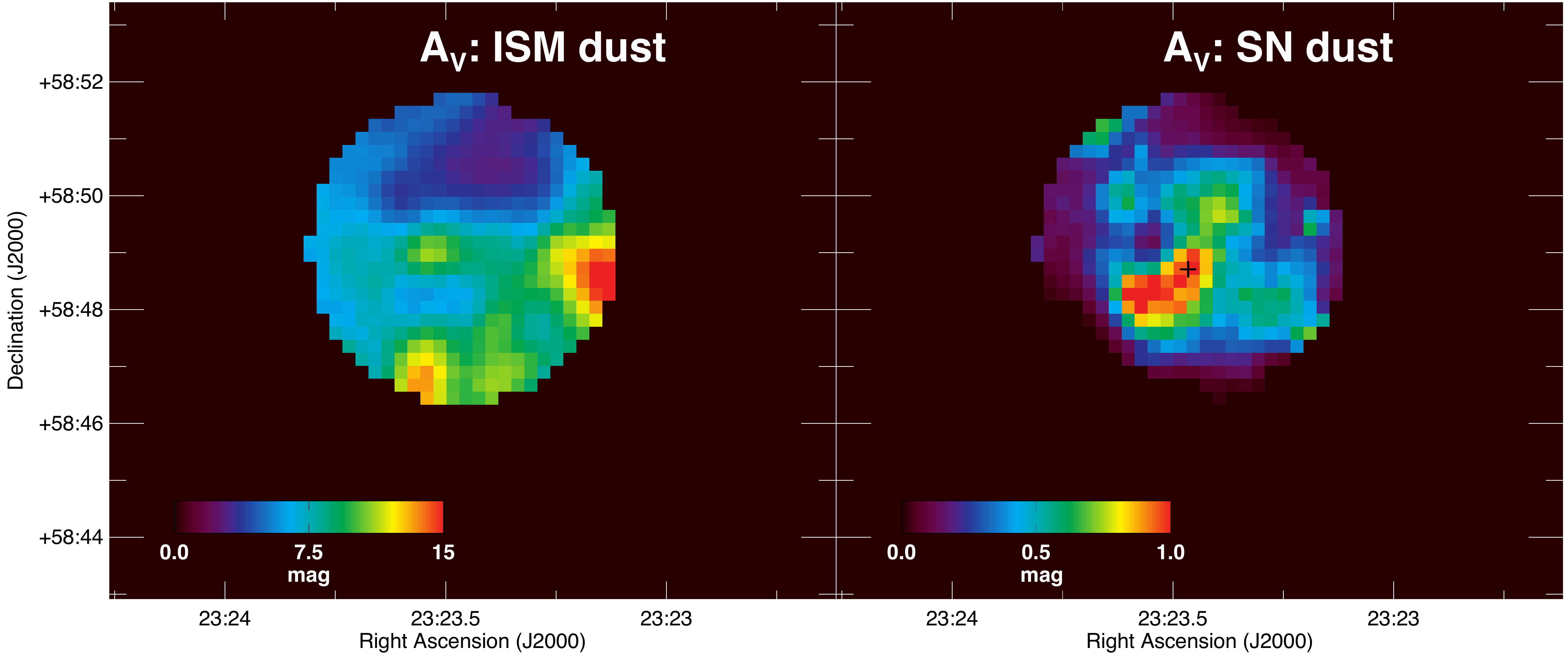}
    \caption{Map of the visual extinction, $A_{\text{V}}$, for the ISM dust projected against Cas\,A (left) and for SN dust (right), derived from the multi-component SED modelling presented in Section \ref{DustSED}. The black cross in the right panel indicates the position of the remnant of the supernova.}
    \label{Ima_CasA_AV}
\end{figure*}

Figure \ref{Ima_CasA_AV} presents the $A_{\text{V}}$ maps for interstellar and SN dust extinction. The position of the point-like X-ray source identified by \citet{2000ApJ...531L..53P} is indicated with a black cross in the right panel. The interstellar visual extinction varies from $A_{\text{V}}$~=~3\,mag in the north-west of Cas\,A, where the extinction is lowest, to peaks of $A_{\text{V}}$~=~15\,mag in the south-east and west, with an average $A_{\text{V}}$ between 6 and 8\,mag. These values are consistent with the visual extinctions derived by \citet{1996ApJ...469..246H} ($A_{\text{V}}$ $\sim$ 4-6\,mag) and \citet{2002ApJ...575..871R} ($A_{\text{V}}$ up to 8\,mag) based on optical and near-infrared spectra, and H$_{2}$CO absorption, respectively. While the H$_{2}$CO-derived extinction maps have a resolution of 6$\arcsec$ (i.e., six times better compared to the SPIRE\,500\,$\mu$m resolution with FWHM $\sim$ 36$\arcsec$), our interstellar extinction map shows the same peaks in $A_{\text{V}}$ towards the west, south-east and near the centre of Cas\,A. We are therefore confident that the ISM dust emission was modelled accurately, and the remaining dust emission can be attributed to SN dust emission. Based on our $A_{\text{V}}$ maps for the ISM and SN dust, we can estimate the probability to detect a possible binary companion which was argued to be responsible for the stripping of hydrogen gas from the progenitor of Cas\,A during the evolved stellar phase \citep{2006ApJ...640..891Y}. The high levels of visual extinction ($A_{\text{V}}$ $\sim$ 8-10\,mag) towards the central regions make it very difficult to detect the supernova remnant or the presence of a binary star at ultraviolet and optical wavelengths but near-IR imaging could be more effective.

\begin{figure}
	\includegraphics[width=8.5cm]{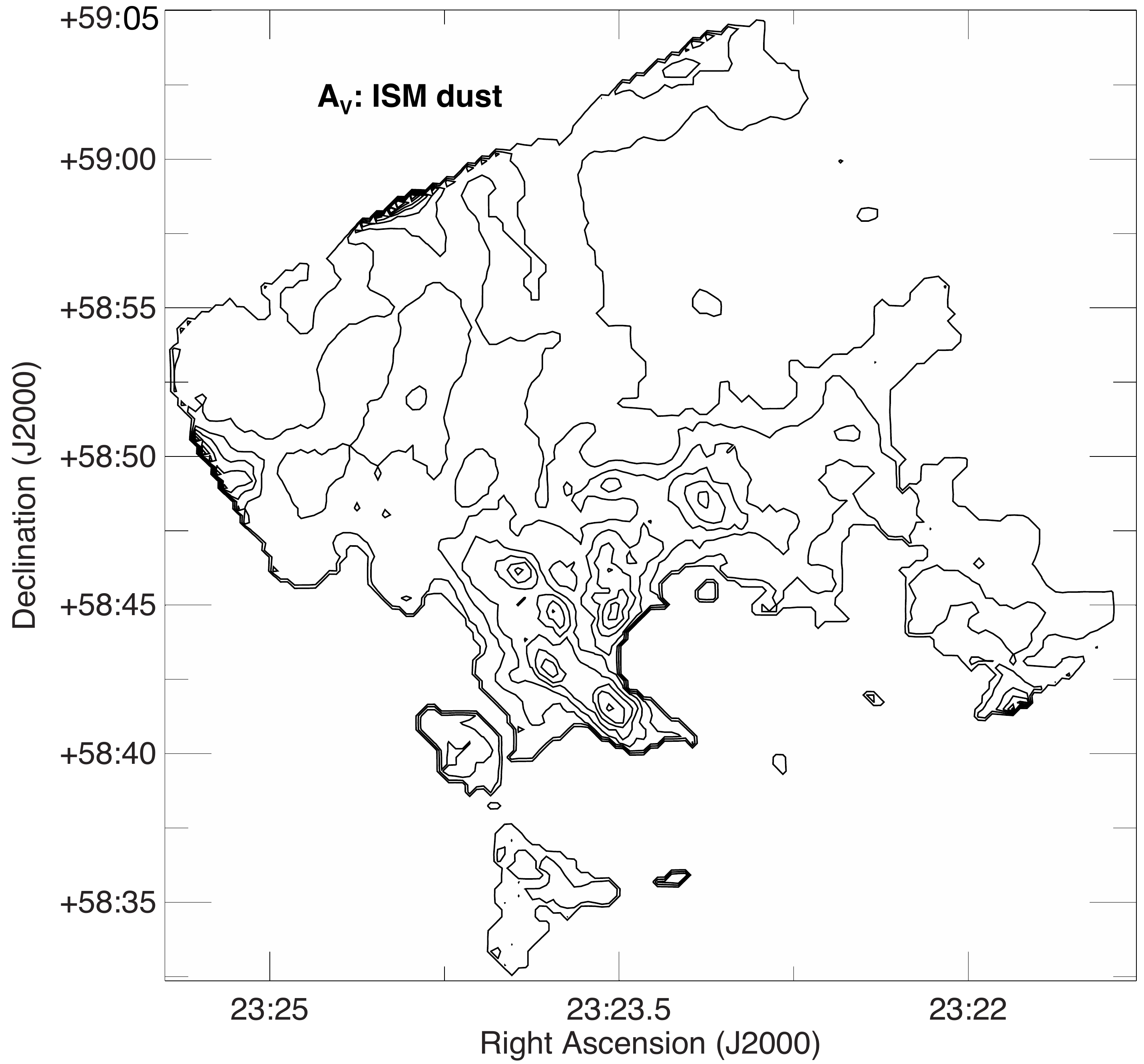}
    \caption{Contour map of the visual extinction, $A_{\text{V}}$, due to ISM dust for the entire field observed with \textit{Herschel} PACS and SPIRE instruments, as derived from the SED modelling of the ISM dust emission presented in Section \ref{Sect_ISdust} and the multi-component SED modelling presented in Section \ref{DustSED}. The contour levels start at $A_{\text{V}}$=2.5 increasing in steps of 2.5 up to $A_{\text{V}}$=20.}
    \label{Ima_CasA_AV_all}
\end{figure}

Figure \ref{Ima_CasA_AV_all} presents a contour map of the visual extinction derived for the entire field of Cas\,A observed with \textit{Herschel}, where the results of the multi-component SED fitting along the line of sight of Cas\,A (see Section \ref{Sect_ISdust}) have been combined with the SED modelling of the ISM emission in the field of Cas\,A (see Appendix \ref{Sect_ISdust}). The regions without any $A_{\text{V}}$ contours correspond to regions with insufficient signal-to-noise detections in the \textit{Herschel} images, which prevented us from modelling the ISM dust emission spectrum. Across the mapped area of 0.5$^{\circ}$, we identify regions with visual extinctions ranging from  $A_{\text{V}}$ = 2 to 20\,mag (the lower values are limited by the sensitivity of our \textit{Herschel} observations). The latter $A_{\text{V}}$ variations correspond to column density variations up to a factor of 18, showing that the amount of material that is positioned along different lines of sight can vary significantly on scales of a few pc. We observe a peak in $A_{\text{V}}$ across the Perseus arm which is positioned just south-east of Cas\,A. 

% Don't change these lines
\bsp	% typesetting comment
\label{lastpage}
\end{document}